# Advancing Differential Privacy: Where We Are Now and Future Directions for Real-World Deployment

Rachel Cummings[1] Damien Desfontaines[2] David Evans[3]
Roxana Geambasu[1] Yangsibo Huang[4] Matthew Jagielski[5]
Peter Kairouz[6] Gautam Kamath[7] Sewoong Oh[6,8] Olga Ohrimenko[9]
Nicolas Papernot[10] Ryan Rogers[11] Milan Shen[12] Shuang Song[10]
Weijie Su[13] Andreas Terzis[10] Abhradeep Thakurta[10]
Sergei Vassilvitskii[14] Yu-Xiang Wang[15] Li Xiong[16] Sergey Yekhanin[17]
Da Yu[18] Huanyu Zhang[19] Wanrong Zhang[20]

[1]Columbia University, New York, New York, United States of America,
[2]Tumult Labs, Zurich, Switzerland,
[3]University of Virginia, Charlottesville, Virginia, United States of America,
[4]Princeton University, Princeton, New Jersey, United States of America,
[5]Google Inc, United States of America,
[6]Google Inc., Seattle, Washington, United States of America,
[7]University of Waterloo, Waterloo, Ontario, Canada,
[8]University of Washington, Seattle, Washington, United States of America,
[9]The University of Melbourne, Melbourne, Australia,

[10]Google Inc, Mountain View, California, United States of America,

[11]LinkedIn, Sunnyvale, California, United States of America,

[12]Meta, Menlo Park, California, United States of America,

[13]University of Pennsylvania, Philadelphia, Pennsylvania, United States of America,

[14]Google Inc, New York, New York, United States of America,

[15]University of California Santa Barbara, Santa Barbara, California, United States of America,

[16]Emory University, Atlanta, Georgia, United States of America,

[17]Microsoft, Redmond, Washington, United States of America,

[18]Sun Yat-sen University, Guangzhou, Guangdong, China,

[19]Meta, New York, New York, United States of America,

[20]OpenDP, Harvard University, Cambridge, Massachusetts, United States of America

**ABSTRACT**

In this article, we present a detailed review of current practices and state-of-the-art methodologies in the field of differential privacy (DP), with a focus of advancing DP's deployment in real-world applications. Key points and high-level contents of the article were originated from the discussions from "Differential Privacy (DP): Challenges Towards the Next Frontier," a workshop held in July 2022 with experts from industry, academia, and the public sector seeking answers to broad questions pertaining to privacy and its implications in the design of industry-grade systems.

This article aims to provide a reference point for the algorithmic and design decisions within the realm of privacy, highlighting important challenges and potential research directions. Covering a wide spectrum of topics, this article delves into the infrastructure needs for designing private systems, methods for achieving better privacy/utility trade-offs, performing privacy attacks and auditing, as well as communicating privacy with broader audiences and stakeholders.

# 1. Introduction

Differential privacy (DP) [(Dwork, Kenthapadi, et al., 2006](); [Dwork, McSherry, et al., 2006)]() has become the leading concept for privacy in statistical data analysis and machine learning applications over the past decade. It is also frequently termed 'formal privacy' due to its robust theoretical underpinnings, which establish a broad and profound basis for research and development across various applications. Although DP has been widely adopted in academia, public services [(Abowd et al., 2022](); [Abowd, 2018)](), and several industrial deployments more recently [(Apple Differential Privacy Team, 2017](); [B. Ding et al., 2017](); [Erlingsson et al., 2014](); [Hartmann & Kairouz, 2023](); [Kairouz, McMahan, Song, et al., 2021](); [Rogers et al., 2021)](), it has yet to become a standard for data sharing in companies or institutions where privacy protection is of utmost importance. This raises two fundamental questions:

1. What are the challenges that institutions face when adopting or attempting to adopt DP, and how can they best be mitigated?
2. What new challenges should we expect as we move toward the next frontier, where DP is a mature technology, and adoption is growing?

In July 2022, we hosted a workshop titled "Differential Privacy (DP): Challenges towards the Next Frontier" with experts from industry, academia, and the public sector to discuss and find solutions to the challenges of differential privacy. This article is revised from a document that is the only publicly available summary of the discussions that took place during the workshop, which represents the collective views of the authors mentioned in the [acknowledgments section](#) and does not reflect the personal opinions of any individual.

We aspire for this article, with its in-depth overview and discussions, to furnish academia with innovative research ideas while simultaneously offering industry practitioners and adopters valuable references for best practices. In other words, we hope this document can be a useful educational tool, inspiring both the current workforce and the next generation to engage in research and development in the field of privacy. In the following, we provide a brief overview of the topics covered in this article.

**Building privacy infrastructure for DP:** There are many high-profile deployments of DP in industries and in the public sector. However, there are several common challenges that arise during these deployments. These include: a) Deciding on the specific DP definition to use and the threat model to consider. b) Selecting privacy parameters and mechanism details that ensure the DP algorithm provides a reasonable trade-off between privacy and utility. c) Integrating DP mechanisms into existing infrastructure that people are already familiar with (i.e., end-to-end). d) Communicating to the user what DP provides and its effects. e) Ensuring that computation and storage costs do not increase disproportionately compared to the non-DP baseline system. [Section 2](#) discusses each of these challenges and provides research agendas around them. This includes developing methods and tools for selecting privacy parameters in a way that non-experts can easily understand and interact with the system. It also involves exposing well-documented implementations of DP algorithms, along with the chain of trust, for external scrutiny. This could potentially take the form of an open-source release of the implemented algorithm.

**Improving privacy/utility trade-offs:** While DP has been used in several large-scale deployments (mentioned above), many of these deployments have ended up with privacy loss parameters that provide little if any meaningful privacy, and the challenge that comes with maintaining a high level of utility while preserving DP has limited its wider adoption. In the recent past, there have been research efforts to bridge this gap, especially, by using one of the following: a) public data to achieve better utility [(Alon et al., 2019](#); [Bassily, Cheu, et al., 2020](#); [Bassily, Moran, & Nandi, 2020](#); [Bie et al., 2022](#); [(Kairouz, Ribero, et al., 2021)](#); [T. Liu et al., 2021](#); [Zhou et al., 2021)](#), b) designing algorithms that are data-adaptive [(Asi, Duchi, et al., 2021)](#); [Feldman & Steinke, 2018](#); [X. Li, Liu, et al., 2022](#); [Singhal & Steinke, 2021)](#), meaning the algorithms can achieve good utility if the data satisfies some 'nice' properties, c) designing personalized DP learning algorithms, which allows a part of the training procedure to be performed without adding any noise/randomness [(Altschuler & Talwar, 2022)](#); [Jain et al., 2021](#); [Kearns et al., 2014](#); [Y.-X. Wang et al., 2015)](#), and d) designing ML pipelines that are tailored specifically to learning with DP [(Iyengar et al., 2019)](#); [Jayaraman et al., 2018](#); [Papernot et al., 2017, 2021](#); [D. Yu, Zhang, Chen, & Liu, 2021)](#). Later in the document, we delve deeper into each of these aspects. When it comes to assessing privacy-utility trade-offs, note that sometimes there do not exist non-DP utility baselines to compare against. This is because certain data analyses are simply not permitted without privacy [(Dwork & Ullman, 2018)](#). In these cases, the ability to provide DP guarantees enables novel analyses—such as the application of ML—in settings where it was previously impossible. Later in [Section 3](), we explore each of these challenges, and discuss potential research questions around them. Along with addressing the challenges mentioned here, one thing that will significantly facilitate progress is the design of benchmark experiments [(Tramèr, Kamath, & Carlini, 2022)](). We will briefly touch on this aspect later in the document.

**Attacks and auditing for privacy protection:** Achieving a theoretical DP bound has its strong benefits but also presents certain limitations. In [Section 4]() we discuss in detail these limitations and the motivations and goals for developing practical privacy attacks and auditing techniques to complement DP guarantees. Currently, most of these auditing methods are based on simulating attacks. These methods can provide lower bounds on the information disclosed by the system [(Zanella-Béguelin et al., 2022)](). DP lower bounds achieved via the lines of state-of-the-art privacy attacks can provide guidance on the acceptable DP parameters, which may be significantly better than what is currently provable. On the other hand, they are lower bounds, so they do not offer any guarantees when it comes to bounding the worst-case privacy risk faced by individuals participating in such machine learning (ML) pipelines. Hence, one has to be careful in interpreting the results of such audits, that is, the attack failure rate does not say anything about the susceptibility of the system to other attacks. We also provide examples from the literature to outline the open challenges in large-scale systems and several research directions.

**Communicating DP guarantees:** In this article, we do not focus on policy aspects of privacy protection, but briefly mention here some of the interesting questions that came up during the workshop in the context of comparing various privacy notions alongside DP. First, it is a well-accepted fact that privacy is a multifaceted quantity, of which DP quantifies only a part. For example, methods such as data minimization, transparency to users on how the data is being used, and data anonymization capture critical aspects of privacy protection of a system that may be complementary to the protection afforded by DP. When designing a privacy-preserving system, and comparing it against other systems, it is important (and often legally required) to disclose what data is being collected and how it will be used. With privacy attacks and auditing measures described below, it will also hopefully help compare currently incomparable syntactic privacy approaches like $k$-anonymity [(Samarati & Sweeney, 1998)](), $\ell$-diversity [(Machanavajjhala et al., 2007)](), and semantic privacy notions like DP. An aspect one should be careful with defining and standardizing how privacy is documented and communicated, is that they should be composable across various sensitive data pipelines, in the traditional sense of composition from the cryptography literature. In [Section 5](), we further specify information about DP mechanisms that can be captured when communicating the guarantees of the system relying on DP to end-users and other stakeholders (e.g., local vs. centralized DP and the granularity of protection). We outline questions that need to be addressed when integrating this information in an end-to-end system, including interaction with other protection methods (e.g., anonymization) and assumptions (e.g., attacker's knowledge), composability, and how these guarantees can address regulation/policy notions of privacy.

**A non-exhaustive exposition to DP Deployments:** Most of the challenges/questions discussed in this document stem from real-world use cases of DP. Desfontaines (2021) provides a (non-exhaustive) survey of publicly documented deployments of DP (specifically from Google, Apple, U.S. Census Bureau, Microsoft, Meta, and LinkedIn). We encourage the readers to follow Desfontaines (2021), and the citations there for the details.

## 2. Building DP Tooling and Infrastructure

For any technology to be widely adopted, it must have a robust infrastructure that provides reasonable solutions to various use cases. Differential privacy is no exception. Multiple existing deployments have used open-source DP tools to generate aggregate statistics or synthetic data, and we can expect even more deployments once infrastructure improvements are made. This section outlines some of the challenges we have seen while deploying DP. We propose a series of desiderata for DP infrastructure and suggest open research directions that could lead to concrete improvements in tooling for implementing differentially private systems.

To clarify, we will use the terms 'user' and 'adopter' to differentiate between the subject who owns or contributes data for which we aim to provide privacy protection and the adoption, development, and usage of the privacy-preserving system.

### 2.1. Challenges in Deploying DP

In this section, we summarize a few common challenges we have encountered as applied researchers, developers, or designers of the systems when DP is deployed in real applications.

**Understanding the problem:** Many organizations and individuals face complex data privacy concerns and may require privacy technology to address them. However, they often need help determining which approach is best suited for their specific problem. DP practitioners are familiar with the issues that DP can resolve, such as releasing aggregate statistics on sensitive data, providing internal access to trusted analysts to run queries on sensitive data, building and deploying ML models trained on sensitive data, and safely collecting telemetry data. Nonetheless, developers, policymakers, and business leaders who need to decide what to do often need help understanding the differences in assumptions and threat models between these use cases.

Adding to the complexity, some specific data analyses, such as outlier detection, are incompatible with DP. Other applications, such as location-based services (e.g., contact tracing for pandemic prevention), require 'precise' *individual* locations instead of collecting aggregate statistics from mobility data. Furthermore, some applications used in economics currently do not have a differentially private method. At the same time, data analytics requires accurate aggregate information such as frequency and mobility patterns. Consequently, having comprehensive definitions of DP (e.g., local differential privacy [local DP], geo-indistinguishability, and variants [Andrés et al., 2013; Xiao & Xiong, 2015]) and specialized perturbation mechanisms are necessary to ensure meaningful privacy and utility guarantees (H. Wang et al., 2022). Potential DP adopters may require assistance in understanding these nuances, mapping their threat model to a specific approach, and choosing the appropriate tooling for their use case.

**Demonstrating the privacy risks of ad hoc approaches: Data** privacy researchers and practitioners have an in-depth understanding of how seemingly harmless statistics computed on a data set or released models can be used to reconstruct individual records. However, the risks associated with such data leaks are not yet widely recognized or understood. Many people believe that removing personally identifiable information (PII) is enough to prevent the leakage of private data. Privacy regulations, such as the U.S. medical privacy Health Insurance Portability and Accountability Act (HIPAA) law codify such assumptions, leading to prevalent ad hoc masking techniques in various industries. Therefore, privacy advocates often have to demonstrate the need for DP by showing that alternative approaches may not be as safe as assumed. However, it can be challenging since adopting DP can lead to worse utility for the released data. This motivates increased research on auditing of privacy leakage via privacy attacks, which we discuss in Section 4. It also encourages DP practitioners to highlight cases where DP unlocks previously inaccessible data, as these demonstrate the opportunity for privacy technologies to enable new applications, not just provide new privacy-utility trade-offs for existing ones.

**Mapping the threat model:** In order to establish a reliable privacy infrastructure, it is imperative to consider trust assumptions carefully, which holds for DP systems as well. Before deploying DP systems, it is essential to address several questions, such as: should we place trust in a central server (Boenisch et al., 2021; Fowl et al., 2021; L. Zhu et al., 2019), how strong an adversary can potentially be, or what kind of data requires protection? It is worth noting that threat vectors may only sometimes be apparent and can come from different sources, such as government actors who may demand existing data or compel organizations to implement new technical measures. Furthermore, insider risk is also a relevant factor, and mitigating it requires additional protection mechanisms.

**Communicating the impact of noise:** When a system deliberately adds random noise to results to satisfy DP, it can be challenging for non-experts to understand its reasoning.[1] However, it can also lead to some unexpected outcomes. For instance, adding noise to a count can generate negative values, and noise may create results not present in the original data set, or some outcomes may be excluded due to counts not reaching a certain threshold. In machine learning with DP, a reduction in model accuracy is expected when compared to learning without DP noise. While some use cases can tolerate a certain level of noise, incorporating this concept can be difficult as it may not align with the goals of some systems. It is crucial to distinguish this aspect of data privacy from typical security practices, where data utility is preserved but restricted to a limited group of authorized developers.

It is, therefore, imperative to have utility metrics to map out the trade-off between privacy and utility. All stakeholders should be involved in determining appropriate noise levels and the overall privacy loss levels. It is essential to note that despite dedicating additional resources to privacy, there will still be a loss in utility compared to not implementing DP. Further discussion on this topic can be found in Section 5.2.

**Controlling computational costs:** DP can sometimes lead to significant computational overhead, which can add complex engineering requirements to a product. This is especially true for machine learning use cases. For instance, the DP-Stochastic gradient descent (DP-SGD) algorithm—which is commonly used for training ML models with DP guarantees (Abadi et al., 2016; Bassily et al., 2014; Song et al., 2013)—includes a clipping step where individual per-example gradients are clipped to have a predefined maximum norm. However, if this step is implemented naively (Hoory et al., 2021), it prevents DP-SGD from leveraging the acceleration opportunities provided by specialized hardware like GPUs or TPUs.[2] This is because averaging the gradient of multiple training examples must happen after the clipping operation, and minibatch sizes can no longer be adjusted to maximally use available hardware. Nonetheless, it is possible to design implementations that work around these limitations with additional work and expertise (Goodfellow, 2015; He et al., 2023; J. Lee & Kifer, 2021; X. Li et al., 2021).

**Enabling exploratory data analysis:** Statisticians, data scientists, and other researchers often want to perform exploratory data analysis to understand the data that they are working with, and choose which analyses to perform or which statistics to release. This typically involves practices like looking at small snippets of the data, which are not straightforward, to adapt to DP as it only naturally facilitates exposing frequent items (Gopi et al., 2020) such as n-grams (Kim et al., 2021) and top-k results (Durfee & Rogers, 2019). If the data scientist does not have direct access to the data, this can lead to a bootstrapping issue: even though they could gain access to it via a DP interface, they cannot determine which queries to run. There is some work going in the direction of designing DP tools for exploratory data analysis (Dwork & Ullman, 2018; Nuñez von Voigt et al., 2020), and existing interfaces like private data sharing interface (PSI) (Gaboardi, Honaker, et al., 2016) could be a good starting point to build tooling for this use case. Further, differentially private synthetic data generation techniques (Jordon et al., 2019; D. Lee et al., 2020; H. Wang, Zhang, et al., 2023; Xie et al., 2018; J. Zhang et al., 2017) could also partially address this problem. But this remains a largely unaddressed problem space.

## 2.2. Desiderata for DP Infrastructure

Learning from the above challenges, we enumerate essential design criteria to develop DP infrastructure that addresses adopter needs and contributes to practical deployments.

**Focus on usability:** To achieve widespread adoption and effectively address real-world use cases, it is crucial to design DP infrastructure with end-users in mind, namely, developers and data analysts. The initial step toward this is understanding who the infrastructure is meant for, their objectives, what they require to achieve them, and their prior level of knowledge. It is often better to design with non-experts in mind since existing infrastructure that can only be used by DP experts severely limits its deployment. To be user-friendly, tooling should have clear documentation, simple and elegant interfaces, automatic hyperparameter tuning, smooth onboarding experiences, and tutorials or training videos. From picking up a new tool to creating a first prototype, the process should be as straightforward as possible. Adopters should not have to comprehend the math behind DP or manually select mechanisms or hyperparameters before obtaining their first results.

**Build trust among adopters:** Maintaining trust is crucial for privacy-sensitive software. All the elements involved in the trust chain must be made available as open-source software to achieve this. The community of DP implementers should aim to unify their tools, pool resources to audit and test these tools, and recommend them to potential adopters. Additionally, this community should strive to be a welcoming and inclusive space where new members can ask questions and receive support.

**Support diverse units and levels of privacy:** The DP literature often assumes that every user contributes only one data point to a data set. However, this assumption is not always true. In large data sets that record user interactions with a service, each user can be associated with multiple data points (Amin, Kulesza, et al., 2019; Augenstein et al., 2019; Epasto et al., 2020; Levy et al., 2021; McMahan et al., 2018). Use cases also require different privacy guarantees, such as preaggregated data, social network data where each record has multiple user identifiers, or data about heterogeneous entities that require different privacy levels (Jorgensen et al., 2015; J. Liu et al., 2021; Y. Liu et al., 2023). Moreover, some DP use cases require quantifying the privacy guarantee associated with multiple privacy units simultaneously, like reporting privacy budget values at the user and attribute levels. Therefore, supporting these possible privacy units is essential to address the growing number of potential use cases.

**Build for scale:** Many people interested in using DP want to apply it to large data sets that cannot be accommodated in memory on a single machine. Therefore, general-purpose DP tools should consider this limitation. DP mechanisms should be designed to be massively parallelizable whenever possible (Amin, Gillenwater, et al., 2022), and the overall architecture of DP software should be compatible with deployments on large clusters.

**Design secure software:** Small bugs in DP software can easily compromise the desired guarantees. To mitigate known issues, like floating-point vulnerabilities (Haney et al., 2022; Ilvento, 2020; Jin et al., 2022; Mironov, 2012), production tools should carefully implement primitives. Additionally, if side-channel attacks (Andrysco et al., 2015; Haeberlen et al., 2011; Jin et al., 2022) are part of the threat model, they should also be mitigated, especially for interactive systems. Generally, tooling should follow best practices in secure software design, such as modular design, systematic code reviews, comprehensive test coverage, regular audits, and effective vulnerability management.

**Modular design:** Several DP tools in the market allow for end-to-end solutions, such as IBM's privacy package, PSI, and Privacy Integrated Queries (PINQ). However, separating the DP accounting from the algorithm implementations that add noise would be more beneficial. Issues related to the quality of randomness and floating-point arithmetic should be addressed in the implementation phase, while (autodp contributors, 2019) can be used for algorithm design and prototyping. This approach ensures that the verification of the actual implementation can be performed independently of the verification of the mathematical proof, making the process more efficient.

## 2.3. Research Directions

In this section, we propose research directions to meet adopters' needs for differentially private infrastructure. Many of the directions below can be advanced the most through close collaborations between the industry, academia, and open-sourced communities. Especially for aspects closer to the end-users and adopters, such as benchmarks and user inputs, we encourage institutions in both public and private sectors to contribute more. In the meantime, students and researchers could play a larger role in the theoretical and algorithmic development front.

**Benchmarks:** Benchmarks are a useful way to showcase DP mechanisms' utility and performance to end-users and encourage further research. However, onboarding examples provided with DP tooling are often limited, lacking nuances like privacy budget management and different privacy units. Benchmarks can provide a more comprehensive set of examples for adopters to explore and use for inspiration and encourage the research community to contribute to and improve the existing infrastructure over time. The benchmarks should reflect typical usage patterns of DP tooling and be open to enable reproducible and transparent research.

**Incorporating inputs from users and adopters:** In order to establish a more holistic perspective and define a 'good-enough' threshold for a system, it is important to incorporate inputs from both users and adopters. In real-world applications, despite strong benchmarks on utility-privacy trade-offs being established, it can still be difficult to determine how much privacy and noise is necessary case by case, and there is no easy way to capture the requirements of adopters in terms of utility. Gathering these inputs can help align benchmarks with practical usability and ensure that all perspectives are taken into account.

**Hyperparameter-free mechanisms:** DP algorithms often require hyperparameters to be specified, such as clipping or truncation bounds, granularity levels, and the fraction of privacy budget for each step. These hyperparameters can be a significant hurdle for non-experts using DP tooling and engineers implementing these tools. The primary goal of DP infrastructure is to abstract away some of the complexity of the underlying mechanisms, including the need for manual hyperparameter setting. Therefore, we encourage the authors of newly developed DP mechanisms to suggest good default values for hyperparameters or to develop additional DP algorithms that can automatically and privately determine optimal values for these hyperparameters.

**Visualization and error metrics:** Data curators and practitioners often struggle to quantify the empirical error of the DP mechanisms and choose appropriate strategies and parameters. They need tools that provide helpful error metrics in a visual way that is intuitive and easy to understand. There is no consensus among DP experts regarding which error metrics are the most important or useful to optimize when designing new algorithms.

This leads to error metrics often not the most relevant to practitioners in a significant portion of the DP literature. Therefore, user research focusing on understanding and visualizing error and the uncertainty associated with DP is crucial to bridging the gap between theory and practice.

**Usability research:** DP infrastructure is developed by experts with in-depth knowledge of the fundamental concepts and familiarity with scientific literature. However, for widespread adoption, the resulting tools must be user-friendly for non-experts. To achieve this, tooling authors must conduct regular tests with developers and perform usability research to improve the tools over time and ensure that the assumptions are accurate. Such formal user research studies can help identify potential adopters for DP tooling while highlighting the gaps between adopter needs, feature availability, and usability.

**System design:** Designing systems to support DP data analysis raises several questions. How can we incorporate DP infrastructure into existing data processing systems in a way that is easy to maintain and extend over time? How does privacy budget tracking work across many use cases, which may use different privacy units or privacy analysis techniques? How can DP be integrated with other privacy-enhancing technologies and processes, such as data governance, access control and monitoring, or insider risk measures? Further exploratory research is required to understand the infrastructure requirements for such improvements. Additionally, designing and prototyping such systems is a worthwhile research direction.

**Fully adaptive and concurrent composition:** The topic of composition theorems has been extensively researched in the DP literature. However, most privacy accounting techniques assume the composed mechanisms have a predetermined privacy budget. Recently, new and more complex notions of compositions have been introduced, such as fully adaptive composition (Rogers et al., 2016; Whitehouse et al., 2023), which allows the privacy budget of subsequent queries to depend on the output of previous queries and concurrent composition (Vadhan & Wang, 2021; Vadhan & Zhang, 2022), which enables multiple analysts to interact with a single DP engine in parallel. These theoretical questions directly impact the implementation of privacy accounting techniques in tooling, and there are still many open questions in this area of research.

## 3. Improving Privacy/Utility Trade-offs for DP Algorithms

From a theoretical perspective, there has been a significant amount of research in the field of DP empirical risk minimization (DP-ERM) (Bassily et al., 2014; Chaudhuri et al., 2011; Feldman et al., 2018; Iyengar et al., 2019; Jayaraman et al., 2018; Kairouz, Ribero, et al., 2021; Mangold, Bellet, et al., 2022; D. Wang et al., 2017; Wu et al., 2017; D. Yu et al., 2019) and DP stochastic convex optimization (DP-SCO) (Asi, Duchi, et al., 2021; Asi, Feldman, et al., 2021; Bassily et al., 2019, 2021; Damaskinos et al., 2021; Kamath, Liu, & Zhang, 2022; D. Wang et al., 2020) for convex losses. However, there has been relatively less exploration of nonconvex models, which are more commonly encountered in real-world applications. A few exceptions to this trend include some research in nonconvex models such as D. Wang et al. (2017), Zhou et al. (2020), Q. Zhang et al.

(2021), Arora et al. (2022), Ganesh, Liu, et al. (2023), Gao and Wright (2023). The hope is that theoretical insights in the space of nonconvex models will drive the design of real systems.

On the practical side, there have been recent works exploring the use of public data (Abadi et al., 2016; Cattan et al., 2022; De et al., 2022; Devlin et al., 2019; Kurakin et al., 2022; Y. Liu et al., 2019; Mehta et al., 2022; Papernot et al., 2019; Radford et al., 2019; Tramèr & Boneh, 2021) and adding correlated noise to the DP-SGD training procedure (DP-follow-the-regularized-leader or DP-FTRL) (Kairouz, McMahan, Song, et al., 2021) to improve privacy, utility, and compute trade-offs. These improvements aim to provide a better optimization profile for DP-SGD or improve computational complexity by allowing training over smaller mini-batches without relying on privacy amplification. However, the exploration of taking these algorithmic improvements and the improvements in the usage of public data into the design of the DP infrastructure remains largely unexplored. Some important questions need to be addressed, such as the safety protections required when using public data along the lines of what is mentioned in Tramèr, Kamath, & Carlini (2022), whether all data available on the internet can be considered public, and the units of privacy that should be considered and designed for in the context of privacy protection. For example, in the context of language models, should a token, a paragraph, or the complete interaction of a single user be considered as a single unit for privacy protection (H. Brown et al., 2022)? Furthermore, it is important to extend the current DP guarantees from a single training run to a system-level guarantee for the complete DP infrastructure.

## 3.1. Public Data in Private Data Analysis

DP often results in a significant loss of utility. While some of this loss is inherent, it can be exacerbated by being overly cautious about what data is considered private or not. For instance, suppose one wants to diagnose patients based on a textual description of their symptoms. Some aspects of this data may indeed be privacy-sensitive, such as characteristics of the patient. However, basic aspects like grammar and syntax are not privacy-sensitive. By training a model privately from scratch, we are treating everything as private. To overcome these obstacles, it is possible to use *public data*.

We will refer to 'public data' as data that is not subject to privacy constraints. This type of data can be obtained from users who have opted out of privacy protection or scraped from publicly available sources on the internet. Therefore, the characteristics of a public data set may vary. Some examples include: the data set may be labeled or unlabeled, small or large, and in- or out-of-distribution with respect to the private data set. The most effective method of utilizing public data will naturally depend on these attributes.

**Table 1. Selected summary of work on DP-Machine Learning with public data.[3]**

| Unlabeled | Labeled | In dist. | Out of dist | Pre-trained | Theory | Experiments | Context |
|---|---|---|---|---|---|---|---|
| ● | | N/A | N/A | | | ● | PATE learning [(Papernot et al., 2018)](#) |
| ● | | ● | | N/A | ● | | PATE theory [(Bassily et al., 2018)](#) |
| ● | | ● | | N/A | ● | ● | PATE theory [(C. Liu et al., 2021)](#) |
| ● | | | | N/A | ● | | Semi-private PAC [(Beimel et al., 2013)](#) |
| ● | ● | ● | | N/A | ● | | Semi-private PAC [(Alon et al., 2019)](#) |
| | ● | | ● | N/A | ● | | Mixture Priv/Pub [(Bassily, Moran, & Nandi, 2020)](#) |
| ● | | ● | ● | feature | | ● | Private-kNN [(Y. Zhu et al., 2020)](#) |
| ● | | N/A | N/A | | | ● | Improve DP-SGD [(D. Yu, Zhang, Chen, & Liu, 2021)](#) |
| N/A | N/A | N/A | N/A | finetune | | ● | Finetune NLP [(X. Li et al., 2021)](#) |
| N/A | N/A | N/A | N/A | finetune | | ● | Finetune NLP [(D. Yu et al., 2022)](#) |
|  | ● | ● |  |  | ● | ● | Mirror Descent (Amid et al., 2021) |
|  | ● | ● | ● | finetune | ● | ● | Mixed DP in CV (Golatkar et al., 2022) |
| N/A | N/A | ● | ● | N/A | ● |  | Mean estimation (Bie et al., 2022) |
| N/A | N/A | ● |  | N/A | ● |  | Query Release (Bassily, Cheu, et al., 2020) |
| N/A | N/A | ● | ● | N/A | ● | ● | Query Release (T. Liu et al., 2021) |

Whenever public data is labeled and in-distribution, the size of private data is much larger than that of public data, or the problem is trivial from a theory perspective (that is, training on only public data works nearly optimally). On the other hand, when the public data are not labeled, it could be either big or small (unspecified). If a paper fine-tunes from a pretrained foundation model, we do not consider it to be using out-of-distribution public data. In Table 1 above, any bullet with out-of-distribution either empirically evaluates the method under out-of-distribution public data or theoretically studies it. The 'theory' column indicates whether a characterization of the privacy-utility trade-off is provided. All DP papers, by default, have a theory for the privacy part, thus not separated categorized.

### 3.1.1. Public Pretraining and Private Finetuning

If there is a large amount of public data available, it can be used to pretrain a model. This pretrained model can then be fine-tuned using a private optimizer on a sensitive data set. This approach, called transfer learning, has been very effective in the nonprivate setting. It has given rise to the concept of *foundation models* (Bommasani et al., 2021), which are large pretrained models that can be adapted to various downstream tasks. Although this approach requires a significant amount of data, the data may be out-of-distribution or even unlabeled, which can be handled through self-supervised learning (X. Li, Tramèr, Kulkarni, & Hashimoto, 2022). Some works have privately fine-tuned large language models like BERT (Devlin et al., 2019), RoBERTa (Y. Liu et al., 2019), and GPT2 (Radford et al., 2019), achieving only modest drops in utility compared to the nonprivate setting (X. Li, Tramèr, Liang, & Hashimoto, 2022; D. Yu, Zhang, Chen, Yin, & Liu, 2021; D. Yu et al., 2022). Others have focused on image classification, using large-scale pretraining data sets such as Places365, ImageNet, and JFT3[4], and fine-tuning on CIFAR-10 or ImageNet (Abadi et al., 2016; Cattan et al., 2022; De et al., 2022; Kurakin et al., 2022; Mehta et al., 2022; Papernot et al., 2019; Tramèr & Boneh, 2021). Recent studies have investigated when and why public data helps. One method proposed to predict how useful a public data set will be is by Gu et al. (2023), while Ganesh, Haghifam, et al. (2023) explore instances where public pretraining is necessary to obtain good utility.

### 3.1.2. Private Learning Assisted by Public Data

Another family of techniques involves incorporating public data into the private training process itself (independent of whether pretrained weights or features are used). These generally assume that the public data set is small, and thus pretraining would be ineffective. A prototypical example computes a low-dimensional principal component analysis (PCA) of the public gradients and projects the private gradients onto this subspace (Golatkar et al., 2022; Kairouz, Ribero, et al., 2021; D. Yu, Zhang, Chen, & Liu, 2021; Zhou et al., 2021). As gradients have been empirically observed to be approximately low-dimensional, this projection preserves most of the signal while reducing the amount of noise introduced by DP-SGD. As another approach, Amid et al. (2021) perform mirror descent with the loss on the public data as the mirror map that results in provable benefit with the size of private data proportional to the dimension d. An approximation of this approach is more simply stated as performing gradient descent with both the sensitive gradients (appropriately privatized) and the public gradients simultaneously (equivalently, using the public loss as a regularizer). A slightly different approach (Golatkar et al., 2022) shows that using public data to obtain an initialization in combination with an aggressive regularization centered around that initialization results in provably improvement even if a constant number of public data points are available. Finally, T Li et al. (2022) investigate private adaptive optimizers, where the gradient moments are estimated using the public gradients.

The PATE framework (Private Aggregation of Teacher Ensembles) employs unlabeled public data (Papernot et al., 2017, 2018) using the sample-and-aggregate paradigm (Nissim et al., 2007). Specifically, PATE trains an ensemble of models nonprivately, and then privately aggregates their predictions on unlabeled public data, which is used to train a (private) student model.

Public data can also be employed for the important task of *hyperparameter selection*. Technically speaking, if one trains privately multiple models on a sensitive data set for the purpose of hyperparameter selection, the privacy analysis must account for the number of models trained. Nonetheless, the vast majority of research papers in the private machine learning literature disregard the privacy cost of determining hyperparameters and only account for training the best model. Some works highlight and offer strategies to reduce the cost of private hyperparameter selection (J. Liu & Talwar, 2019; Mohapatra et al., 2022; Papernot & Steinke, 2022; H. Wang, Gao, et al., 2023), though these generally incur constant factor overheads in the privacy cost. Instead, one can freely use public data to select suitable hyperparameters and transfer them to the private setting of interest; see, for example, Ramaswamy et al. (2020).

Conceptually related to public data, one can also employ 'public priors,' in which we make choices about the training procedure that have better inductive bias for the settings of interest. A notable example is the work of Tramèr and Boneh (2021). By using data-independent ScatterNet features (Oyallon & Mallat, 2015), which are known to perform well on image classification for ImageNet-like data in the nonprivate setting, they show improved utility for similar tasks under DP. This fits into a broader line of work in private ML that makes substitutions to components of traditional nonprivate training pipelines, ranging from activation functions, to pooling functions, and normalization layers (Cheng et al., 2022; Nasirigerdeh et al., 2023; Papernot et al., 2021).

While most of the work above is empirical in nature, there are studies that have focused on the theoretical advantages of public data. A line of works studies the sample complexity of probably approximately correct (PAC) learning with access to public data (Alon et al., 2019; Bassily, Moran, & Nandi, 2020; Bassily et al., 2018; Beimel et al., 2015), giving a reasonably tight characterization. They show that a smaller amount of public data is needed, and even unlabeled public data generally suffice. Bassily et al. (2018) and C. Liu et al. (2021) in particular focus on a PATE-like setting (Papernot et al., 2017), where private predictions are made online for an unlabeled public data set. Bounds on the accuracy of online predictions are quantified, as well as the accuracy of models trained using this privately labeled public data, under independent and identically distributed (i.i.d.) (Bassily et al., 2018) and active query (C. Liu et al., 2021) assumptions. Most of these works focus on the case where the public and private distributions are identical, though Bassily, Moran, and Nandi (2020) introduce a mixture-based setting where the distributions may differ.

### 3.1.3. Public Data in Private Query/Statistics Release

Another canonical setting for differentially private data analysis is the problem of private query release. In this setting, the data analyst is tasked with privately releasing the answers to a set of statistical queries computed with respect to a (sensitive) data set. This general problem captures a number of important use cases. For instance, releasing private summary data for the 2020 US Census (Abowd et al., 2022) can be seen as an instance of private query release. Some works study the effect of public data in these settings(Bassily, Cheu, et al., 2020; Ji & Elkan, 2013; T. Liu et al., 2021; M. Wang et al., 2017), both theoretically and empirically. A recent challenge organized by NIST (National Institute of Standards and Technology) focuses on temporal map data (NIST, 2020), and provides participants with public data from a different year to help make predictions.

### 3.1.4. Partially Public Data

One can imagine settings where certain aspects of the data are public, while others are private. For example, in computational advertising, impressions are considered public information, and only conversions are sensitive as they may reveal users' preferences. This inspired a relaxation of DP called *label differential privacy* (Beimel et al., 2013; Chaudhuri & Hsu, 2011; Ghazi et al., 2021), where only the labels are sensitive, while the feature vectors are not considered to be private. This nonsensitive information allows one to obtain a better prior on the label of a point, in comparison to the uniform prior in the standard privacy setting. Under this model, Ghazi et al. (2021) proposes a multistage training framework that progressively refines these priors.

### 3.1.5. Pitfalls of Public Data

There are a number of challenges that arise when we introduce public data into the private machine learning pipeline (H. Brown et al., 2022; Tramèr, Kamath, & Carlini, 2022). One difficulty involves the fact that publicly available data may not be appropriate to treat as fully 'public.' Some examples include when data is posted online illegally, without the original owner's knowledge or consent, or for use exclusively in one particular context; concerns of *contextual integrity* (Nissenbaum, 2004). It is a risk to consider models that treat such content as public. Another issue pertains to the actual utility of public data in models used for privacy-sensitive settings. By their very nature, such applications often involve data that is not well-represented in public data sets. Since many benchmarks used for private machine learning are imported from the nonprivate setting and resemble data that is available on the internet (e.g., ImageNet), results on these benchmarks may be overrepresenting the usefulness of public data for the settings in which privacy is actually relevant.

## 3.2. Data-Adaptive Differentially Private Algorithms

Currently, most of the widely used DP mechanisms are not adaptive to the data set. For example, common use cases of the Laplace and Gaussian mechanisms add noise calibrated to the *global sensitivity* of the given query as determined by the worst possible pairs of neighboring input data sets. In many real-life problems, however, it is often possible to adapt to a particular data set and add *less noise* when the actual input is 'nice.' Classical examples of such data-adaptive approaches include smooth sensitivity (Nissim et al., 2007) and propose-test-release (PTR) (Dwork & Lei, 2009). These approaches aim at calibrating the noise to the *local sensitivity*, that is, the maximum possible change to the query when we change a given input data set to its neighbor. Recent work generalizes these approaches beyond local sensitivities and noise addition mechanisms. Other properties of the data such as eigenvalue conditions, sparsity, and bounded support have been shown to unlock higher utility at the same privacy budget. In the context of deep learning, a number of adaptive variants of DP-SGD were proposed to reduce the noise. We survey existing results on data-adaptive DP algorithms and outline future directions in making these algorithms practical for large-scale deployment.

### 3.2.1. Smooth Sensitivity

For several queries of interest, such as the median of scalar values, the global sensitivity of the query can be significantly larger than the local sensitivity, that is, how much the query output can change by modifying one data point *for the given data set*. However, naively adding noise proportional to the local sensitivity is not differentially private because the noise itself could reveal sensitive information, especially when the local sensitivity could change drastically as we add / remove data points (e.g., in median queries). Smooth sensitivity (Nissim et al., 2007) is the oldest and one of the most elegant ideas for designing data-adaptive DP algorithms in these situations. It involves computing exponentially smoothed upper bounds of the local sensitivity and choosing heavier-tailed noise satisfying certain dilation properties according to the smoothed local sensitivity. It remains the primary approach for obtaining data-adaptive DP algorithms with pure DP. Moreover, the smooth sensitivity approach can be generically implemented for many problems through a sample-and-aggregate scheme, which has led to the first family of asymptotically efficient statistical estimators (Smith, 2011).

On the other hand, the smooth sensitivity framework is also limited in several ways. It is computationally inefficient due to the need to search all neighboring data sets for many hops (though useful classes of exceptions exist). The heavy-tailed noise and overhead from sample-and-aggregate often make the method less practical on moderate-sized data sets. In addition, known smooth-sensitivity-based mechanisms also have poor dimension dependence. One notable open problem is to derive fine-grained privacy accounting for smooth sensitivity-based mechanisms with Rényi DP (Mironov, 2017), $f$-DP (Bu et al., 2020; J. Dong et al., 2022), or privacy profiles, so they become more compatible with modern privacy accounting tools. The recent work of Bun and Steinke (2019) takes an important step in this direction by deriving a new set of qualifying noise values that come with concentrated DP (zCDP) guarantees. Another open problem is to characterize the dimension dependence in smooth-sensitivity-based mechanisms.

### 3.2.2. Propose-Test-Release

PTR (Dwork & Lei, 2009) is a viable alternative to the smooth sensitivity framework with the same goal of calibrating the noise to local sensitivity. This has been adopted to solve several problems including private selection (Dwork & Ullman, 2018; Y. Zhu & Wang, 2022) and private semi-supervised learning (Redberg et al., 2022). The main challenge is again that the local sensitivity itself is data-dependent. PTR gets around this issue by (1) proposing an upper bound on the local sensitivity, (2) privately testing whether it is a valid upper bound on the local sensitivity for the given data set, and then (3) releasing the private query output only if the data set passes the test. As in the median example, this framework is powerful for computing robust statistics of samples drawn i.i.d. from a distribution, where the local sensitivity can be significantly smaller than the global sensitivity. However, designing robust estimators for standard statistical estimation problems, such as mean estimation, covariance estimation, linear regression, and principal component analysis, is challenging in high-dimensions. To this end, X. Liu, Kong, and Oh (2022) propose a framework for solving these problems via a high-dimensional PTR algorithm.

The PTR framework has several challenges to be widely adopted. First, it needs to propose a bound, rather than computing one from the data. Second, the generic construction of the 'private test' in step (2) is a 'distance test' that requires computing the nearest data set (in terms of number of data point modifications) that violates the proposed local sensitivity bound, and thus is not computationally efficient in general. For statistical estimation problems, the computational complexity of PTR often grows exponentially in the ambient dimension. For these reasons, typical adoption of PTR is either for low-dimensional problems—Dwork and Lei (2009) studies trimmed mean, median, and short-cut regression and Avella-Medina and Brunel (2019) and Brunel and Avella-Medina (2020) study robust mean and median—or have exponential run-times, for example, mean estimation (X. Liu et al., 2021), covariance-aware mean estimation (G. Brown et al., 2021), depth-based medians (Ramsay & Chenouri, 2021; Ramsay et al., 2022), and general statistical estimation (X. Liu, Kong, & Oh, 2022). For the special case of covariance-aware mean estimation, recent advances in G. Brown et al. (2023) and Duchi et al. (2023) managed to achieve the optimal sample complexity with an efficient algorithm, which previously was only possible with exponential-time algorithms based on PTR in G. Brown et al. (2021) and X. Liu, Kong, and Oh (2022). Several other private statistical estimation problems exhibit computational gap, where exponential time algorithms achieve strictly better sample complexity compared to efficient counterparts. It remains open whether these gaps can be closed with the tools from G. Brown et al. (2021) and Duchi et al. (2023).

### 3.2.3. PTR-like Mechanisms

There are a number of variants of data-adaptive mechanism design that leverage PTR-like ideas, but do not suffer from the aforementioned limitations. They are less widely applicable than the vanilla (distance-test-based) PTR, but are computationally efficient when they qualify. These approaches include *privately bounding local sensitivity*, which avoids the 'propose' step of PTR, and *privately testing the stability margin* [NO_PRINTED_FORM], which avoids the part that adds noise, and thus also works for non-numeric queries. A concise treatment of these methods is given in Vadhan (2017). These methods allow PTR to be applied with less overhead to learning algorithms. Notable example applications include private topic models (LDA) with spectral methods (Decarolis et al., 2020) and model agnostic private learning (a variant of PATE) (Bassily et al., 2018). Recently, PTR and PTR-like mechanisms were generalized to cover data-adaptive algorithm design beyond noise-adding mechanisms (Redberg et al., 2022). Their approach, known as generalized PTR, modifies the three steps of the vanilla PTR as follows: (1) it proposes native parameters $\phi$ of a randomized algorithm $\mathcal{M}_\phi$ rather than an upper bound of the local sensitivity; (2) it privately tests whether the resulting data-dependent privacy parameter ($\varepsilon$ as a function of a particular input data set and $\phi$) is below a prescribed budget; (3) it executes the algorithm with the proposed parameters only if it passes the test. This allows a PTR-like procedure to be derived for mechanisms such as posterior sampling, objective perturbation, PATE, and so on.

### 3.2.4. Data-Adaptive DP Algorithms via Data-Dependent DP Losses

Recall that data-adaptive DP algorithms aim at 'adding a smaller amount of noise' when the data set is 'nice.' The flipside of data-adaptive DP algorithm design is what is known as *data-dependent DP losses* $\varepsilon$ (Data), which aim at quantifying how much smaller the privacy loss incurred by a 'nice' data set is when we add a fixed amount of noise (Papernot et al., 2018; Y. Zhu et al., 2020). The latter scheme fixes the accuracy while allowing the privacy loss to vary, and thus can be appealing in applications where preserving utility is more important than ensuring (a fixed level of) privacy (Ligett et al., 2017; Whitehouse et al., 2022). Different from the standard 'accuracy first' setting, the privacy losses are data-dependent now, thus considered sensitive information. To address this issue, Papernot et al. (2018) proposes a smooth sensitivity-based method to release the data-dependent Rényi DP parameters. Strictly speaking, even if data-dependent DP losses are released privately, the overall algorithm is still not a data-adaptive DP algorithm. Redberg et al. (2022) shows that one can always construct such a data-adaptive DP algorithm with any prescribed privacy budget by a simple postprocessing procedure that *abstains* unless a high-probability private upper bound of the data-dependent DP loss is smaller than the given privacy budget.

Related to this idea, Feldman and Zrnic (2021) propose to maintain a privacy accountant for each individual in the data set and to remove data points whose *personalized* DP budget is used up. This approach, known as an individual Rényi filter, ensures a worst-case DP bound while allowing the algorithm to run longer than the worst case. Unlike the PTR and smooth sensitivity frameworks, this approach does not need to explicitly test any data-adaptive properties, and thus could be a promising approach for more applications. An interesting open problem is to design the equivalent of the individual Rényi filter for per-instance DP losses (Y.-X. Wang, 2019) (instead of the personalized DP losses).

## 3.2.5. Statistical Estimation and Inference

Private statistical estimation problems are widely considered as one of the most important areas where data-adaptive DP algorithms are applied. These kinds of tasks, such as private mean estimation, are generally impossible to solve for worst-case data sets. This is because the mean of a data set can be highly sensitive to the addition or removal of even a single extreme outlier. Furthermore, the worst-case sensitivity is large for every data set, even for otherwise 'well-behaved' data sets. In contrast, the median is insensitive for well-concentrated data sets. To address this issue, it is common to make *distributional* assumptions, such as sub-Gaussianity or bounded moments of the underlying distribution (Barber & Duchi, 2014; Bun & Steinke, 2019; Kamath et al., 2019; Karwa & Vadhan, 2018). Some works require minimal knowledge of underlying distribution parameters, such as which moment of a distribution is bounded, and instead adapt to them (W. Dong & Yi, 2021).

Exploring the relationship between robust and private statistics is an important line of work. Both types of estimators should be unaffected by changes in parts of the data set. However, formalizing the connections between the two has been challenging. Dwork and Lei (2009) introduced the aforementioned PTR framework, highlighting that robust statistics are especially suitable for privatization. Chaudhuri and Hsu (2012) leveraged ideas from robust statistics to obtain optimal rate bounds using smooth sensitivity. In the multivariate setting, some works have used similar approaches, either by privatizing multivariate medians like the Tukey median (Tukey, 1960), or by employing PTR-based ideas (G. Brown et al., 2021; X. Liu, Kong, & Oh, 2022; X. Liu et al., 2021; Ramsay et al., 2022). Other works have applied recent advances in efficient multivariate robust statistics (Diakonikolas et al., 2016; Lai et al., 2016) to the privacy domain (Alabi et al., 2023; Hopkins et al., 2022, 2023; Kothari et al., 2021). Some recent works have employed the inverse-sensitivity mechanism (Asi &

Duchi, 2020), which combines the exponential mechanism with robust statistics to convert a robust estimator to a private one (Asi et al., 2023; Bun, Kamath, et al., 2019; Hopkins et al., 2023; Kamath et al., 2020). While it is also known that privacy implies robustness (Georgiev & Hopkins, 2022), reductions between robustness and privacy are far from an equivalence, and there are some technical caveats. Therefore, further research is required to fully understand the relationship between robustness and privacy.

Most work in this area is largely theoretical. Some works attempt to produce practical tools for private estimation (Amin, Joseph, et al., 2022; Biswas et al., 2020; W. Du et al., 2020; Tsfadia et al., 2022). Finally, a recent line of work shows how to privately learn critical parameters and tuning the bias-variance trade-off from data to improve overall utility (Amin, Dick, et al., 2019, 2022). Proper tuning, while time-consuming, leads to large empirical utility gains. Building better and more practical estimators, particularly based on recent advances through connections with robust statistics, is another interesting direction forward for the field.

As for statistical hypothesis testing, early research on private methods can be traced back to the works of Uhlerop et al. (2013) and Vu and Slavkovic (2009). A significant body of work has since emerged, with one line of research focusing on the design of differentially private versions of popular test statistics for testing goodness-of-fit, closeness, and independence, as well as private ANOVA (Campbell et al., 2018; Couch et al., 2019; Gaboardi, Lim, et al., 2016; Kakizaki et al., 2017; Kifer & Rogers, 2017; Swanberg et al., 2019; Y. Wang et al., 2015). This line of research specifically looks at the performance of these methods when dealing with small sample sizes. Another complementary direction of research examines the minimax sample complexity of private hypothesis testing (Acharya, Kamath, et al., 2018; Acharya, Sun, & Zhang, 2018; Aliakbarpour et al., 2018; B. Cai et al., 2017). Other works in this vein focus on testing of simple hypotheses (Canonne et al., 2019; Cummings et al., 2018). In addition, there is a related area of research that studies hypothesis testing in the Local DP setting. Works such as Acharya, Canonne, Freitag, and Tyagi (2019), Duchi et al. (2013), Garg et al. (2018), and Sheffet (2018) focus on developing tests that are simpler to reason about than central-model tests. Overall, these research efforts aim to improve the effectiveness and practicality of private hypothesis testing.

### 3.2.6. Data-Adaptive Algorithms in DP-SGD and Its Variants

Attempts to obtain data-adaptive DP algorithms for deep learning tasks are often variants of DP-SGD that involve data-adaptive choices of its hyperparameters, for example, learning rate, weight decay, pre-conditioner (Asi, Duchi, et al., 2021; Zhou et al., 2021), clipping threshold (Andrew et al., 2021; Thakkar et al., 2019), privacy budget allocation (J. Lee & Kifer, 2018; L. Yu et al., 2019), and so on, sometimes in every iteration. These methods are not traditionally considered data-adaptive DP algorithms, but we believe they not only are but also might be among the best of such algorithms in practice whenever they are applicable. Some of these data-adaptive choices are consequences of adaptive composition (D. Yu, Zhang, Chen, Yin, & Liu, 2021). Others require an additional privacy budget to be allocated for making these choices (Andrew et al., 2021)—much like most other recipes for data-adaptive DP algorithms that we have seen earlier. We will discuss a few representative methods below.

Private gradient descent methods require clipping the norm of each sample gradient in order to control the sensitivity. However, the ideal norm-clipping threshold is challenging to know a priori and may even change dramatically over the iterations. Andrew et al. (2021) and Thakkar et al. (2019) overcome such challenges by finding a threshold adaptive to the distribution of the norm of the gradients in each mini-batch, while spending a small fraction of the privacy budget in the process. Theoretically showing the gain of such adaptive clipping requires some statistical assumptions on the data, such as those typically assumed in statistical estimation problems. For example, for linear regression, Varshney et al. (2022) use adaptive norm-clipping to achieve improved sample complexities compared to gradient descent with nonadaptive clipping in T. T. Cai et al. (2019) and other approaches that use adaptive regularizers but are not based on gradient descent (Milionis et al., 2022; Y.-X. Wang, 2018). The state-of-the-art sample complexity for DP linear regression is achieved in X. Liu et al. (2023) using a novel adaptive norm-clipping, which also provides robustness against label corruption. Beyond adapting to the gradient norm only, a more sophisticated clipping method that adapts to the geometry of the gradients is critical in achieving optimal sample complexity in principal component analysis (X. Liu, Kong, Jain, & Oh, 2022). A natural question to ask is if such methods can be applied to private structured estimation problems to give similar gains, such as sparse linear regression (T. T. Cai et al, 2019, 2020; Hu et al., 2022; Kifer et al., 2012; Talwar et al., 2015; Thakurta & Smith, 2013; L. Wang & Gu, 2019) and sparse principal component analysis (Ge et al., 2018; Hu et al., 2023).

In practice, spending privacy budget on adaptive clipping can be costly, especially if one desires more sophisticated adaptive methods. A series of works achieves significant gains by using in-distribution public data, that is, examples sampled from the same distribution as the training data but do not require privacy. Amid et al. (2021), Asi, Duchi, et al. (2021), Golatkar et al. (2022), and Nasr et al. (2022) use more sophisticated clipping that adapts to the geometry of the gradients, Asi, Duchi, et al. (2021), Kairouz et al. (2020), D. Yu, Zhang, Chen, and Liu (2021), and Zhou et al. (2021) project the gradient to a lower dimensional subspace, and Asi, Duchi, et al. (2021) and T. Li et al. (2022) use public data to compute the statistics of past gradients in the Adam optimizer. The main idea is to find better adaptive methods using the public data to avoid spending the privacy budget. However, these approaches suffer significantly when there is a distribution shift in the public data. It remains an important open question whether one can harness the benefit of public data with adaptive DP-SGD using out-of-distribution public data. We refer to Section 3.1 for an extensive summary of the use of public data.

A line of work improves the DP-SGD by carefully allocating the privacy budget per iteration. J. Lee and Kifer (2018) is the first to propose the adaptive privacy budget strategy, where it uses a smaller privacy budget for gradients with large norms and a larger privacy budget for gradients with small norms. L. Yu et al. (2019) provides a dynamic privacy budget allocation method with the assistance of a public validation data set. Specifically, it checks the validation accuracy periodically during the training process, and when the validation accuracy stops increasing, it triggers the privacy budget to increase for subsequent epochs. However, this line of work critically relies on the convexity of the problem. It remains an interesting open question whether similar adaptive schemes can be applied to improve more general nonconvex optimization.

We conclude the discussion on data-adaptive DP algorithms by remarking that the design of such algorithms is still an art that needs to be done on a case-by-case basis. It requires domain knowledge in identifying what 'nice' properties of a data set to exploit. It is also more delicate and error-prone than nonadaptive DP algorithms. Nevertheless, going data-adaptive is a crucial step in bringing DP algorithms to an acceptable level of utility in applications. We believe more work is needed on this problem before we can converge on a handful of best practices.

## 3.3. Mitigating Heterogeneous Privacy/Utility Trade-offs

The noise added to achieve DP could have a larger impact on the utility of marginalized subpopulations because they are more prone to the uncertainty induced by noise. When making decisions based on differentially private data summaries, underrepresented groups may suffer from a larger utility loss (Pujol et al., 2020; Steed et al., 2022; Tran, Fioretto, et al., 2021). A similar phenomenon also exists in differentially private machine learning applications. Machine learning algorithms are known to have unfair performance for groups that are underrepresented in training data (Bolukbasi et al., 2016; Buolamwini & Gebru, 2018; Caliskan et al., 2017). Unfortunately, learning with DP could exacerbate such unfairness. The accuracy drop associated with DP turns out to be higher for underrepresented groups, that is, 'the poor get poorer' (Bagdasaryan et al., 2019; Farrand et al., 2020; Ganev et al., 2022; Noe et al., 2022; Suriyakumar et al., 2021; Tran, Dinh, & Fioretto, 2021). Underrepresented groups thus suffer from worse privacy/utility trade-offs.

Mitigating heterogeneous privacy/utility trade-offs is an active research topic (Fioretto et al., 2022; Mangold, Perrot, et al., 2022). Pujol et al. (2020) propose a postprocessing 'repair' algorithm to improve the fairness of decision-making. Esipova et al. (2022) and Xu et al. (2020) show that the gradient bias in private learning is an important source of disproportionate impacts and propose mitigations correspondingly. Uniyal et al. (2021) compare PATE (Papernot et al., 2017) with DP-SGD (Abadi et al., 2016) and show that PATE has less impact than DP-SGD.

Although the above line of work has considerably mitigated the disproportionate impacts under various settings, another line of work theoretically shows that unfairness in accuracy is inevitable in some data distributions. Agarwal (2020) and Cummings et al. (2019) show there exist data distributions on which enforcing fairness for a private algorithm would necessarily result in trivial accuracy. Feldman (2020) and Sanyal et al. (2022) further study this tension under the assumption that the data has a long-tailed structure. Their work suggests that, for a nontrivial target accuracy, one has to trade off between privacy and fairness. Sanyal et al. (2022) validate the theory on several real-world data sets and show that relaxing the target overall accuracy is an effective method to improve fairness.

## 3.4. Building Information Theoretic Understanding of DP

In addition to developing better algorithms, it is crucial to understand the fundamental limits arising from imposing DP constraints. Several methods have been proposed to lower bound the accuracy of algorithms for central DP. Fingerprinting (Bun et al., 2014; Bun, Steinke, & Ullman, 2017; T. T. Cai et al., 2019; Dwork et al., 2015; Kamath, Mouzakis, & Singhal (2022); Kamath et al., 2019; Steinke & Ullman, 2015, 2017a) is a versatile approach under approximate DP that has established strong lower bounds for various problems, such as attribute mean estimation (Kamath et al., 2019), ERMs (Bassily et al., 2014; Levy et al., 2021; Talwar et al., 2015), and private selection (Steinke & Ullman, 2017b). Private Assouad's method (Acharya, Sun, & Zhang, 2021) is particularly useful for problems with discrete domains, such as distribution estimation (Y. Liu et al., 2020). For the special case of $\delta = 0$ (i.e., pure DP), the packing argument (Beimel et al., 2014; Bun & Steinke, 2016; Hardt & Talwar, 2010; Vadhan, 2017), as well as its probabilistic version, private Fano's inequality (Acharya, Sun, & Zhang, 2021), is a geometric approach that can provide tight lower bounds for ($\varepsilon, 0$)-DP. Notably, the proof of this approach is relatively straightforward and aligns well with nonprivate proof techniques (Bun, Kamath, et al., 2019; Kamath et al., 2021; H. Zhang et al., 2020).

Local DP (Evfimievski et al., 2003), in contrast to a global or central setting, removes the assumption that the data curator or aggregator collecting raw data is trusted by the end-users, hence will add noise to the individual samples before sending to the central aggregator. Local DP has also been a focus of research, with initial work (Duchi et al., 2013, 2018) proposing private versions of Le Cam, Fano, and Assouad, subsequently improved and generalized upon by Błasiok et al. (2019), Duchi and Rogers (2019), Duchi and Ruan (2018), and Ye and Barg (2018). Additionally, Acharya, Canonne, Freitag, and Tyagi (2019) and Acharya, Canonne, and Tyagi (2019) proposed a lower-bounding technique based on chi-squared contraction, which yielded tight lower bounds for distribution estimation and identity testing (Acharya et al., 2022).

Despite the progress made in understanding DP lower bounds, several challenges remain to be addressed, which we discuss next.

**More complicated scenarios:** Despite significant progress in proving lower bounds, most research has focused on relatively straightforward scenarios such as mean estimation (Kamath & Ullman, 2020; Karwa & Vadhan, 2018), hypothesis testing (Acharya, Sun, & Zhang, 2018; Canonne et al., 2019, 2020), and ERM (Asi, Feldman, et al., 2021; D. Liu & Lu, 2021; Song et al., 2021). Consequently, the issue of deriving strong lower bound results for more intricate scenarios, such as graph problems (Fichtenberger et al., 2021; Imola et al., 2021) and online learning (Shariff & Sheffet, 2018), persists as a compelling challenge.

**Interacting with other constraints:** Previous literature has typically examined privacy as the sole constraint and investigated the trade-off between privacy and utility. However, in real-world applications, privacy is not the only constraint and often interacts with other limitations such as computational complexity (Dwork et al., 2009; Ullman, 2016), interactivity (Acharya, Canonne, et al., 2021; Acharya et al., 2022; Joseph et al., 2019), communication (Acharya, Kairouz, et al., 2021; Acharya & Sun, 2019), fairness (J. Ding et al., 2020), interpretability (Harder et al., 2020), robustness of statistical estimation (Acharya, Sun, & Zhang, 2021; Cheu et al., 2021; X. Liu, Kong, & Oh, 2022; X. Liu et al., 2021), robustness against data poisoning (M. Du et al., 2020; Ma et al., 2019), and robustness against adversarial examples (Phan et al., 2020). Understanding how privacy interacts with these other constraints remains a complex task.

**Instance-optimal lower bounds:** The minimax framework is usually the first thing that comes to mind when measuring the utility of statistical estimation tasks. However, this worst-case optimality is practically meaningless and results in unnecessary noise being added to the system when the global sensitivity is far larger than the local sensitivity. Therefore, privacy researchers have begun exploring instance optimality (Asi & Duchi, 2020; Błasiok et al., 2019; Dick et al., 2023; Z. Huang et al., 2021; Nikolov & Tang, 2023), a framework that seeks to optimize the loss for each individual data set. While this approach has demonstrated potential, there is still a necessity for algorithm and lower bound advancements, making it an exciting and demanding area of research.

**Choosing the right privacy metric:** The DP framework is characterized by its concrete privacy semantics, derived from the its 'differential' interpretation. Notwithstanding, there exists a variety of DP variants, encompassing pure DP and $(\varepsilon, \delta)$-DP, alongside divergence-based DP definitions such as Rényi DP (Mironov, 2017) and truncated concentrated DP (Bun et al., 2018), and $f$-DP (J. Dong et al., 2022), which are interpreted directly through a hypothesis-testing viewpoint. Among these, some DP metrics excel in terms of interpretability, while others possess more information-theoretic characteristics, making them well-suited for applications requiring tight privacy bounds, such as hyperparameter tuning (Papernot & Steinke, 2022) and information-theoretic trade-offs between privacy and utility (J. Dong et al., 2021). We posit that an advantageous research direction entails developing a framework for determining the most suitable privacy definition for a given task.

**Balancing privacy parameters:** A widely acknowledged challenge in implementing DP mechanisms stems from the necessity to use a large $\varepsilon$ parameter to ensure an exceedingly small $\delta$ parameter, in order to preclude the 'release-one-at-random' mechanism (Dwork, 2008). In particular, it is common practice to set the $\delta$ parameter significantly smaller than the reciprocal of the data set size. Contrary to the 'release-one-at-random' mechanism, numerous practical mechanisms allow for a smooth trade-off between the two privacy parameters. This observation implies that the need for such a small $\delta$ might not be essential and, if so, a considerably smaller $\varepsilon$ parameter could be employed, potentially fostering a wider appreciation of DP within practitioners and stakeholders. As such, it is imperative to rigorously and comprehensively address the selection of the $\delta$ parameter as an important research direction.

**Better toolbox:** Lastly, even though lower bound proof has seen significant accomplishments, we still need better tools to address existing limitations. For instance, the fingerprinting lemma's inability to handle discrete inputs, and the private Assouad's method is imprecise for certain typical problems, such as Gaussian mean estimation. Moreover, for packing lower bounds, it only applies when $\delta$ is exceptionally small. Furthermore, the majority of prior tools have focused on asymptotic analysis, which overlooks the significance of constants. As our comprehension of the asymptotic case has advanced, there is a pressing need for tools with improved constants.

## 4. Privacy Attacks and Auditing as a Measure of Protection

The main objective of DP is to safeguard private data from being leaked to an adversary. DP accomplishes this goal by providing a provable guarantee, which is a generic bound on privacy leakage that makes few assumptions about the goals and capabilities of an adversary. However, DP has some limitations that need to be considered.

Firstly, the DP bound may not provide stakeholders with sufficient information about the degree to which the system is initially vulnerable to specific attacks against the privacy of data. This may make it challenging to assess privacy risks before deploying a DP algorithm, and also to measure the gains and benefits of deploying an algorithm that provides DP guarantees.

Secondly, the DP bound may be too conservative or inaccurate in certain scenarios as it is not tailored to the adversary's capabilities. For instance, this is the case if the threat model for a particular system does not allow for some of the adversaries that are captured by the DP definition. In some cases, the DP algorithm's bound may even be incorrect due to the complexity of its implementation. There are documented instances (Jin et al., 2022; Mironov, 2012; Tramèr, Terzis, et al., 2022) where the DP algorithm's implementation did not match the algorithmic description, making all the formal guarantees provided by DP proofs meaningless.

Thirdly, DP only captures a specific notion of privacy, defined by the neighboring data sets used in the inference bound. DP does not provide any guarantees for arbitrary data release (Cormode, 2010), where the predefined neighborhood notion can become vacuous, and can reveal data set and population statistics (Ateniese et al., 2015; W. Zhang et al., 2021) and properties of data distributions (Jayaraman & Evans, 2022; Suri & Evans, 2022). This may be a significant privacy consideration when the training data set or distribution itself is not public. Hence, additional definitions, audits, or analyses may be necessary to ensure that a model does not violate disclosure requirements in ways that are not captured by DP. For example, Pufferfish privacy (Kifer & Machanavajjhala, 2014) can be used, which is a more general privacy framework that has been utilized to protect correlated data (Song et al., 2017) and data set–level information (W. Zhang et al., 2022).

These limitations motivate the need for an evaluation of privacy that is based on empirical analyses of the underlying systems. The pioneering work of Z. Ding et al. (2018) initiated the study of empirical evaluation of privacy for simple mechanisms, which is subsequently analyzed and improved in Gilbert and McMillan (2018) and X. Liu and Oh (2019). The important issue of which two neighboring data sets to use in the empirical measurements has not been systematically addressed in these works. Today, these analyses are often done in the form of simulated attacks where an analyst defines an attacker with goals and resources and attempts to execute an attack that can measure leakage from the target system. Such attack-based evaluations or auditing provide results that are complementary to the theoretical DP guarantees.

In this section, we distinguish privacy attacks and auditing by their intent. A privacy attack is conducted by an adversary to reveal private information, while a privacy audit is conducted by the system designer to anticipate the privacy leakage present in the system. An audit can be seen as a way of anticipating the threat of privacy attacks, and techniques used for privacy attacks can also be used to conduct audits.

## 4.1. Attack Algorithms

There are several types of attacking algorithms that are often used for privacy auditing, here we give a brief overview of two of those. We encourage the readers to explore some of the classic attacks in the DP literature (Dinur & Nissim, 2003; Dwork et al., 2017) and their implications in the context of US census (Garfinkel et al., 2019) for a broader exposure.

**Membership inference attacks (MIA):** MIA aims to infer whether a data point was used to train a target machine learning model or not. These attacks can lead to substantial privacy concerns for the individuals involved. For instance, upon ascertaining that a clinical record was used in training a model associated with a specific disease, membership inference attacks may deduce, with a high probability, that the individual linked to the clinical record has the disease. Prior works (Carlini, Chien, et al., 2021; Jayaraman et al., 2020; Sablayrolles et al., 2019; Shokri et al., 2017; Watson et al., 2021; Yeom et al., 2018) have shown that attackers can effectively launch MIA by solely relying on the prediction $vector$ of a data point from a machine learning model. These attacks typically exploit subtle differences in the model's performance on in-sample (training) data versus out-of-sample (nontraining) data. Recent studies (Choquette-Choo et al., 2021; Z. Li & Zhang, 2021) further demonstrate the possibility of MIA in more constrained settings, where the target model provides only the predicted $label$ to the attacker. Other works demonstrated MIA in aggregated data release settings (Pyrgelis et al., 2017).

**Data extraction attacks:** Data extraction attacks have become an increasing threat in machine learning, as adversaries seek to undermine the security and privacy of training data by retrieving the original examples or uncovering sensitive information embedded within them. Model inversion attack (Fredrikson et al., 2015) demonstrates that a 'representative' face image corresponding to a person (a class) can be reconstructed from a face recognition model given the name of the person (the class label) and white-box access to the model. While the membership inference attack recovers the 'existence' information of a target data point, model inversion attack recovers the visual property or features of a target class (which can correspond to a person).

Point-wise reconstruction of an individual training data point is an even more severe type of leakage than recovering the 'representative' data. Recent studies have showcased the feasibility of extracting individual private image data from trained image classifiers (Balle et al., 2022; Haim et al., 2022). However, these findings rely on strong assumptions concerning model architectures (e.g., simple convolutional neural network (CNN) models with a few layers) or the attacker's capabilities (e.g., the attacker knows all but one training example and aims to reconstruct the remaining one). In the context of generative models, the attacks can be made more potent. Prior work (Carlini, Tramèr, et al., 2021; Carlini et al., 2020, 2023) demonstrates the ability to extract high-fidelity text or images by prompting a trained language model or diffusion model. These findings emphasize the potential security and privacy risks associated with the memorization of training data in generative models.

Data extraction attacks could become increasingly powerful when side-channel leakages are present. For instance, it has been demonstrated that private image (Geiping et al., 2020; Y. Huang et al., 2021; Yin et al., 2021; Zhao et al., 2020; L. Zhu et al., 2019) and text (Dimitrov et al., 2022; Fowl et al., 2022; Gupta et al., 2022) data can be extracted during federated learning, assuming that the attacker has white-box access to the model being trained and the exchanged gradients. This further highlights the importance of addressing potential security vulnerabilities of data leakage in distributed learning environments.

## 4.2. Interpretation of Attack and Auditing Results

As mentioned earlier, attacks provide complementary metrics to evaluate the privacy of an algorithm in addition to the $\varepsilon$ parameter obtained when analyzing the DP guarantees provided by the algorithm. However, the success or failure of an attack is not a direct replacement for criteria based on a value of $\varepsilon$ for a DP bound. Instead, attacks provide lower bounds on privacy. If an attack fails, it is a necessary condition to obtain a privacy-preserving algorithm. Nonetheless, the failure of attacks is not a sufficient condition, as a new attack may be proposed later, resulting in a higher lower bound. This shows that the privacy leakage of the algorithm had previously been underestimated. The bound provided by a DP analysis provides an upper bound, and the value of $\varepsilon$ being close to 0 provides a sufficient condition for the algorithm to preserve privacy. However, attacks do provide insights into the sources of leakage in a system that handles data and can inform the calibration of mechanisms that provide DP guarantees, with the aforementioned caveats.

**Mitigating attacks through non-DP methods:** Ensuring DP for any mechanism involves limiting the sensitivity of the mechanism and introducing calibrated randomness into it. However, some recent studies (Carlini et al., 2019; W. R. Huang et al., 2022; Thakkar et al., 2020) have shown that by simply bounding the sensitivity of the mechanism, state-of-the-art attacks can be significantly reduced. Although this approach alone does not guarantee DP, it can be combined with the use of randomness to provide 'weak' DP while remaining resilient to the best-known auditing techniques (e.g., Ramaswamy et al., 2020). One significant advantage of sensitivity-bounded approaches over ad hoc measures for privacy protection is that they provide a clear path toward achieving DP through amenable randomization procedures, which can then provide provable protection against MIA.

**Bounds on classes of attacks:** There are also approaches that attempt to find an intermediate point between the evaluation of a particular attack strategy and solely relying on the DP bound. Recently, a line of work has been exploring to derive bounds on the success of a particular class of attacks, regardless of the particular attack strategy employed. For example, if we want to reason about membership inference adversaries without having to instantiate a particular attack, we could derive a bound on the success of any membership inference adversary (Mahloujifar et al., 2022; Thudi et al., 2022). These bounds can be derived from DP bounds themselves (Erlingsson et al., 2019; Jayaraman et al., 2020; Sablayrolles et al., 2019; Yeom et al., 2018), and they provide a way to interpolate between the DP analysis and auditing through concrete attack instantiations. This approach does not introduce an arms race as new attack strategies are developed and can be particularly useful in situations where evaluating a specific attack strategy is not feasible.

**'Creepy' vs DP:** It is important to note the distinction between DP and other types of attacks on data sets. While DP can be effective to protect against specific types of attacks, such as MIA, it is not designed to defend against several other attacks such as property inference attacks or attribute inference attacks. Property inference attacks involve an adversary trying to learn statistical information about a model's data set (Ganju et al., 2018), while attribute inference attacks involve an adversary trying to use a model to learn a sensitive feature from a nonsensitive one (Fredrikson et al., 2015). While there has been debate about the severity of privacy violations in attribute inference attacks (Bun et al., 2021; Kenny et al., 2021; McSherry, 2016), it is essential to not confuse the success of these attacks with a violation of the privacy guarantees provided by DP. Future research may investigate how users perceive different types of privacy violations and techniques to mitigate them.

## 4.3. Benchmarking Data Sets

While performing various attack algorithms, the community has realized that many data sets that are commonly used as benchmarking data sets for ML task performance are not necessarily relevant to the privacy problem. For example, privacy methods and attacks performed on typical ML data sets, such as CIFAR-10, may be difficult to generalize to other domains, such as tabular data. Hence there is a need to identify and surface more accessible data sets that can enable us to compare different kinds of attacks or auditing exercises, facilitating an understanding of improved attacks or privacy-preserving methods, ideally with diverse sources to represent a broader set of applications.

There exists tabular data like the US Census,[5] Locations,[6] Purchase-100,[7] and Texas hospitals.[8] These are public data sets that contain private features, making them appropriate for testing risk of releasing models trained on similar sensitive data sets. For specialized domains, there is the MIMIC data set,[9] which is a critical care data set containing a large number of patients and commonly used in health research, epidemiology, and ML. Mobility data is also highly sensitive, and there exist public mobility data sets such as GeoLife[10] from Microsoft Research Asia, which covers trajectories of 178 different users. However, there does not exist a large-scale public mobility data set.

We call for the community to contribute more toward these data sets for both ML and non-ML problems, helping us to establish a common ground to quickly improve both DP algorithms and attacks.

## 4.4. Research Directions

Although privacy attacks and auditing have gained increasing interest from academia and industry, due to their more intuitive interpretation and practical benefits, it is still a relatively new domain and there are many important open challenges that need more research.

**Usage of auditing results in real deployments:** While privacy attacks and audits have the possibility of offering rich information about the privacy of the system, a large-scale deployment requires more research. As discussed above in Section 4.2, we believe that auditing through privacy attacks is a good complement to DP guarantees but does not offer a complete replacement. This leaves us with a number of open questions: can we rely on the outputs of current attacks to make important decisions, such as deciding not to require DP for a certain component? Can we leverage the practical attacks to decide on an 'acceptable' privacy unit or privacy parameter values? If yes, what are the suitable metrics and thresholds? If not, how could we develop attacks that provide some solid basis for such decisions? Some attempts were discussed in Nasr et al. (2021) and Tramèr, Terzis, et al. (2022) to derive an empirical epsilon value as an outcome of the auditing results. With these kinds of metrics, the question remains whether we should leverage them and how to leverage them properly for decision-making.

Moreover, in real-life applications, we face more scaling challenges. On one hand, powerful state-of-art attacking algorithms are often too expensive in computation cost, thus difficult to deploy to a system with high requirements for freshness and latency. At the same time, if we would like to audit or attack systems or ML models that are processing a large amount of data (e.g., high-capacity models trained on gigantic data sets), we may only afford to do it once or, for its equivalence, with only a subset of data, which then undermines the strength of the auditing results, since as we have discussed attack results only represent lower bounds.

There are several promising directions to pursue here. Is there a possibility of understanding models that are trained on incremental data to identify subsets of future training data or specific types of leakage that we can focus on rather than looking at all possible leakage profiles? Are there potential approximations to robust attacks that we think meaningfully work for models trained on 'reasonable' inputs that we can reuse across model training and recalibrate with heavier attacks periodically? Some practical approaches entail deploying simpler, less computationally heavy attacks, which may not be the strongest, while calibrating them using state-of-the-art (SOTA) attacks in a less frequent way. This is not a perfect solution, and if the goal is to increase the adoption of attacks as auditing measures, we should continue to research more effective attacks with cost constraints in mind.

**Auditing ML pipelines:** For ML pipelines specifically, an additional difficulty with auditing is that auditing supposes that the inputs and outputs of the system accessible to the adversary are clearly identified. This may not always be the case in ML pipelines. For example,

- Continual learning and recurrent training: In typical production settings, several ML models are trained and released daily or more frequently. These are often trained in an incremental fashion and we are constantly updating models with new data (as it is collected). It can also be the case when we need to delete data from previously trained models, as it may be required for privacy-related reasons such as 'the right to be forgotten' as required by General Data Protection Regulation (GDPR) and discussed in the literature on machine unlearning (Bourtoule et al., 2021; Cao & Yang, 2015; Neel et al., 2021). This has been shown to increase privacy leakage compared to a single-model release (Chen et al., 2021; Jagielski, Wu, et al., 2022; Salem et al., 2020; Zanella-Béguelin et al., 2020).
- Hyperparameter tuning: This is a common component of machine learning pipelines, which as we discussed previously, has the potential to increase privacy leakage because of repeated accesses to the same training data required to compare the model's performance for different candidate values of the hyperparameters that need to be tuned. Some techniques have been designed to mitigate this issue (J. Liu & Talwar, 2019; Papernot & Steinke, 2022).

These different scenarios can result in the adversary having access to intermediate states of the model, or to periodic releases of the model trained on related data sets. While DP's postprocessing and composition properties make it particularly attractive for use in such systems, auditing generally does not benefit from the same properties. Changes in the adversary's access to the system may thus invalidate the results of prior audits. Thus, research needs to answer a number of questions such as (1) how leakage measured by audits of individual components of a broader ML pipeline need to be composed, (2) how frequently these audits need to be repeated, (3) do we need to develop attacks that specifically target more complex releases, such as continuous/periodic releases of models/statistics, or releases of complex ML bundles that include multiple versions of a model and summary statistics computed on the same data set?

Take the example of unlearning mentioned above. Recent research shows how the vulnerability of a particular data point to membership inference is relative to other points contained in the training set of the model (Carlini et al., 2022). This means that unlearning some of the training points can alter a particular data point's vulnerability to membership inference attacks. Put another way, audits based on membership inference need to be repeated regularly if the model's training set is updated.

**Can Privacy Come From Obscurity in ML Pipelines?:** While the prior were examples of pipelines that can make privacy worse, there are also situations where limited access to the system may be conjectured to improve privacy leakage. Consider two examples of limited information access:

- 'Label only access' restricts an adversary to only having the label outputs of a query to the target model. This is common in many deployments; in spam detection, for example, an adversary can only see whether a spam message is classified as spam or not, and cannot see the spam classifier's predicted probability. Given that many privacy attacks rely on access to the probabilities returned by a model, it is natural to think that privacy attacks can be made impossible through this obscurity. However, it is now known that such restrictions do not prevent leakage: Choquette-Choo et al. (2021) and Z. Li and Zhang (2021) show that attacks are possible in this threat model, by querying in the neighborhood of a target example. This is evidence that such obscurity claims should be justified as well with adaptive attacks.

- 'Final step access' is where a model is produced with a chain of multiple algorithms, and an adversary can only access the output of the final step. This is a common assumption, for example, in nonconvex model training with DP, where DP guarantees are proven for an adversary with access to all iterations of training. It is known that an adversary with all iterates of training is stronger than an adversary with access to only the last iterate, when models are convex (Feldman et al., 2018; Y.-X. Wang et al., 2015). However, there is currently only empirical evidence (Jagielski, Thakkar, et al., 2022; Nasr et al., 2021) that this relaxed threat model improves privacy in the more general nonconvex setting, which may be overturned by stronger attacks.

**Auditing tools beyond attacks:** The current methods of auditing and attacking ML models for privacy issues assume that the auditor has comprehensive access to the model. However, in some cases, an auditor might have limited access to the ML model, such as in the case of a nonprofit organization that needs to audit the privacy claims of a model without direct access to it. In such cases, it becomes challenging to audit the model and determine whether the observed privacy loss or leakage is acceptable, and how to use the results of such audits to inform decisions made by policymakers.

While attacks against the ML model are necessary, they are not sufficient as an auditing tool. Attacks do not always help find bugs in code-implementing algorithms, even when these algorithms come with a privacy proof. Therefore, there is a need to develop techniques that audit implementations and data, rather than just models. For instance, it would be helpful to quantify expected leakage by training models on various data sets and comparing them. Additionally, it would be useful to develop auditing tools that provide explanations as to how privacy is protected by the corresponding system, rather than merely providing a quantitative assessment of privacy.

In conclusion, the goal of auditing is to gather evidence to decide whether it is safe to release a model, show if an adjustment to improve privacy reduced leakage and by how much, and produce a report for a policy committee to make decisions on privacy parameters and privacy-utility trade-offs. Therefore, research needs to develop techniques that audit the implementation of models and data, rather than just models, and provide explanations on how privacy is protected by the corresponding system.

## 5. Beyond DP: Modeling Privacy for Real-World Scenarios

We have so far focused on the importance of preserving user privacy by applying DP to statistical data analyses and machine learning applications. As discussed in prior sections, DP sets a formal limit on any user's influence on the outcome of a computation.

In an ideal world, each actor in the system would learn nothing more than the differentially private output of the computation needed to play their role. For example, if an analyst only needs to determine whether a particular quality metric exceeds a desired threshold in order to authorize deploying the model to end-users, then in an idealized world, only a differentially private version of that bit of information would be available to the analyst; such an analyst would need access to neither the training data nor the model parameters, for instance. Similarly, end-users enjoying the user experiences powered by the trained model might only require differentially private predictions from the model and nothing else. More importantly, in such a world, each user contributing data would know how their data is being used, what it is being used for, and what level of DP is being enforced, and they would have the ability to verify or falsify these claims.

However, it is clear that DP alone is not enough to achieve the above idealized world. Indeed, DP does not describe where the user data is stored, how it is accessed, or who has access to it. Further, DP does not specify how the privacy guarantees are communicated with or verified by the user. In fact, DP serves a particular privacy principle, *the data anonymization principle*, [11] but as the above idealized scenario indicates, there are other important privacy principles including transparency and consent into how data is used, or data minimization approaches that appropriately restrict access to raw data and intermediate computations [(Bonawitz et al., 2022)](#). In this section, we will overview these privacy principles and describe how they combine with DP to provide 'privacy in-depth' guarantees for real-world scenarios.

## 5.1. Data Minimization

We start by discussing a privacy principle focused on data minimization: collecting whatever is necessary for a specific computation (*focused collection*), processing a user's data as early as possible (*early aggregation*), limiting access to data at all stages (*access controls*), and discarding both collected and processed data as soon as possible (*minimal retention*). That is, data minimization implies restricting access to all data to the smallest set of people possible, often accomplished via security mechanisms, such as encryption at rest and on the wire, access-control lists, and more nascent technologies such as secure multiparty computation and trusted execution environments, to be discussed later.

These processes are perhaps best embodied via federated learning and analytics [(Kairouz, McMahan, Avent, et al., 2021](#); [McMahan et al., 2017)](#). Critically, data collection and aggregation are inseparable in the federated approach—purpose-specific transformations of client data are collected for immediate aggregation, with analysts having no access to per-user messages. Federated learning (FL) can be combined with DP and empirical privacy auditing, treated in detail in [Section 4](#), to ensure released aggregates are sufficiently anonymous.

## 5.2. Communicating the Guarantees of DP

DP has seen wide deployment across industry and government organizations as the gold standard of privacy-preserving data analysis, but its mathematical precision makes its privacy guarantees difficult to understand.

Simply describing it as 'the gold standard' instills confidence but does not provide information about the nature of the privacy guarantees. On the other hand, describing it as 'a bound on the worst-case ratio of the probability of a particular output of a randomized algorithm across two neighboring databases' is equally meaningless to those unfamiliar with the definition or without a mathematical background. How, then, should DP be explained to end-users and other stakeholders who may not already be experts in DP? How can these people be empowered to make informed choices about the data protections afforded to their data and the data of others, in relation to the use of DP systems?

### 5.2.1. Communication to End-Users

End-users are those who contribute data toward the systems and participate in the data collection process, who are important stakeholders but often may not have a mathematical background and require extremely brief descriptions. It can be challenging to explain the guarantees of DP without using math. To make things more complex, the protections provided by DP are not absolute and require contextualization to understand their exact meaning. For instance, the $\varepsilon$ parameter quantifies privacy loss in the *worst-case* over all pairs of neighboring databases, but it does not provide *instance-specific* guarantees that match the decisions faced by users in practice. This can be particularly challenging with complex data such as mobility trajectory or health data where the concepts of 'what' is being protected or the 'neighboring pair' that defines the indistinguishability are very nuanced.

Therefore, it is important to effectively communicate the technical guarantees of DP to the individuals whose data will be used in DP systems, and this is a subject of ongoing work that will be a critical challenge to solve in the future of DP systems deployed in practice.

**DP definition:** Recent work of Cummings et al. (2021) and Xiong et al. (2022) showed through a series of user studies that end-users of DP systems care about the kinds of information leaks against which DP protects and are more willing to share their private information when the risks of these leaks are less likely to happen. However, Cummings et al. (2021) also showed that current in-the-wild descriptions of DP led users to have haphazard and incorrect privacy expectations about protections from these leaks. These incorrect expectations persisted when comparing against the ground truth protections in both the local and central models of DP.

Since most existing in-the-wild descriptions of DP can apply equally well to both the local and central models —for example, describing DP as 'the gold standard' or saying that it 'injects statistical noise'—it makes sense that end-users would not be able to distinguish between the two models based only on these descriptions. Future work is needed to develop new DP descriptions that help end-users accurately distinguish between the privacy guarantees offered by these different models of DP, such as that proposed in Smart et al. (2020).

**Explaining $\varepsilon$:** Nontechnical methods for explaining the $\varepsilon$ value are critical for meaningful explanations of the privacy guarantees, since ranging $\varepsilon$ from 0 to $\infty$ can respectively range the privacy guarantees from 'perfect privacy' to 'no privacy.' Simply stating that a system satisfies DP without specifying the $\varepsilon$ value used provides no meaningful information to end-users. However, simply informing end-users of the $\varepsilon$ value without providing guidance on the meaning or interpretation of this parameter is equally meaningless. End-users typically do not reason about privacy in the way quantified by $\varepsilon$, and people are known to struggle with reasoning about risks from low-probability events [(Tversky & Kahneman, 1974)](#).

Recent work of [Nanayakkara et al. (2023)](#) developed and evaluated methods for explaining the probabilistic privacy guarantees to end-users: two that communicate the relative odds of various outcomes with and without the user's data, and one offering concrete examples of outputs sampled from DP mechanisms with the appropriate $\varepsilon$. Odds-based explanation methods were found to be more effective than output-based methods, as well as existing DP explanation methods that do not explicitly describe $\varepsilon$.

This initial work suggests a promising path for future explanatory tools. Future work should extend this line of inquiry to both develop improved methods for communicating the privacy guarantees, as well as methods for evaluating efficacy of new methods. Equipped with effective explanations of DP that enable end-users to understand the privacy guarantees they are being offered through DP systems, privacy researchers can also elicit their preferences over the $\varepsilon$ values in different contexts of data use to understand the trade-off faced by consumers between privacy (as measured by $\varepsilon$), accurate analysis of their data, and utility (as measured by money). Such understanding would be extremely valuable to DP practitioners, who must tune the $\varepsilon$ to balance privacy needs of users with accuracy needs of analysis, and can possibly compensate end-users for their privacy loss, either directly with payments or indirectly through improved services.

**Other implementation hyperparameters:** Beyond the $\varepsilon$ value used in a DP system, other details of the implementation affect the privacy guarantees provided to users, such as the $\delta$ value, the algorithm used, and how the privacy budget is composed across multiple accesses of the user's data. For example, one call to the randomized response mechanism with $\varepsilon = 10$ may be different in the eyes of an end-user from 10 calls to the Gaussian mechanism, each with $\varepsilon = 1$.

These description tools may draw influence from existing work on *privacy nutrition labels* [(Kelley et al., 2009)](#) that will highlight salient privacy features of a technical system. In the context of DP, this may include general privacy properties of DP, as well as details of a specific DP implementation, such as local/central privacy, the DP algorithm and its privacy parameter ($\varepsilon, \delta$), and information about privacy budget composition across multiple queries.

Disclosing the $\varepsilon$ value and other relevant hyperparameters of their DP system will allow companies and other organizations to improve transparency to end-users, such as the 'Epsilon registry' proposed in [Dwork et al. (2019)](#); such disclosures will only be meaningful when end-users are also given means of interpreting the parameters and their impact on real-world privacy guarantees. Beyond simply informing end-users of the privacy they will receive, organizations should provide end-users with meaningful choices so that users can have more control over their own data. These *user-centric controls* may be related to DP—such as choosing the $\varepsilon$ value that their data will receive, or choosing whether their data are collected under the local or central model—or beyond the scope of DP—such as opting out of data collection without opting out of the service, or invoking the GDPR's 'right to be forgotten.'

## 5.2.2. Communication to Other Stakeholders

In addition to communicating to end-users, there are many other stakeholders involved in the implementation of DP in practice. These other stakeholders represent the metaphorical DP supply chain, including *engineers* who must implement and test DP systems, *data curators* who must manage access to databases through DP systems, *executives* who must decide whether DP systems should be used in their organizations, *lawyers* who must determine which DP systems are compliant with laws and regulations, and *policymakers* who have the power to enact new regulations surrounding the use of DP systems. While each of these stakeholders has their own relevant areas of expertise, they are not assumed to be experts in DP, and will need explanatory tools to aid them in making informed decisions surrounding the use of DP. Suggestions for future explanatory methods tailored to each audience are briefly described below.

**Engineers and technical managers:** Engineers—and their managers—must implement and test DP algorithms for the private analysis needs of their organizations and teams. This involves both ensuring that the algorithms and their privacy guarantees are correct, as well as making more nuanced decisions, such as managing privacy composition across multiple queries and choosing the per-query $\varepsilon$ based on use-case specific privacy and accuracy needs. These engineers are expected to have a high level of technical knowledge, coming both from formal academic training and on-the-job experience. However, unless they have previously worked on privacy products, most of these engineering teams are unfamiliar with both verifying correctness of DP guarantees and the privacy-accuracy consequences of various parameter choices.

One partial solution is developing open source DP libraries and repositories that are vetted by experts, such as [Google (2023)](#), [Holohan et al. (2019)](#), ["OpenDP" (2022)](#), and [Yousefpour et al. (2021)](#). While these libraries will not be able to provide code for every possible DP algorithm, they will provide a foundation of verified-correct DP algorithms that can be used as building blocks in larger, more complex DP systems. Additionally, the DP community should develop a visual analysis tool that enables the engineer to visualize the privacy and accuracy guarantees that result from different $\varepsilon$ values. A partial version of this tool already exists in [Hay et al. (n.d.)](#), but adding features, such as different types of queries with different sensitivities, composition across multiple queries, and allowing the engineer to specify their own data format (such as number of features and range of data values) will bring the visualization closer to the needs of practice.

**Data curators:** Data curators are charged with facilitating access to data sets when such data are needed, and also governing appropriate use of the data sets. These may be internal-facing curators, such as a committee inside an organization allocating access to data for other teams within the organization, or external-facing curators, such as the US Census Bureau's Federal Statistical Research Data Centers [(United States Census](#)

Bureau, n.d.), which provide regulated access to confidential data collected by federal statistical agencies. Data curators are accustomed to determining when access to protected data set should be allowed or not, which typically include administrative application, review, and approval processes. However, recent work (Sarathy et al., 2023) showed the novel challenges that arise when data curators integrate DP into their process.

For example, when engineers or data analysts are interested in identifying a data set that is useful for their analysis task, the curator may wish to provide a summary of the metadata associated with each potential data set. Since these metadata may be provided many times to many different analysts, several questions arise: what metadata should be provided and at what level of privacy in order to provide useful information about the applicability of the data set to the specific task, while not exhausting the privacy budget? Additionally, as multiple analysts access the same data set, the curator may be required to manage the overall privacy budget of the data set, and choose the privacy budget allocated to each analyst. These privacy budgets could be uniform to reduce the curator's overhead, or they could vary based on the relative importance of the analysis task, or based on the level of trust (e.g., higher levels of oversight may enable larger privacy budgets). For these and other real-world challenges faced by data curators adopting DP into their data management process, the DP community can provide explanatory methods and best practice documents as guidance for data curators.

**Executives:** Even when engineering and research teams are convinced of the virtues of DP, high-level executives are the ones who are empowered to decide whether or not DP systems should be used in their organizations. The incentives, goals, and perspectives of these executives may differ substantially from those of their more technical employees. Executives are not typically focused on adopting cutting-edge privacy tools like DP, but instead on improving performance of their organization along key corporate metrics, such as revenue, market share, or consumer satisfaction.

For this audience, the language of *privacy-accuracy trade-off*—which is commonly used in technical descriptions of DP—may be unappealing, as it implies that introducing DP necessarily harms accuracy and potentially key performance metrics that rely on data. Instead, one can expect a more amenable response from language that emphasizes the benefits of DP, such as improved trust from users (Lovejoy, 2017), improved quality of data collected from users (Cummings et al., 2016), and protection against practical attacks (Salem et al., 2022) (as discussed in Section 4) and data leaks (Carlini et al., 2019). Additionally, executives will be more likely to choose to adopt novel privacy tools if there is clear demand from the organization's users, and if it is clear that doing so will be compliant with current and future privacy laws. Aside from new notable exceptions, privacy is not a dimension along which companies compete. Increased pressure from users, perhaps as a result of the steps described in Section 5.2.1, may change this. Steps for improving the connection between DP tools and privacy laws are discussed next.

**Lawyers:** Before an organization can adopt DP, corporate and privacy lawyers are left with the difficult task of determining whether a particular DP system—or perhaps DP in general—is compliant with all privacy laws. Unfortunately, current privacy laws do not provide explicit guidance on this question, so legal scholarship is required to interpret the language of each relevant privacy law. This introduces several challenges, including: (1) the mathematically nuanced language of DP is very different from the language of the law, (2) the continuous $\varepsilon$ privacy parameter must be reconciled with binary compliance requirements of the law, and (3) the legal landscape of privacy is a patchwork of geographic- and domain-specific regulations, rather than a single consistent privacy law.

Some progress has been made on the first challenge by Wood et al. (2018), which provides a minimally technical survey of DP basics in language that is familiar to legal scholars. Other partial progress has been made in clearly articulating that DP satisfies the privacy requirements of specific privacy laws (Nissim et al., 2018) and that it provides protections commonly required in privacy laws (Cohen & Nissim, 2020). Much more work is needed in this direction to provide comprehensive and easily available answers to lawyers for the questions they face surrounding the use of DP in real-world systems.

**Policymakers:** If current privacy laws are determined to be unclear or underspecified—or alternatively, as new technological advances are made in the field of privacy—then regulators and policymakers are charged with developing new regulations to guide and govern the use of DP in practice. One major challenge is that there is not consistent guidance from the DP community on best practices for the use of DP. The most notable issue is the question of appropriate choice of $\varepsilon$ (see Section 5.2.1 for more discussion). It is widely agreed that the choice of $\varepsilon$ should vary with the context of use, and should balance privacy and accuracy needs of the application domain. However, there is no concrete legal or numerical guidance on how these considerations should translate into a choice of appropriate $\varepsilon$ value (Dwork et al., 2019).

The DP community can contribute to improved technical guidance for policymakers by engaging more directly with the policy and regulatory communities. This may include work that is typically considered to be outside the workload of an academic or industry researcher, such as writing white papers or responding to government requests for public comments, Advance Notice of Proposed Rulemaking, or requests for information. This may also include developing educational programs targeted specifically at regulators, to equip them with the necessary knowledge to make policy decisions that are informed and consistent with the best technical practices.

## Acknowledgments

We thank the two anonymous referees for their constructive comments that helped improve the presentation of this work. We also thank all the workshop participants who contributed significant discussions for the paper. Full list of participants beyond the authors: Borja Balle, Will Bullock, Nicholas Carlini, Graham Cormode, Kamalika Chaudhuri, Seyi Feyisetan, Miguel Guevara, Monika Henzinger, James Honaker, Huseyin A. Inan, Krishnaram Kenthapadi, Aleksandra Korolova, Sara Krehbiel, Janardhan Kulkarni, Ravi Kumar, Mathias Lecuyer, Chen-Kuei Lee, Brendan McMahan, Gerome Miklau, Ilya Mironov, Milad Nasr, John Nguyen, Oana

Niculaescu, Ananth Raghunathan, Aaron Roth, Mikhail Rudoy, Michael Shoemate, Adam Smith, Thomas Steinke, Mukund Sundararajan, Om Thakkar, Florian Tramèr, Praneeth Vepakomma, and Steven Wu.

## Disclosure Statement

Rachel Cummings, Damien Desfontaines, David Evans, Roxana Geambasu, Yangsibo Huang, Matthew Jagielski, Peter Kairouz, Gautam Kamath, Sewoong Oh, Olga Ohrimenko, Nicolas Papernot, Ryan Rogers, Milan Shen, Shuang Song, Weijie Su, Andreas Terzis, Abhradeep Thakurta, Sergei Vassilvitskii, Yu-Xiang Wang, Li Xiong, Sergey Yekhanin, Da Yu, Huanyu Zhang, and Wanrong Zhang have no financial or non-financial disclosures to share for this article.

## References

NIST. (2022, December 30) 2020 differential privacy temporal map challenge. https://www.nist.gov/ctl/pscr/open-innovation-prize-challenges/past-prize-challenges/2020-differential-privacy-temporal

Abadi, M., Chu, A., Goodfellow, I., McMahan, H. B., Mironov, I., Talwar, K., & Zhang, L. (2016). Deep learning with differential privacy. In *CCS '16: Proceedings of the 2016 ACM Conference on Computer and Communications Security* (pp. 308–318). Association for Computing Machinery. https://doi.org/10.1145/2976749.2978318

Abowd, J., Ashmead, R., Cumings-Menon, R., Garfinkel, S., Heineck, M., Heiss, C., Johns, R., Kifer, D., Leclerc, P., Machanavajjhala, A., Moran, B., Sexton, W., Spence, M., & Zhuravlev, P. (2022). The 2020 Census disclosure avoidance system TopDown Algorithm. *Harvard Data Science Review*, (Special Issue 2). https://doi.org/10.1162/99608f92.529e3cb9

Abowd, J. M. (2018). The U.S. Census Bureau adopts differential privacy. In *Proceedings of the 24th ACM SIGKDD International Conference on Knowledge Discovery & Data Mining* (p. 2867). Association for Computing Machinery. https://doi.org/10.1145/3219819.3226070

Acharya, J., Canonne, C. L., Freitag, C., & Tyagi, H. (2019). Test without trust: Optimal locally private distribution testing. In K. Chaudhuri, & M. Sugiyama (Eds.), *Proceedings of the 22nd International Conference on Artificial Intelligence and Statistics* (Vol. 89, pp. 2067–2076). Proceedings of Machine Learning Research. https://proceedings.mlr.press/v89/acharya19b.html

Acharya, J., Canonne, C. L., Liu, Y., Sun, Z., & Tyagi, H. (2021). Interactive inference under information constraints. *IEEE Transactions on Information Theory*, *68*(1), 502–516. https://doi.org/10.1109/TIT.2021.3123905

Acharya, J., Canonne, C. L., Sun, Z., & Tyagi, H. (2022). The role of interactivity in structured estimation. In P.-L. Loh, & M. Raginsky (Eds.), *Proceedings of Thirty Fifth Conference on Learning Theory* (Vol. 178, pp. 1328–1355. Proceedings of Machine Learning Research. https://proceedings.mlr.press/v178/acharya22b.html

Acharya, J., Canonne, C. L., & Tyagi, H. (2019). Inference under information constraints: Lower bounds from chi-square contraction. In A. Beygelzimer, & D. Hsu (Eds.), *Proceedings of the 32nd Annual Conference on Learning Theory* (Vol. 99, pp. 3–17). Proceedings of Machine Learning Research. http://proceedings.mlr.press/v99/acharya19a.html

Acharya, J., Kairouz, P., Liu, Y., & Sun, Z. (2021). Estimating sparse discrete distributions under privacy and communication constraints. In V. Feldman, K. Ligett, & S. Sabato (Eds.), *Proceedings of the 32nd International Conference on Algorithmic Learning Theory* (Vol. 132, pp. 79–98). Proceedings of Machine Learning Research. https://proceedings.mlr.press/v132/acharya21b.html

Acharya, J., Kamath, G., Sun, Z., & Zhang, H. (2018). INSPECTRE: Privately estimating the unseen. In J. Dy, & A. Krause (Eds.), *Proceedings of the 35th International Conference on Machine Learning* (Vol. 80, pp. 30–39). Proceedings of Machine Learning Research. https://proceedings.mlr.press/v80/acharya18a.html

Acharya, J., & Sun, Z. (2019). Communication complexity in locally private distribution estimation and heavy hitters. In K. Chaudhuri, & R. Salakhutdinov (Eds.), *Proceedings of the 36th International Conference on Machine Learning* (Vol. 97, pp. 51–60). Proceedings of Machine Learning Research https://proceedings.mlr.press/v97/acharya19c.html

Acharya, J., Sun, Z., & Zhang, H. (2018). Differentially private testing of identity and closeness of discrete distributions. In S. Bengio, H. Wallach, H. Larochelle, K. Grauman, N. Cesa-Bianchi, & R. Garnett (Eds.), *Advances in neural information processing systems* (Vol. 31, pp. 6878–6891). Curran Associates. https://proceedings.neurips.cc/paper/2018/hash/7de32147a4f1055bed9e4faf3485a84d-Abstract.html

Acharya, J., Sun, Z., & Zhang, H. (2021). Robust testing and estimation under manipulation attacks. In M. Meila & T. Zhang (Eds.), *Proceedings of the 38th International Conference on Machine Learning* (Vol. 139, pp. 43–53). Proceedings of Machine Learning Research. https://proceedings.mlr.press/v139/acharya21a.html

Agarwal, S. (2020). *Trade-offs between fairness, interpretability, and privacy in machine learning* (Master's thesis, University of Waterloo). http://hdl.handle.net/10012/15861

Alabi, D., Kothari, P. K., Tankala, P., Venkat, P., & Zhang, F. (2023). Privately estimating a Gaussian: Efficient, robust, and optimal. In *Proceedings of the 55th Annual ACM Symposium on the Theory of Computing* (pp. 483–496). Association for Computing Machinery. https://doi.org/10.1145/3564246.3585194

Aliakbarpour, M., Diakonikolas, I., & Rubinfeld, R. (2018). Differentially private identity and equivalence testing of discrete distributions. In J. Dy, & A. Krause (Eds.), *Proceedings of the 35th International Conference* on *Machine Learning* (Vol. 80, pp. 169–178). Proceedings of Machine Learning Research. https://proceedings.mlr.press/v80/aliakbarpour18a.html

Alon, N., Bassily, R., & Moran, S. (2019). Limits of private learning with access to public data. In H. Wallach, H. Larochelle, A. Beygelzimer, F. d'Alché-Buc, E. Fox, & R. Garnett (Eds.), *Advances in neural information processing systems* (Vol. 32, pp. 10342–10352). Curran Associates. https://papers.nips.cc/paper_files/paper/2019/hash/9a6a1aaafe73c572b7374828b03a1881-Abstract.html

Altschuler, J. M., & Talwar, K. (2022). *Privacy of noisy stochastic gradient descent: More iterations without more privacy loss*. ArXiv. https://doi.org/10.48550/arXiv.2205.13710

Amid, E., Ganesh, A., Mathews, R., Ramaswamy, S., Song, S., Steinke, T., Suriyakumar, V. M., Thakkar, O., & Thakurta, A. (2021). *Public data-assisted mirror descent for private model training*. ArXiv. https://doi.org/10.48550/arXiv.2112.00193

Amin, K., Dick, T., Khodak, M., & Vassilvitskii, S. (2022). *Private algorithms with private predictions*. ArXiv. https://doi.org/10.48550/arXiv.2210.11222

Amin, K., Dick, T., Kulesza, A., Munoz, A., & Vassilvitskii, S. (2019). Differentially private covariance estimation. In H. Wallach, H. Larochelle, A. Beygelzimer, F. d'Alché-Buc, E. Fox, & R. Garnett (Eds.), *Advances in neural information processing systems* (Vol. *32,* pp. 14190–14199). Curran Associates. https://papers.nips.cc/paper_files/paper/2019/hash/4158f6d19559955bae372bb00f6204e4-Abstract.html

Amin, K., Gillenwater, J., Joseph, M., Kulesza, A., & Vassilvitskii, S. (2022). *Plume: Differential privacy at scale.* ArXiv. https://doi.org/10.48550/arXiv.2201.11603

Amin, K., Joseph, M., Ribero, M., & Vassilvitskii, S. (2022). *Easy differentially private linear regression*. ArXiv. https://doi.org/10.48550/arXiv.2208.07353

Amin, K., Kulesza, A., Munoz, A., & Vassilvtiskii, S. (2019). Bounding user contributions: A bias-variance trade-off in differential privacy. In K. Chaudhuri, & R. Salakhutdinov (Eds.), *Proceedings of the 36th International Conference on Machine Learning* (Vol. 97, pp. 263–271). Proceedings of Machine Learning Research. https://proceedings.mlr.press/v97/amin19a.html

Andrés, M. E., Bordenabe, N. E., Chatzikokolakis, K., & Palamidessi, C. (2013). Geo-indistinguishability: Differential privacy for location-based systems. In *CCS'13: Proceedings of the 2013 ACM SIGSAC Conference on Computer & Communications Security*, (pp. 901–914). Association for Computer Machinery. https://doi.org/10.1145/2508859.2516735

Andrew, G., Thakkar, O., McMahan, B., & Ramaswamy, S. (2021). Differentially private learning with adaptive clipping. In M. Ranzato, A. Beygelzimer, Y. Dauphin, P.S. Liang, & J. Wortman Vaughan (Eds.),

*Advances in neural information processing systems* (Vol. 34, pp. 17455–17466). Curran Associates. https://proceedings.neurips.cc/paper/2021/hash/91cff01af640a24e7f9f7a5ab407889f-Abstract.html

Andrysco, M., Kohlbrenner, D., Mowery, K., Jhala, R., Lerner, S., & Shacham, H. (2015). On subnormal floating point and abnormal timing. In *2015 IEEE Symposium on Security and Privacy* (pp. 623–639). IEEE. https://doi.org/10.1109/SP.2015.44

Apple Differential Privacy Team (2017, December 6). Learning with privacy at scale. *Apple Machine Learning Research*. https://machinelearning.apple.com/research/learning-with-privacy-at-scale

Arora, R., Bassily, R., González, T., Guzmán, C., Menart, M., & Ullah, E. (2022). *Faster rates of convergence to stationary points in differentially private optimization*. ArXiv. https://doi.org/10.48550/arXiv.2206.00846

Asi, H., Duchi, J., Fallah, A., Javidbakht, O., & Talwar, K. (2021). Private adaptive gradient methods for convex optimization. In M. Meila, & T. Zhang (Eds.), *Proceedings of the 38th International Conference on Machine Learning* (Vol. 139, pp. 383–392). Proceedings of Machine Learning Research. https://proceedings.mlr.press/v139/asi21a.html

Asi, H., & Duchi, J. C. (2020). *Near instance-optimality in differential privacy*. ArXiv. https://doi.org/10.48550/arXiv.2005.10630

Asi, H., Feldman, V., Koren, T., & Talwar, K. (2021). *Private stochastic convex optimization: Optimal rates in ℓ1 geometry*. ArXiv. https://doi.org/10.48550/arXiv.2103.01516

Asi, H., Ullman, J., & Zakynthinou, L. (2023). *From robustness to privacy and back*. ArXiv. https://doi.org/10.48550/arXiv.2302.01855

Ateniese, G., Mancini, L. V., Spognardi, A., Villani, A., Vitali, D., & Felici, G. (2015). Hacking smart machines with smarter ones: How to extract meaningful data from machine learning classifiers. *International Journal of Security and Networks*, *10*(3), 137–150. https://doi.org/10.1504/IJSN.2015.071829

Augenstein, S., McMahan, H. B., Ramage, D., Ramaswamy, S., Kairouz, P., Chen, M., Mathews, R., et al. (2019). *Generative models for effective ML on private, decentralized datasets*. ArXiv. https://doi.org/10.48550/arXiv.1911.06679

autodp contributors. (2019). *Autodp: Automating differential privacy computation*. GitHub. https://github.com/yuxiangw/autodp

Avella-Medina, M., & Brunel, V.-E. (2019). *Differentially private sub-Gaussian location estimators*. ArXiv. https://doi.org/10.48550/arXiv.1906.11923

Bagdasaryan, E., Poursaeed, O., & Shmatikov, V. (2019). Differential privacy has disparate impact on model accuracy. In H. Wallach, H. Larochelle, A. Beygelzimer, F. d'Alché-Buc, E. Fox, & R. Garnett (Eds.), *Advances in neural information processing systems* (Vol. 32, pp. 15479–15488). Curran Associates. https://proceedings.neurips.cc/paper/2019/hash/fc0de4e0396fff257ea362983c2dda5a-Abstract.html

Balle, B., Cherubin, G., & Hayes, J. (2022). Reconstructing training data with informed adversaries. *2022 IEEE Symposium on Security and Privacy* (pp. 1138–1156). IEEE. https://doi.org/10.1109/SP46214.2022.9833677

Barber, R. F., & Duchi, J. C. (2014). *Privacy and statistical risk: Formalisms and minimax bounds*. ArXiv. https://doi.org/10.48550/arXiv.1412.4451

Bassily, R., Cheu, A., Moran, S., Nikolov, A., Ullman, J., & Wu, S. (2020). Private query release assisted by public data. In H. Daumé III, & A. Singh (Eds.), *Proceedings of the 37th International Conference on Machine Learning* (Vol. 119, pp. 695–703). Proceedings of Machine Learning Research. https://proceedings.mlr.press/v119/bassily20a.html

Bassily, R., Feldman, V., Talwar, K., & Thakurta, A. G. (2019). Private stochastic convex optimization with optimal rates. In H. Wallach, H. Larochelle, A. Beygelzimer, F. d'Alché-Buc, E. Fox, & R. Garnett (Eds.), *Advances in neural information processing systems* (Vol. 32, pp. 11282–11291). Curran Associates. https://proceedings.neurips.cc/paper_files/paper/2019/hash/3bd8fdb090f1f5eb66a00c84dbc5ad51-Abstract.html

Bassily, R., Guzmán, C., & Menart, M. (2021). Differentially private stochastic optimization: New results in convex and non-convex settings. In M. Ranzato, A. Beygelzimer, Y. Dauphin, P.S. Liang, & J. Wortman Vaughan (Eds.), *Advances in neural information processing systems* (Vol. 34, pp. 9317–9329). Curran Associates. https://proceedings.neurips.cc/paper/2021/hash/4ddb5b8d603f88e9de689f3230234b47-Abstract.html

Bassily, R., Moran, S., & Nandi, A. (2020). Learning from mixtures of private and public populations. In H. Larochelle, M. Ranzato, R. Hadsell, M.F. Balcan, & H. Lin (Eds.), *Advances in neural information processing systems* (Vol. 33, pp. 2947–2957). Curran Associates. https://proceedings.neurips.cc/paper/2020/hash/1ee942c6b182d0f041a2312947385b23-Abstract.html

Bassily, R., Smith, A., & Thakurta, A. (2014). Private empirical risk minimization: Efficient algorithms and tight error bounds. In *Proceedings of the 55th Annual IEEE Symposium on Foundations of Computer Science* (pp. 464–473). IEEE. https://doi.org/10.1109/FOCS.2014.56

Bassily, R., Thakkar, O., & Guha Thakurta, A. (2018). Model-agnostic private learning. In S. Bengio, H. Wallach, H. Larochelle, K. Grauman, N. Cesa-Bianchi, & R. Garnett (Eds.), *Advances in neural information* *processing systems* (Vol. 31, pp. 7102–7112). Curran Associates. https://papers.nips.cc/paper_files/paper/2018/hash/aa97d584861474f4097cf13ccb5325da-Abstract.html

Beimel, A., Brenner, H., Kasiviswanathan, S. P., & Nissim, K. (2014). Bounds on the sample complexity for private learning and private data release. *Machine Learning*, *94*(3), 401–437. https://doi.org/10.1007/s10994-013-5404-1

Beimel, A., Nissim, K., & Stemmer, U. (2013). Private learning and sanitization: Pure vs. approximate differential privacy. In *Lecture notes in computer science: Vol. 8096 Approximation, randomization, and combinatorial optimization. Algorithms and techniques* (pp. 363–378). Springer. https://doi.org/10.1007/978-3-642-40328-6_26

Beimel, A., Nissim, K., & Stemmer, U. (2015). Learning privately with labeled and unlabeled examples. In *Proceedings of the 26th Annual ACM-SIAM Symposium on Discrete Algorithms* (pp. 461–477). Society for Industrial and Applied Mathematics. https://dl.acm.org/doi/abs/10.5555/2722129.2722161

Bie, A., Kamath, G., & Singhal, V. (2022). *Private estimation with public data*. ArXiv. https://doi.org/10.48550/arXiv.2208.07984

Biswas, S., Dong, Y., Kamath, G., & Ullman, J. (2020). Coinpress: Practical private mean and covariance estimation. In H. Larochelle, M. Ranzato, R. Hadsell, M.F. Balcan, & H. Lin (Eds.), *Advances in neural information processing systems* (Vol. 33, pp. 14475–14485). Curran Associates. https://proceedings.neurips.cc/paper_files/paper/2020/hash/a684eceee76fc522773286a895bc8436-Abstract.html

Błasiok, J., Bun, M., Nikolov, A., & Steinke, T. (2019). Towards instance-optimal private query release. In *Proceedings of the Thirtieth Annual ACM-SIAM Symposium on Discrete Algorithms* (pp. 2480–2497). Society for Industrial and Applied Mathematics. https://doi.org/10.1137/1.9781611975482.152

Boenisch, F., Dziedzic, A., Schuster, R., Shamsabadi, A. S., Shumailov, I., & Papernot, N. (2021). *When the curious abandon honesty: Federated learning is not private*. ArXiv. https://doi.org/10.48550/arXiv.2112.02918

Bolukbasi, T., Chang, K.-W., Zou, J. Y., Saligrama, V., & Kalai, A. T. (2016). Man is to computer programmer as woman is to homemaker? In D. Lee, M. Sugiyama, U. Luxburg, I. Guyon, & R. Garnett (Eds.), Debiasing word embeddings. *Advances in Neural Information Processing Systems 29* (pp. 4349–4357). https://papers.nips.cc/paper_files/paper/2016/hash/a486cd07e4ac3d270571622f4f316ec5-Abstract.html

Bommasani, R., Hudson, D. A., Adeli, E., Altman, R., Arora, S., von Arx, S., Bernstein, M. S., Bohg, J., Bosselut, A., Brunskill, E., Brynjolfsson, E., Buch, S., Card, D., Castellon, R., Chatterji, N., Chen, A., Creel, K., Davis, J. Q., Demszky, D., . . . Liang, P. (2021). *On the opportunities and risks of foundation models*. ArXiv. https://doi.org/10.48550/arXiv.2108.07258

Harvard Data Science Review • Issue 6.1, Winter 2024    Advancing Differential Privacy: Where We Are Now and Future Directions for Real-World DeploymentBonawitz, K., Kairouz, P., Mcmahan, B., & Ramage, D. (2022). Federated learning and privacy. *Communications of the ACM*, *65*(4), 90–97. https://doi.org/10.1145/3500240

Bourtoule, L., Chandrasekaran, V., Choquette-Choo, C. A., Jia, H., Travers, A., Zhang, B., Lie, D., & Papernot, N. (2021). Machine unlearning. In *2021 IEEE Symposium on Security and Privacy* (pp. 141–159). IEEE. https://doi.org/10.1109/SP40001.2021.00019

Brown, G., Gaboardi, M., Smith, A., Ullman, J., & Zakynthinou, L. (2021). Covariance-aware private mean estimation without private covariance estimation. In M. Ranzato, A. Beygelzimer, Y. Dauphin, P.S. Liang, & J. Wortman Vaughan (Eds.), *Advances in neural information processing systems* (Vol. 43, pp. 7950–7964). Curran Associates. https://proceedings.neurips.cc/paper/2021/hash/42778ef0b5805a96f9511e20b5611fce-Abstract.html

Brown, G., Hopkins, S. B., & Smith, A. (2023). *Fast, sample-efficient, affine-invariant private mean and covariance estimation for subgaussian distributions*. ArXiv. https://doi.org/10.48550/arXiv.2301.12250

Brown, H., Lee, K., Mireshghallah, F., Shokri, R., & Tramèr, F. (2022). What does it mean for a language model to preserve privacy? In *Proceedings of the 2022 ACM Conference on Fairness, Accountability, and Transparency* (pp. 2280–2292). Association for Computer Machinery. https://doi.org/10.1145/3531146.3534642

Brunel, V.-E., & Avella-Medina, M. (2020). *Propose, test, release: Differentially private estimation with high probability*. ArXiv. https://doi.org/10.48550/arXiv.2002.08774

Bu, Z., Dong, J., Long, Q., & Su, W. J. (2020). Deep learning with Gaussian differential privacy. *Harvard Data Science Review*, *2*(3). https://doi.org/10.1162/99608f92.cfc5dd25

Bun, M., Desfontaines, D., Dwork, C., Naor, M., Nissim, K., Roth, A., Smith, A., Steinke, T., Ullman, J., & Vadhan, S. (2021, June 3). *Statistical inference is not a privacy violation*. DifferentialPrivacy.org. https://differentialprivacy.org/inference-is-not-a-privacy-violation/.

Bun, M., Dwork, C., Rothblum, G. N., & Steinke, T. (2018). Composable and versatile privacy via truncated CDP. *Proceedings of the 50th Annual ACM SIGACT Symposium on Theory of Computing* (pp. 74–86). Association for Computer Machinery. https://doi.org/10.1145/3188745.3188946

Bun, M., Kamath, G., Steinke, T., & Wu, Z. S. (2019). Private hypothesis selection. In H. Wallach, H. Larochelle, A. Beygelzimer, F. d'Alché-Buc, E. Fox, & R. Garnett (Eds.), *Advances in neural information processing systems* (Vol. 32, pp. 156–167). Curran Associates. https://proceedings.neurips.cc/paper_files/paper/2019/hash/9778d5d219c5080b9a6a17bef029331c-Abstract.html
45

Harvard Data Science Review • Issue 6.1, Winter 2024    Advancing Differential Privacy: Where We Are Now and Future Directions for Real-World DeploymentBun, M., & Steinke, T. (2016). Concentrated differential privacy: Simplifications, extensions, and lower bounds. In M. Hirt, & A. Smith (Eds), *Lecture notes in computer science: Vol. 9985. Theory of Cryptography* (pp. 635–658). https://doi.org/10.1007/978-3-662-53641-4_24

Bun, M., & Steinke, T. (2019). Average-case averages: Private algorithms for smooth sensitivity and mean estimation. In H. Wallach, H. Larochelle, A. Beygelzimer, F. d'Alché-Buc, E. Fox, & R. Garnett (Eds.), *Advances in neural information processing systems* (Vol. 32, 181–191). Curran Associates. https://proceedings.neurips.cc/paper_files/paper/2019/hash/3ef815416f775098fe977004015c6193-Abstract.html

Bun, M., Steinke, T., & Ullman, J. (2019). Make up your mind: The price of online queries in differential privacy. *Journal of Privacy and Confidentiality, 9*(1).. https://doi.org/10.29012/jpc.655

Bun, M., Ullman, J., & Vadhan, S. (2014). Fingerprinting codes and the price of approximate differential privacy. In *Proceedings of the 46th Annual ACM Symposium on the Theory of Computing* (pp. 1–10). Association for Computer Machinery. https://doi.org/10.1145/2591796.2591877

Buolamwini, J., & Gebru, T. (2018). Gender shades: Intersectional accuracy disparities in commercial gender classification. In S. A. Friedler, & C. Wilson (Eds.), *Proceedings of the 1st Conference on Fairness, Accountability, and Transparency* (Vol. 81, pp. 77–91). Proceedings of Machine Learning Research. https://proceedings.mlr.press/v81/buolamwini18a.html

Cai, B., Daskalakis, C., & Kamath, G. (2017). Priv'IT: Private and sample efficient identity testing. In D. Precup, & Y. W. Teh (Eds.), *Proceedings of the 34th International Conference on Machine Learning* (Vol. 70, pp. 635–644). Proceedings of Machine Learning Research. https://proceedings.mlr.press/v70/cai17a.html

Cai, T. T., Wang, Y., & Zhang, L. (2019). *The cost of privacy: Optimal rates of convergence for parameter estimation with differential privacy*. ArXiv. https://doi.org/10.48550/arXiv.1902.04495

Cai, T. T., Wang, Y., & Zhang, L. (2020). *The cost of privacy in generalized linear models: Algorithms and minimax lower bounds*. ArXiv. https://doi.org/10.48550/arXiv.2011.03900

Caliskan, A., Bryson, J. J., & Narayanan, A. (2017). Semantics derived automatically from language corpora contain human-like biases. *Science, 356*(6334), 183–186. https://doi.org/10.1126/science.aal4230

Campbell, Z., Bray, A., Ritz, A., & Groce, A. (2018). Differentially private ANOVA testing. In *2018 International Conference on Data Intelligence and Security* (pp. 281–285). IEEE. https://doi.org/10.1109/ICDIS.2018.00052

Canonne, C. L., Kamath, G., McMillan, A., Smith, A., & Ullman, J. (2019). The structure of optimal private tests for simple hypotheses. In *Proceedings of the 51st Annual ACM Symposium on the Theory of Computing*
46

(pp. 310–321). Association for Computer Machinery. https://doi.org/10.1145/3313276.3316336

Canonne, C. L., Kamath, G., McMillan, A., Ullman, J., & Zakynthinou, L. (2020). Private identity testing for high-dimensional distributions. In H. Larochelle, M. Ranzato, R. Hadsell, M.F. Balcan, & H. Lin (Eds.), *Advances in neural information processing systems* (Vol. 33 pp. 10099–10111). Curran Associates. https://proceedings.neurips.cc/paper_files/paper/2020/hash/72b32a1f754ba1c09b3695e0cb6cde7f-Abstract.html

Cao, Y., & Yang, J. (2015). Towards making systems forget with machine unlearning. In *2015 IEEE Symposium on Security and Privacy* (pp. 463–480). IEEE. https://doi.org/10.1109/SP.2015.35

Carlini, N., Chien, S., Nasr, M., Song, S., Terzis, A., & Tramer, F. (2021). *Membership inference attacks from first principles*. ArXiv. https://doi.org/10.48550/arXiv.2112.03570

Carlini, N., Hayes, J., Nasr, M., Jagielski, M., Sehwag, V., Tramèr, F., Balle, B., Ippolito, D., & Wallace, E. (2023). *Extracting training data from diffusion models*. ArXiv. https://doi.org/10.48550/arXiv.2301.13188

Carlini, N., Jagielski, M., & Mironov, I. (2020). Cryptanalytic extraction of neural network models. In D. Micciancio, & T. Ristenpart (Eds.), *Lecture notes in computer science: Vol. 12172. Advances in Cryptology – CRYPTO 2020* (pp. 189–218). Springer. https://doi.org/10.1007/978-3-030-56877-1_7

Carlini, N., Jagielski, M., Papernot, N., Terzis, A., Tramer, F., & Zhang, C. (2022). *The privacy onion effect: Memorization is relative*. ArXiv. https://doi.org/10.48550/arXiv.2206.10469

Carlini, N., Liu, C., Erlingsson, Ú., Kos, J., & Song, D. (2019). The secret sharer: Evaluating and testing unintended memorization in neural networks. In *Proceedings of the 28th USENIX Conference on Security Symposium* (pp. 267–284). USENIX Association. https://www.usenix.org/system/files/sec19-carlini.pdf

Carlini, N., Tramèr, F., Wallace, E., Jagielski, M., Herbert-Voss, A., Lee, K., Roberts, A., Brown, T. B., Song, D., Erlingsson, Ú., Oprea, A., Raffel, C. (2021). Extracting training data from large language models. In *Proceedings of the 30th USENIX Security Symposium* (pp. 2633–2650). USENIX Association. https://www.usenix.org/system/files/sec21-carlini-extracting.pdf

Cattan, Y., Choquette-Choo, C. A., Papernot, N., & Thakurta, A. (2022). *Fine-tuning with differential privacy necessitates an additional hyperparameter search*. ArXiv. https://doi.org/10.48550/arXiv.2210.02156

Chaudhuri, K., & Hsu, D. (2011). Sample complexity bounds for differentially private learning. In S. M. Kakade, & U. von Luxburg (Eds.), *Proceedings of the 24th Annual Conference on Learning Theory* (Vol. 19, pp.155–186). Proceedings of Machine Learning Research. https://proceedings.mlr.press/v19/chaudhuri11a.html

Chaudhuri, K., & Hsu, D. (2012). Convergence rates for differentially private statistical estimation. In J. Langford, & J. Pineau (Eds.), *Proceedings of the 29th International Conference on International Conference on Machine Learning* (pp. 1715–1722). Omnipress.

Chaudhuri, K., Monteleoni, C., & Sarwate, A. D. (2011). Differentially private empirical risk minimization. *Journal of Machine Learning Research*, *12*(29), 1069–1109. https://www.jmlr.org/papers/v12/chaudhuri11a.html

Chen, M., Zhang, Z., Wang, T., Backes, M., Humbert, M., & Zhang, Y. (2021). When machine unlearning jeopardizes privacy. In *CCS '21: Proceedings of the 2021 ACM SIGSAC Conference on Computer and Communications Security* (pp. 896–911). Association for Computing Machinery. https://doi.org/10.1145/3460120.3484756

Cheng, A., Wang, J., Zhang, X. S., Chen, Q., Wang, P., & Cheng, J. (2022). DPNAS: Neural architecture search for deep learning with differential privacy. *Proceedings of the AAAI Conference on Artificial Intelligence*, *36*(6), 6358–6366. hhtps://doi.org/10.1609/aaai.v36i6.20586

Cheu, A., Smith, A., & Ullman, J. (2021). Manipulation attacks in local differential privacy. In *2021 IEEE Symposium on Security and Privacy* (pp. 883–900). IEEE. https://doi.org/10.1109/SP40001.2021.00001

Choquette-Choo, C. A., Tramer, F., Carlini, N., & Papernot, N. (2021). Label-only membership inference attacks. In M Meila, & T Zhang (Eds.), *Proceedings of the 38th International Conference on Machine Learning* (Vol. 139, pp.1964–1974). Proceedings of Machine Learning Research. https://proceedings.mlr.press/v139/choquette-choo21a.html

Cohen, A., & Nissim, K. (2020). Towards formalizing the GDPR's notion of singling out. *Proceedings of the National Academy of Sciences*, *117*(15), 8344–8352. https://doi.org/10.1073/pnas.1914598117

Cormode, G. (2010). *Individual privacy vs population privacy: Learning to attack anonymization*. ArXiv. https://doi.org/10.48550/ARXIV.1011.2511

Couch, S., Kazan, Z., Shi, K., Bray, A., & Groce, A. (2019). Differentially private nonparametric hypothesis testing. In *CCS '19: Proceedings of the 2019 ACM Conference on Computer and Communications Security* (pp. 737–751). Association for Computing Machinery. https://doi.org/10.1145/3319535.3339821

Cummings, R., Gupta, V., Kimpara, D., & Morgenstern, J. (2019). On the compatibility of privacy and fairness. In *Adjunct Publication of the 27th Conference on User Modeling, Adaptation and Personalization* (pp. 309–315). Association for Computing Machinery. https://doi.org/10.1145/3314183.3323847

Cummings, R., Kaptchuk, G., & Redmiles, E. M. (2021). "I need a better description": An investigation into user expectations for differential privacy. In *CCS '21: Proceedings of the 2021 ACM SIGSAC Conference on*

*Computer and Communications Security* (pp. 3037–3052). Association for Computing Machinery. https://doi.org/10.1145/3460120.3485252

Cummings, R., Krehbiel, S., Mei, Y., Tuo, R., & Zhang, W. (2018). Differentially private change-point detection. In S. Bengio, H. Wallach, H. Larochelle, K. Grauman, N. Cesa-Bianchi, & R. Garnett (Eds.), *Advances in neural information processing systems* (Vol. 31, pp. 10825–10834). Curran Associates. https://papers.nips.cc/paper_files/paper/2018/hash/f19ec2b84181033bf4753a5a51d5d608-Abstract.html

Cummings, R., Ligett, K., Pai, M. M., & Roth, A. (2016). The strange case of privacy in equilibrium models. In *Proceedings of the 2016 ACM Conference on Economics and Computation* (pp. 659–659). Association for Computer Machinery. https://doi.org/10.1145/2940716.2940740

Damaskinos, G., Mendler-Dünner, C., Guerraoui, R., Papandreou, N., & Parnell, T. (2021). Differentially private stochastic coordinate descent. *Proceedings of the AAAI Conference on Artificial Intelligence*, *35*(8), 7176–7184. https://doi.org/10.1609/aaai.v35i8.16882

De, S., Berrada, L., Hayes, J., Smith, S. L., & Balle, B. (2022). *Unlocking high-accuracy differentially private image classification through scale*. ArXiv. https://doi.org/10.48550/arXiv.2204.13650

Decarolis, C., Ram, M., Esmaeili, S., Wang, Y.-X., & Huang, F. (2020). An end-to-end differentially private latent Dirichlet allocation using a spectral algorithm. In H. Daumé III, & A. Singh (Eds.), *Proceedings of the 37th International Conference on Machine Learning* (Vol. 119, pp. 2421–2431). Proceedings of Machine Learning Research. https://proceedings.mlr.press/v119/decarolis20a.html

Desfontaines, D. (2021, October, 1). A list of real-world uses of differential privacy. *Ted is writing things*. https://desfontain.es/privacy/real-world-differential-privacy.html

Devlin, J., Chang, M.-W., Lee, K., & Toutanova, K. (2019). BERT: Pre-training of deep bidirectional transformers for language understanding. In J. Burstein, C. Doran, & T. Solorio (Eds.), *Proceedings of the 2019 Conference of the North American Chapter of the Association for Computational Linguistics: Human Language Technologies, Volume 1 (Long and Short Papers)* (pp. 4171–4186). Association for Computational Linguistics. https://doi.org/10.18653/v1%2FN19-1423

Diakonikolas, I., Kamath, G., Kane, D. M., Li, J., Moitra, A., & Stewart, A. (2016). Robust estimators in high dimensions without the computational intractability. In *Proceedings of the 57th Annual IEEE Symposium on Foundations of Computer Science* (pp. 655–664). IEEE. https://doi.org/10.1109/FOCS.2016.85

Dick, T., Kulesza, A., Sun, Z., & Suresh, A. T. (2023). *Subset-based instance optimality in private estimation*. ArXiv. https://doi.org/10.48550/arXiv.2303.01262

Dimitrov, D. I., Balunović, M., Jovanović, N., & Vechev, M. (2022). *LAMP: Extracting text from gradients with language model priors*. ArXiv. https://doi.org/10.48550/arXiv.2202.08827

Ding, B., Kulkarni, J., & Yekhanin, S. (2017). Collecting telemetry data privately. In I. Guyon, U. Von Luxburg, S. Bengio, H. Wallach, R. Fergus, S. Vishwanathan, & R. Garnett (Eds.), *Advances in neural information processing systems* (Vol. 30, pp. 3571–3580). Curran Associates. https://papers.nips.cc/paper_files/paper/2017/hash/253614bbac999b38b5b60cae531c4969-Abstract.html

Ding, J., Zhang, X., Li, X., Wang, J., Yu, R., & Pan, M. (2020). Differentially private and fair classification via calibrated functional mechanism. *Proceedings of the AAAI Conference on Artificial Intelligence*, *34*(1), 622–629. https://doi.org/10.1609/aaai.v34i01.5402

Ding, Z., Wang, Y., Wang, G., Zhang, D., & Kifer, D. (2018). Detecting violations of differential privacy. In *CCS '18: Proceedings of the 2018 ACM SIGSAC Conference on Computer and Communications Security* (pp. 475–489). Association for Computing Machinery. https://doi.org/10.1145/3243734.3243818

Dinur, I., & Nissim, K. (2003). Revealing information while preserving privacy. In *Proceedings of the 22nd ACM SIGMOD-SIGACT-SIGART Symposium on Principles of Database Systems* (pp. 202–210). Association for Computer Machinery. https://doi.org/10.1145/773153.773173

Dong, J., Roth, A., & Su, W. J. (2022). Gaussian differential privacy. *Journal of the Royal Statistical Society: Series B (Statistical Methodology)*, *84*(1), 3–37. https://doi.org/10.1111/rssb.12454

Dong, J., Su, W., & Zhang, L. (2021). A central limit theorem for differentially private query answering. In M. Ranzato, A. Beygelzimer, Y. Dauphin, P.S. Liang, & J. Wortman Vaughan (Eds.), *Advances in neural information processing systems* (Vol. 34, pp. 14759–14770). Curran Associates. https://proceedings.neurips.cc/paper/2021/hash/7c2c48a32443ad8f805e48520f3b26a4-Abstract.html

Dong, W., & Yi, K. (2021). *Universal private estimators*. ArXiv. https://doi.org/10.48550/arXiv.2111.02598

Du, M., Jia, R., & Song, D. (2020, April 26–May 1). *Robust anomaly detection and backdoor attack detection via differential privacy* [Poster session]. ICLR 2020: The Eighth International Conference on Learning Representations, Virtual Event. https://openreview.net/forum?id=SJx0q1rtvS

Du, W., Foot, C., Moniot, M., Bray, A., & Groce, A. (2020). *Differentially private confidence intervals*. ArXiv. https://doi.org/10.48550/arXiv.2001.02285

Duchi, J., Haque, S., & Kuditipudi, R. (2023). *A fast algorithm for adaptive private mean estimation*. ArXiv. https://doi.org/10.48550/arXiv.2301.07078

Duchi, J., & Rogers, R. (2019). Lower bounds for locally private estimation via communication complexity. In A. Beygelzimer, & D. Hsu (Eds.), *Proceedings of the 32nd Annual Conference on Learning Theory* (Vol. 99, pp. 1161–1191). Proceedings of Machine Learning Research. https://proceedings.mlr.press/v99/duchi19a.html

Duchi, J. C., Jordan, M. I., & Wainwright, M. J. (2013). Local privacy and statistical minimax rates. In *Proceedings of the 54th Annual IEEE Symposium on Foundations of Computer Science* (pp. 429–438). IEEE. https://doi.org/10.1109/FOCS.2013.53

Duchi, J. C., Jordan, M. I., & Wainwright, M. J. (2018). Minimax optimal procedures for locally private estimation. *Journal of the American Statistical Association, 113*(521), 182–201. https://doi.org/10.1080/01621459.2017.1389735

Duchi, J. C., & Ruan, F. (2018). *The right complexity measure in locally private estimation: It is not the Fisher information*. ArXiv. https://doi.org/10.48550/arXiv.1806.05756

Durfee, D., & Rogers, R. M. (2019). Practical differentially private top-k selection with pay-what-you-get composition. In H. Wallach, H. Larochelle, A. Beygelzimer, F. d'Alché-Buc, E. Fox, & R. Garnett (Eds.), *Advances in neural information processing systems* (Vol. 32, pp. 3532–3542). Curran Associates. https://papers.nips.cc/paper_files/paper/2019/hash/b139e104214a08ae3f2ebcce149cdf6e-Abstract.html

Dwork, C. (2008). Differential privacy: A survey of results. In M. Agrawal, D. Du, Z. Duan, A. Li, (Eds), *Lecture notes in computer science: Vol. 4978. Theory and Applications of Models of Computation* (pp. 1–19). Springer. https://doi.org/10.1007/978-3-540-79228-4_1

Dwork, C., Kenthapadi, K., McSherry, F., Mironov, I., & Naor, M. (2006). Our data, ourselves: Privacy via distributed noise generation. In S. Vaudenay (Ed.), *Proceedings of the 24th Annual International Conference on the Theory and Applications of Cryptographic Techniques* (pp. 486–503). Springer. https://doi.org/10.1007/11761679_29

Dwork, C., Kohli, N., & Mulligan, D. (2019). Differential privacy in practice: Expose your epsilons! *Journal of Privacy and Confidentiality*, *9*(2). https://doi.org/10.29012/jpc.689

Dwork, C., & Lei, J. (2009). Differential privacy and robust statistics. In *Proceedings of the 41st Annual ACM Symposium on the Theory of Computing* (pp. 371–380). Association for Computer Machinery. https://doi.org/10.1145/1536414.1536466

Dwork, C., McSherry, F., Nissim, K., & Smith, A. (2006). Calibrating noise to sensitivity in private data analysis. In S. Halevi, & T. Rabin (Eds.), *Proceedings of the 3rd Conference on Theory of Cryptography* (pp. 265–284). Springer. https://doi.org/10.1007/11681878_14

Dwork, C., Naor, M., Reingold, O., Rothblum, G. N., & Vadhan, S. (2009). On the complexity of differentially private data release: Efficient algorithms and hardness results. In *Proceedings of the 41st Annual ACM Symposium on the Theory of Computing* (pp. 381–390). Association for Computer Machinery. https://doi.org/10.1145/1536414.1536467

Dwork, C., Smith, A., Steinke, T., & Ullman, J. (2017). Exposed! A survey of attacks on private data. *Annual Review of Statistics and Its Application*, *4*(1), 61–84. https://doi.org/10.1146/annurev-statistics-060116-054123

Dwork, C., Smith, A., Steinke, T., Ullman, J., & Vadhan, S. (2015). Robust traceability from trace amounts. In *Proceedings of the 56th Annual IEEE Symposium on Foundations of Computer Science* (pp. 650–669). IEEE. https://doi.org/10.1109/FOCS.2015.46

Dwork, C., & Ullman, J. (2018). The Fienberg Problem: How to allow human interactive data analysis in the age of differential privacy. *Journal of Privacy and Confidentiality*, *8*(1). https://doi.org/10.29012/jpc.687

Epasto, A., Mahdian, M., Mao, J., Mirrokni, V., & Ren, L. (2020). Smoothly bounding user contributions in differential privacy. In H. Larochelle, M. Ranzato, R. Hadsell, M.F. Balcan, & H. Lin (Eds.), *Advances in neural information processing systems* (Vol. *33*, pp. 13999–14010). Curran Associates. https://proceedings.neurips.cc/paper_files/paper/2020/hash/a0dc078ca0d99b5ebb465a9f1cad54ba-Abstract.html

Erlingsson, Ú., Mironov, I., Raghunathan, A., & Song, S. (2019). *That which we call private*. ArXiv. https://doi.org/10.48550/arXiv.1908.03566

Erlingsson, Ú., Pihur, V., & Korolova, A. (2014). RAPPOR: Randomized aggregatable privacy-preserving ordinal response. In *CCS '14: Proceedings of the 2014 ACM Conference on Computer and Communications Security* (pp. 1054–1067). Association for Computing Machinery. https://doi.org/10.1145/2660267.2660348

Esipova, M. S., Ghomi, A. A., Luo, Y., & Cresswell, J. C. (2022). *Disparate impact in differential privacy from gradient misalignment*. ArXiv. https://doi.org/10.48550/arXiv.2206.07737

Evfimievski, A., Gehrke, J., & Srikant, R. (2003). Limiting privacy breaches in privacy preserving data mining. In *Proceedings of the 22nd ACM SIGMOD-SIGACT-SIGART Symposium on Principles of Database Systems* (pp. 211–222). Association for Computer Machinery. https://doi.org/10.1145/773153.773174

Farrand, T., Mireshghallah, F., Singh, S., & Trask, A. (2020). Neither private nor fair: Impact of data imbalance on utility and fairness in differential privacy. In *Proceedings of the 2020 Workshop on Privacy-Preserving Machine Learning in Practice* (pp. 15–19). Association for Computing Machinery. https://doi.org/10.1145/3411501.3419419

Feldman, V. (2020). Does learning require memorization? A short tale about a long tail. In *Proceedings of the 52nd Annual ACM Symposium on the Theory of Computing* (pp. 954–959). Association for Computer Machinery. https://doi.org/10.1145/3357713.3384290

Feldman, V., Mironov, I., Talwar, K., & Thakurta, A. (2018). Privacy amplification by iteration. In *2018 IEEE 59th Annual Symposium on Foundations of Computer Science* (pp. 521–532). IEEE. https://doi.org/10.1109/FOCS.2018.00056

Feldman, V., & Steinke, T. (2018). Calibrating noise to variance in adaptive data analysis. In S. Bubeck, V. Perchet, & P. Rigollet (Eds.), *Proceedings of the 31st Conference On Learning Theory* (Vol. 75, pp. 535–544). Proceedings of Machine Learning Research. https://proceedings.mlr.press/v75/feldman18a.html

Feldman, V., & Zrnic, T. (2021). Individual privacy accounting via a Rényi filter. In M. Ranzato, A. Beygelzimer, Y. Dauphin, P.S. Liang, & J. Wortman Vaughan (Eds.), *Advances in neural information processing systems* (Vol. 34, pp. 28080–28091). Curran Associates. https://proceedings.neurips.cc/paper/2021/hash/ec7f346604f518906d35ef0492709f78-Abstract.html

Fichtenberger, H., Henzinger, M., & Ost, W. (2021). *Differentially private algorithms for graphs under continual observation*. ArXiv. https://doi.org/10.48550/arXiv.2106.14756

Fioretto, F., Tran, C., Van Hentenryck, P., & Zhu, K. (2022). Differential privacy and fairness in decisions and learning tasks: A survey. In L. De Raedt (Ed.), *Proceedings of the Thirty-First International Joint Conference on Artificial Intelligence* (pp. 5470-5477). International Joint Conferences on Artificial Intelligence Organization. https://doi.org/10.24963/ijcai.2022/766

Fowl, L., Geiping, J., Czaja, W., Goldblum, M., & Goldstein, T. (2021). *Robbing the fed: Directly obtaining private data in federated learning with modified models*. ArXiv. https://doi.org/10.48550/arXiv.2110.13057

Fowl, L., Geiping, J., Reich, S., Wen, Y., Czaja, W., Goldblum, M., & Goldstein, T. (2022). *Decepticons: Corrupted transformers breach privacy in federated learning for language models*. ArXiv. https://doi.org/10.48550/arXiv.2201.12675

Fredrikson, M., Jha, S., & Ristenpart, T. (2015). Model inversion attacks that exploit confidence information and basic countermeasures. In *CCS '15: Proceedings of the 22nd ACM SIGSAC conference on computer and communications security* (pp. 1322–1333). Association for Computing Machinery. https://doi.org/10.1145/2810103.2813677

Gaboardi, M., Honaker, J., King, G., Murtagh, J., Nissim, K., Ullman, J., & Vadhan, S. (2016). *PSI (Ψ): A private data sharing interface*. ArXiv. https://doi.org/10.48550/arXiv.1609.04340

Gaboardi, M., Lim, H., Rogers, R. M., & Vadhan, S. P. (2016). Differentially private chi-squared hypothesis testing: Goodness of fit and independence testing. In M. F. Balcan, & K. Q. Weinberger (Eds.), *Proceedings of the 33rd International Conference on Machine Learning* (Vol. 48, pp. 1395–1403). Proceedings of Machine Learning Research. https://proceedings.mlr.press/v48/rogers16.html

Ganesh, A., Haghifam, M., Nasr, M., Oh, S., Steinke, T., Thakkar, O., Thakurta, A., & Wang, L. (2023). *Why is public pretraining necessary for private model training?* ArXiv. https://doi.org/10.48550/arXiv.2302.09483

Ganesh, A., Liu, D., Oh, S., & Thakurta, A. (2023). *Private (stochastic) non-convex optimization revisited: Second-order stationary points and excess risks*. ArXiv. https://doi.org/10.48550/arXiv.2302.09699

Ganev, G., Oprisanu, B., & De Cristofaro, E. (2022). Robin Hood and Matthew effects: Differential privacy has disparate impact on synthetic data. In K. Chaudhuri, S. Jegelka, L. Song, C. Szepesvari, G. Niu, & S. Sabato (Eds.), *Proceedings of the 39th International Conference on Machine Learning* (Vol. 162, pp. 6944–6959). Proceedings of Machine Learning Research. https://proceedings.mlr.press/v162/ganev22a.html

Ganju, K., Wang, Q., Yang, W., Gunter, C. A., & Borisov, N. (2018). Property inference attacks on fully connected neural networks using permutation invariant representations. In *CCS '18: Proceedings of the 2018 ACM SIGSAC Conference on Computer and Communications Security* (pp. 619–633). Association for Computing Machinery. https://doi.org/10.1145/3243734.3243834

Gao, C., & Wright, S. J. (2023). *Differentially private optimization for smooth nonconvex ERM*. ArXiv. https://doi.org/10.48550/arXiv.2302.04972

Garfinkel, S., Abowd, J. M., & Martindale, C. (2019). Understanding database reconstruction attacks on public data. *Communications of the ACM*, *62*(3), 46–53. https://doi.org/10.1145/3287287

Garg, S., Raz, R., & Tal, A. (2018). Extractor-based time-space lower bounds for learning. In *Proceedings of the 50th Annual ACM Symposium on the Theory of Computing* (pp. 990–1002). Association for Computer Machinery. https://doi.org/10.1145/3188745.3188962

Ge, J., Wang, Z., Wang, M., & Liu, H. (2018). Minimax-optimal privacy-preserving sparse PCA in distributed systems. In A. Storkey, & F. Perez-Cruz *Proceedings of the Twenty-First International Conference on Artificial Intelligence and Statistics* (Vol. 84, pp. 1589–1598). Proceedings of Machine Learning Research. https://proceedings.mlr.press/v84/ge18a.html

Geiping, J., Bauermeister, H., Dröge, H., & Moeller, M. (2020). Inverting gradients - How easy is it to break privacy in federated learning? In H. Larochelle, M. Ranzato, R. Hadsell, M.F. Balcan, & H. Lin (Eds.), *Advances in neural information processing systems* (Vol. 33, pp. 16937–16947). Curran Associates. https://proceedings.neurips.cc/paper_files/paper/2020/hash/c4ede56bbd98819ae6112b20ac6bf145-Abstract.html

Georgiev, K., & Hopkins, S. B. (2022). Privacy induces robustness: Information-computation gaps and sparse mean estimation. In S. Koyejo, S. Mohamed, A. Agarwal, D. Belgrave, K. Cho, & A. Oh (Eds.), *Advances in neural information processing systems* (Vol. *35*, pp. 6829–6842). Curran Associates. https://proceedings.neurips.cc/paper_files/paper/2022/hash/2d76b6a9f96181ab717c1a15ab9302e1-Abstract-Conference.html

Ghazi, B., Golowich, N., Kumar, R., Manurangsi, P., & Zhang, C. (2021). Deep learning with label differential privacy. In M. Ranzato, A. Beygelzimer, Y. Dauphin, P. S. Liang, & J. Wortman Vaughan (Eds.), *Advances in neural information processing systems* (Vol. 34, pp. 27131–27145). Curran Associates. https://proceedings.neurips.cc/paper/2021/hash/e3a54649aeec04cf1c13907bc6c5c8aa-Abstract.html

Gilbert, A., & McMillan, A. (2018). Property testing for differential privacy. In 2018 *56th Annual Allerton Conference on Communication, Control, and Computing* (pp. 249–258). IEEE. https://doi.org/10.1109/ALLERTON.2018.8636068

Golatkar, A., Achille, A., Wang, Y.-X., Roth, A., Kearns, M., & Soatto, S. (2022). Mixed differential privacy in computer vision. In *2022 IEEE/CVF Conference on Computer Vision and Pattern Recognition* (pp. 8366-8376). IEEE. https://doi.org/10.1109/CVPR52688.2022.00819

Goodfellow, I. (2015). *Efficient per-example gradient computations*. ArXiv. https://doi.org/10.48550/arXiv.1510.01799

Google. (2023). *Google's differential privacy libraries*. GitHub. Retreived March 17, 2023, from https://github.com/google/differential-privacy

Gopi, S. G., Gulhane, P., Kulkarni, J., Shen, J., Shokouhi, M., & Yekhanin, S. (2020). Differentially private set union. In H. Daumé III, & A. Singh (Eds.), *Proceedings of the 37th International Conference on Machine Learning* (Vol. 119, pp. 3627–3636). Proceedings of Machine Learning Research. https://proceedings.mlr.press/v119/gopi20a.html

Gu, X., Kamath, G., & Wu, Z. S. (2023). *Choosing public datasets for private machine learning via gradient subspace distance*. ArXiv. https://doi.org/10.48550/arXiv.2303.01256

Gupta, S., Huang, Y., Zhong, Z., Gao, T., Li, K., & Chen, D. (2022). Recovering private text in federated learning of language models. In S. Koyejo, S. Mohamed, A. Agarwal, D. Belgrave, K. Cho, & A. Oh (Eds.), *Advances in neural information processing systems* (Vol. 35, pp. 8130–8143). Curran Associates. https://proceedings.neurips.cc/paper_files/paper/2022/hash/35b5c175e139bff5f22a5361270fce87-Abstract-Conference.html

Haeberlen, A., Pierce, B. C., & Narayan, A. (2011). Differential privacy under fire. In *Proceedings of the 20th USENIX Conference on Security* (pp. 33). USENIX Association. https://www.usenix.org/conference/usenix-

security-11/differential-privacy-under-fire

Haim, N., Vardi, G., Yehudai, G., Shamir, O., & Irani, M. (2022). *Reconstructing training data from trained neural networks*. ArXiv. https://doi.org/10.48550/arXiv.2206.07758

Haney, S., Desfontaines, D., Hartman, L., Shrestha, R., & Hay, M. (2022). *Precision-based attacks and interval refining: How to break, then fix, differential privacy on finite computers*. ArXiv. https://doi.org/10.48550/arXiv.2207.13793

Harder, F., Bauer, M., & Park, M. (2020). Interpretable and differentially private predictions. *Proceedings of the AAAI Conference on Artificial Intelligence*, *34*(04), 4083–4090. https://doi.org/10.1609/aaai.v34i04.5827

Hardt, M., & Talwar, K. (2010). On the geometry of differential privacy. In *Proceedings of the 42nd Annual ACM Symposium on the Theory of Computing* (pp. 705–714). Association for Computer Machinery. https://doi.org/10.1145/1806689.1806786

Hartmann, F., & Kairouz, P. (2023, March 2). Distributed differential privacy for federated learning. *Google Research*. https://blog.research.google/2023/03/distributed-differential-privacy-for.html

Hay, M., Machanavajjhala, A., Miklau, G., Chen, Y., Zhang, D., & Bissias, G. (n.d.). Dpcomp.org. Retrieved March 17, 2023, from https://www.dpcomp.org/

He, J., Li, X., Yu, D., Zhang, H., Kulkarni, J., Lee, Y. T., Backurs, A., Yu, N., & Bian, J. (2023, May 1–5). *Exploring the limits of differentially private deep learning with group-wise clipping* [Poster session]. ICLR 2023: The Eleventh International Conference on Learning Representations, Kigali, Rwanda. https://openreview.net/pdf?id=oze0clVGPeX

Holohan, N., Braghin, S., Mac Aonghusa, P., & Levacher, K. (2019). Diffprivlib: The IBM differential privacy library. *ArXiv.* https://doi.org/10.48550/arXiv.1907.02444

Hoory, S., Feder, A., Tendler, A., Cohen, A. Erell, S., Laish, I., Nakhost, H., Stemmer, U., Benjamini, A., Hassidim, A., & Matias, Y. (2021). Learning and evaluating a differentially private pre-trained language model. In M.-F. Moens, X. Huang, L. Specia, & S. W.-t. Yih (Eds.), *Findings of the Association for Computational Linguistics: EMNLP 2021* (pp. 1178–1189). Association for Computational Linguistics. https://doi.org/10.18653/v1/2021.findings-emnlp.102

Hopkins, S. B., Kamath, G., & Majid, M. (2022). Efficient mean estimation with pure differential privacy via a sum-of-squares exponential mechanism. In *Proceedings of the 54th Annual ACM Symposium on the Theory of Computing* (pp. 1406–1417). Association for Computer Machinery. https://doi.org/10.1145/3519935.3519947

Hopkins, S. B., Kamath, G., Majid, M., & Narayanan, S. (2023). Robustness implies privacy in statistical estimation. In *Proceedings of the 55th Annual ACM Symposium on the Theory of Computing* (pp. 497–506). Association for Computer Machinery. https://doi.org/10.1145/3564246.3585115

Hu, L., Ni, S., Xiao, H., & Wang, D. (2022). High dimensional differentially private stochastic optimization with heavy-tailed data. In *Proceedings of the 41st ACM SIGMOD-SIGACT-SIGAI Symposium on Principles of Database Systems* (pp. 227–236). Association for Computer Machinery. https://doi.org/10.1145/3517804.3524144

Hu, L., Xiang, Z., Liu, J., & Wang, D. (2023). Privacy-preserving sparse Generalized Eigenvalue Problem. In F. Ruiz, J. Dy, & J.–W. van de Meent (Eds.), *Proceedings of The 26th International Conference on Artificial Intelligence and Statistics* (Vol. 206, pp. 5052–5062). Proceedings of Machine Learning Research. https://proceedings.mlr.press/v206/hu23a.html

Huang, W. R., Chien, S., Thakkar, O. D., & Mathews, R. (2022). Detecting unintended memorization in language-model-fused ASR. In H. Ko & J. H. L. Hansen (Eds.), *Interspeech 2022* (pp. 2808–2812). International Speech Communication Association. https://doi.org/10.21437/Interspeech.2022-10909

Huang, Y., Gupta, S., Song, Z., Li, K., & Arora, S. (2021). Evaluating gradient inversion attacks and defenses in federated learning. In M. Ranzato, A. Beygelzimer, Y. Dauphin, P.S. Liang, & J. Wortman Vaughan (Eds.), *Advances in neural information processing systems* (Vol. 34, pp. 7232–7241). Curran Associates. https://proceedings.neurips.cc/paper/2021/hash/3b3fff6463464959dcd1b68d0320f781-Abstract.html

Huang, Z., Liang, Y., & Yi, K. (2021). Instance-optimal mean estimation under differential privacy. In M. Ranzato, A. Beygelzimer, Y. Dauphin, P.S. Liang, & J. Wortman Vaughan (Eds.), *Advances in neural information processing systems* (Vol. 34, (pp. 25993–26004). Curran Associates. https://proceedings.neurips.cc/paper/2021/hash/da54dd5a0398011cdfa50d559c2c0ef8-Abstract.html

Ilvento, C. (2020). Implementing the exponential mechanism with base-2 differential privacy. In *CCS '20: Proceedings of the 2020 ACM SIGSAC Conference on Computer and Communications Security* (pp. 717–742). Association for Computing Machinery. https://doi.org/10.1145/3372297.3417269

Imola, J., Murakami, T., & Chaudhuri, K. (2021). Locally differentially private analysis of graph statistics. In *Proceedings of the 30th USENIX Security Symposium* (pp. 983–1000). USENIX Association. https://www.usenix.org/conference/usenixsecurity21/presentation/imola

Iyengar, R., Near, J. P., Song, D., Thakkar, O., Thakurta, A., & Wang, L. (2019). Towards practical differentially private convex optimization. In *2019 IEEE Symposium on Security and Privacy* (pp. 299–316). IEEE. https://doi.org/10.1109/SP.2019.00001

Jagielski, M., Thakkar, O., Tramèr, F., Ippolito, D., Lee, K., Carlini, N., Wallace, E., Song, S., Thakurta, A., Papernot, N., & Zhang, C. (2022). Measuring forgetting of memorized training examples.AarXiv. https://doi.org/10.48550/arXiv.2207.00099

Jagielski, M., Wu, S., Oprea, A., Ullman, J., & Geambasu, R. (2022). How to combine membership-inference attacks on multiple updated models. ArXiv. https://doi.org/10.48550/arXiv.2205.06369

Jain, P., Rush, J., Smith, A., Song, S., & Guha Thakurta, A. (2021). Differentially private model personalization. In M. Ranzato, A. Beygelzimer, Y. Dauphin, P.S. Liang, & J. Wortman Vaughan (Eds.), *Advances in neural information processing systems* (Vol. 34, pp. 29723–29735). Curran Associates. https://proceedings.neurips.cc/paper/2021/hash/f8580959e35cb0934479bb007fb241c2-Abstract.html

Jayaraman, B., & Evans, D. (2022). Are attribute inference attacks just imputation? In *Proceedings of the 2022 ACM SIGSAC Conference on Computer and Communications Security* (pp. 1569–1582). Association for Computing Machinery. https://doi.org/10.1145/3548606.3560663

Jayaraman, B., Wang, L., Evans, D., & Gu, Q. (2018). Distributed learning without distress: Privacy-preserving empirical risk minimization. In S. Bengio, H. Wallach, H. Larochelle, K. Grauman, N. Cesa-Bianchi, & R. Garnett (Eds.), *Advances in neural information processing systems* (Vol. 31, pp. 6343–6354). Curran Associates. https://papers.nips.cc/paper_files/paper/2018/hash/7221e5c8ec6b08ef6d3f9ff3ce6eb1d1-Abstract.html

Jayaraman, B., Wang, L., Knipmeyer, K., Gu, Q., & Evans, D. (2020). *Revisiting membership inference under realistic assumptions*. ArXiv. https://doi.org/10.48550/arXiv.2005.10881

Ji, Z., & Elkan, C. (2013). Differential privacy based on importance weighting. *Machine Learning*, *93*(1), 163–183. https://doi.org/10.1007/s10994-013-5396-x

Jin, J., McMurtry, E., Rubinstein, B. I., & Ohrimenko, O. (2022). Are we there yet? Timing and floating-point attacks on differential privacy systems. In *2022 IEEE Symposium on Security and Privacy* (pp. 473–488). IEEE. https://doi.org/10.1109/SP46214.2022.9833672

Jordon, J., Yoon, J., & Van Der Schaar, M. (2019, May 6–9). *PATE-GAN: Generating synthetic data with differential privacy guarantees* [Poster presentation]. ICLR 2019: 7th International Conference on Learning Representations, New Orleans, LA, United States. https://openreview.net/forum?id=S1zk9iRqF7

Jorgensen, Z., Yu, T., & Cormode, G. (2015). Conservative or liberal? personalized differential privacy. In *2015 IEEE 31St International Conference on Data Engineering* (pp. 1023–1034). IEEE. https://doi.org/10.1109/ICDE.2015.7113353

Joseph, M., Mao, J., Neel, S., & Roth, A. (2019). The role of interactivity in local differential privacy. In *Proceedings of the 60th Annual IEEE Symposium on Foundations of Computer Science* (pp. 94–105). IEEE. https://doi.org/10.1109/FOCS.2019.00015

Kairouz, P., McMahan, B., Song, S., Thakkar, O., Thakurta, A., & Xu, Z. (2021). Practical and private (deep) learning without sampling or shuffling. In M. Meila, & T. Zhang (Eds.), *Proceedings of the 38th International Conference on Machine Learning* (Vol. 139, pp. 5213–5225). https://proceedings.mlr.press/v139/kairouz21b.html

Kairouz, P., McMahan, H. B., Avent, B., Bellet, A., Bennis, M., Bhagoji, A. N., Bonawitz, K., Charles, Z., Cormode, G., Cummings, R., D'Oliveira, R. G. L., Eichner, H., El Rouayheb, S., Evans, D., Gardner, J., Garrett, Z., Gascón, A., Ghazi, B., Gibbons, P. B., … Zhao, S. (2021). Advances and open problems in federated learning. *Foundations and Trends® in Machine Learning*, *14*(1–2), 1–210. http://doi.org/10.1561/2200000083

Kairouz, P., Ribero, M., Rush, K., & Thakurta, A. (2020). *Fast dimension independent private AdaGrad on publicly estimated subspaces*. ArXiv. https://doi.org/10.48550/arXiv.2008.06570

Kairouz, P., Ribero, M., Rush, K., & Thakurta, A. (2021). (Nearly) dimension independent private ERM with AdaGrad rates via publicly estimated subspaces. In M. Belkin, & S. Kpotufe (Eds.), *Proceedings of the 34th Annual Conference on Learning Theory* (Vol. 134, pp. 2717–2746). Proceedings of Machine Learning Research. https://proceedings.mlr.press/v134/kairouz21a.html

Kakizaki, K., Sakuma, J., & Fukuchi, K. (2017). Differentially private chi-squared test by unit circle mechanism. In D. Precup, & Y. W. Teh (Eds.), *Proceedings of the 34th International Conference on Machine Learning* (Vol. 40, pp. 1761–1770). Proceedings of Machine Learning Research. https://proceedings.mlr.press/v70/kakizaki17a.html

Kamath, G., Li, J., Singhal, V., & Ullman, J. (2019). Privately learning high-dimensional distributions. In A. Beygelzimer, & D. Hsu (Eds.), *Proceedings of the 32nd Annual Conference on Learning Theory* (Vol. 99, pp. 1853–1902). Proceedings of Machine Learning Research. https://proceedings.mlr.press/v99/kamath19a.html

Kamath, G., Liu, X., & Zhang, H. (2022). Improved rates for differentially private stochastic convex optimization with heavy-tailed data. In K. Chaudhuri, S. Jegelka, L. Song, C. Szepesvari, G. Niu, & S. Sabato (Eds.), *Proceedings of the 39th International Conference on Machine Learning* (Vol. 162, pp. 10633–10660). Proceedings of Machine Learning Research. https://proceedings.mlr.press/v162/kamath22a.html

Kamath, G., Mouzakis, A., & Singhal, V. (2022). New lower bounds for private estimation and a generalized fingerprinting lemma. In S. Koyejo, S. Mohamed, A. Agarwal, D. Belgrave, K. Cho, & A. Oh (Eds.), *Advances in neural information processing systems*, (Vol. 35, pp. 24405–24418). Curran Associates.

https://proceedings.neurips.cc/paper_files/paper/2022/hash/9a6b278218966499194491f55ccf8b75-Abstract-Conference.html

Kamath, G., Mouzakis, A., Singhal, V., Steinke, T., & Ullman, J. (2021). *A private and computationally-efficient estimator for unbounded Gaussians*. ArXiv. https://doi.org/10.48550/arXiv.2111.04609

Kamath, G., Singhal, V., & Ullman, J. (2020). Private mean estimation of heavy-tailed distributions. In J. Abernethy, & S. Agarwal (Eds.), *Proceedings of the 33rd Annual Conference on Learning Theory* (Vol. 125, 2204–2235). Proceedings of Machine Learning Research. https://proceedings.mlr.press/v125/kamath20a.html

Kamath, G., & Ullman, J. (2020). *A primer on private statistics.* ArXiv. https://doi.org/10.48550/arXiv.2005.00010

Karwa, V., & Vadhan, S. (2018). Finite sample differentially private confidence intervals. In A. R. Karlin (Ed.), *Leibniz International Proceedings in Informatics* (Vol. 49). *9th Conference on Innovations in Theoretical Computer Science*, (pp. 44:1–44:9). Dagstuhl Publishing. https://drops.dagstuhl.de/opus/volltexte/2018/8344/pdf/LIPIcs-ITCS-2018-44.pdf

Kearns, M., Pai, M., Roth, A., & Ullman, J. (2014). Mechanism design in large games: Incentives and privacy. *American Economic Review, 104*(5), 431–435.

Kelley, P. G., Bresee, J., Cranor, L. F., & Reeder, R. W. (2009). A "nutrition label" for privacy. In L. F. Cranor (Ed.), *Proceedings of the 5th Symposium on Usable Privacy and Security*. Association for Computing Machinery. https://doi.org/10.1145/1572532.1572538

Kenny, C. T., Kuriwaki, S., McCartan, C., Rosenman, E. T., Simko, T., & Imai, K. (2021). The use of differential privacy for census data and its impact on redistricting: The case of the 2020 US Census. *Science Advances*, *7*(41), Article eabk3283. https://doi.org/10.1126/sciadv.abk3283

Kifer, D., & Machanavajjhala, A. (2014). Pufferfish: A framework for mathematical privacy definitions. *ACM Transactions Database Systems*, *39*(1), Article 3. https://doi.org/10.1145/2514689

Kifer, D., & Rogers, R. M. (2017). A new class of private chi-square tests. In A. Singh, & J. Zhu (Eds.), *Proceedings of the 20th International Conference on Artificial Intelligence and Statistics* (Vol. 54, pp. 991–1000). Proceedings of Machine Learning Research. https://proceedings.mlr.press/v54/rogers17a.html

Kifer, D., Smith, A., & Thakurta, A. (2012). Private convex empirical risk minimization and high-dimensional regression. In S. Mannor, N. Srebro, & R. C. Williamson (Eds.), *Proceedings of the 25th Annual Conference on Learning Theory* (Vol. 23, pp. 25.1–25.40). Proceedings of Machine Learning Research. https://proceedings.mlr.press/v23/kifer12.html

Kim, K., Gopi, S., Kulkarni, J., & Yekhanin, S. (2021). Differentially private n-gram extraction. In M. Ranzato, A. Beygelzimer, Y. Dauphin, P.S. Liang, & J. Wortman Vaughan (Eds.), *Advances in neural information processing systems* (Vol. 34, pp. 5102–5111). Curran Associates. https://proceedings.neurips.cc/paper/2021/hash/28ce9bc954876829eeb56ff46da8e1ab-Abstract.html

Kothari, P. K., Manurangsi, P., & Velingker, A. (2021). *Private robust estimation by stabilizing convex relaxations*. ArXiv. https://doi.org/10.48550/arXiv.2112.03548

Kurakin, A., Chien, S., Song, S., Geambasu, R., Terzis, A., & Thakurta, A. (2022). Toward training at ImageNet scale with differential privacy. ArXiv. https://doi.org/10.48550/arXiv.2201.12328

Lai, K. A., Rao, A. B., & Vempala, S. (2016). Agnostic estimation of mean and covariance. In *Proceedings of the 57th Annual IEEE Symposium on Foundations of Computer Science* (pp. 665–674). IEEE. https://doi.org/10.1109/FOCS.2016.76

Lee, D., Yu, H., Jiang, X., Rogith, D., Gudala, M., Tejani, M., Zhang, Q., & Xiong, L. (2020). Generating sequential electronic health records using dual adversarial autoencoder. *Journal of the American Medical Informatics Association*, *27*(9), 1411–1419. https://doi.org/10.1093/jamia/ocaa119

Lee, J., & Kifer, D. (2018). Concentrated differentially private gradient descent with adaptive per-iteration privacy budget. In *Proceedings of the 24th ACM SIGKDD International Conference on Knowledge Discovery & Data Mining* (pp. 1656–1665). Association for Computer Machinery. https://doi.org/10.1145/3219819.3220076

Lee, J., & Kifer, D. (2021). Scaling up differentially private deep learning with fast per-example gradient clipping. *Proceedings on Privacy Enhancing Technologies*, *2021*(1),128–144 . https://doi.org/10.2478/popets-2021-0008

Levy, D., Sun, Z., Amin, K., Kale, S., Kulesza, A., Mohri, M., & Suresh, A. T. (2021). Learning with user-level privacy. In M. Ranzato, A. Beygelzimer, Y. Dauphin, P.S. Liang, & J. Wortman Vaughan (Eds.), *Advances in neural information processing systems* (Vol. 34, pp. 12466–12479). Curran Associates. https://proceedings.neurips.cc/paper_files/paper/2021/hash/67e235e7f2fa8800d8375409b566e6b6-Abstract.html

Li, T., Zaheer, M., Reddi, S. J., & Smith, V. (2022). Private adaptive optimization with side information. In K. Chaudhuri, S. Jegelka, L. Song, C. Szepesvari, G. Niu, & S. Sabato (Eds.), *Proceedings of the 39th International Conference on Machine Learning* (Vol. 162, pp. 13086–13105). Proceedings of Machine Learning Research. https://proceedings.mlr.press/v162/li22x.html

Li, X., Liu, D., Hashimoto, T., Inan, H. A., Kulkarni, J., Lee, Y. T., & Thakurta, A. G. (2022). *When does differentially private learning not suffer in high dimensions?* ArXiv. https://doi.org/10.48550/arXiv.2207.00160

Li, X., Tramer, F., Liang, P., & Hashimoto, T. (2021). *Large language models can be strong differentially private learners*. ArXiv. https://doi.org/10.48550/arXiv.2110.05679

Li, X., Tramèr, F., Kulkarni, J., & Hashimoto, T. (2022, March 15). *Differentially private deep learning can be effective with self-supervised models*. DifferentialPrivacy.org. https://differentialprivacy.org/dp-fine-tuning/

Li, X., Tramèr, F., Liang, P., & Hashimoto, T. (2022, April 25–29). *Large language models can be strong differentially private learners* [Conference presentation]. ICLR 2022: The Tenth International Conference on Learning Representations, online. https://openreview.net/forum?id=bVuP3ltATMz

Li, Z., & Zhang, Y. (2021). Membership leakage in label-only exposures. In *CCS '21: Proceedings of the 2021 ACM SIGSAC Conference on Computer and Communications Security* (pp. 880–895). Association for Computing Machinery. https://doi.org/10.1145/3460120.3484575

Ligett, K., Neel, S., Roth, A., Waggoner, B., & Wu, S. Z. (2017). Accuracy first: Selecting a differential privacy level for accuracy constrained ERM. In I. Guyon, U. Von Luxburg, S. Bengio, H. Wallach, R. Fergus, S. Vishwanathan, & R. Garnett (Eds.), *Advances in neural information processing systems* (Vol. 30, pp. 1566–2575). Curran Associates. https://papers.nips.cc/paper_files/paper/2017/hash/86df7dcfd896fcaf2674f757a2463eba-Abstract.html

Liu, C., Zhu, Y., Chaudhuri, K., & Wang, Y.-X. (2021). Revisiting model-agnostic private learning: Faster rates and active learning. *The Journal of Machine Learning Research*, *22*(1), 11936–11979. https://jmlr.org/papers/volume22/20-1251/20-1251.pdf

Liu, D., & Lu, Z. (2021). Lower bounds for differentially private ERM: Unconstrained and non-Euclidean. ArXiv. https://doi.org/10.48550/arXiv.2105.13637

Liu, J., & Talwar, K. (2019). Private selection from private candidates. In *Proceedings of the 51st Annual ACM Symposium on the Theory of Computing* (pp. 298–309). Association for Computer Machinery. https://doi.org/10.1145/3313276.3316377

Liu, J., Lou, J., Xiong, L., Liu, J., & Meng, X. (2021). Projected federated averaging with heterogeneous differential privacy. *Proceedings of the VLDB Endowment*, *15*(4), 828–840. https://doi.org/10.14778/3503585.3503592

Liu, T., Vietri, G., Steinke, T., Ullman, J., & Wu, S. (2021). Leveraging public data for practical private query release. In M. Meila & T. Zhang (Eds.), *Proceedings of the 38th International Conference on Machine Learning* (Vol. 139, pp. 6968–6977). Proceedings of Machine Learning Research. https://proceedings.mlr.press/v139/liu21v.html

Liu, X., Jain, P., Kong, W., Oh, S., & Suggala, A. S. (2023). *Near optimal private and robust linear regression*. ArXiv. https://doi.org/10.48550/arXiv.2301.13273

Liu, X., Kong, W., Jain, P., & Oh, S. (2022). DP-PCA: Statistically optimal and differentially private PCA. In S. Koyejo, S. Mohamed, A. Agarwal, D. Belgrave, K. Cho, & A. Oh (Eds.), *Advances in neural information processing systems* (Vol. 35, pp. 29929–29943). Curran Associates https://proceedings.neurips.cc/paper_files/paper/2022/hash/c150ebe1b9d1ca0eb61502bf979fa87d-Abstract-Conference.html

Liu, X., Kong, W., Kakade, S., & Oh, S. (2021). Robust and differentially private mean estimation. In M. Ranzato, A. Beygelzimer, Y. Dauphin, P.S. Liang, & J. Wortman Vaughan (Eds.), *Advances in neural information processing systems* (Vol. 34, pp. 3887–3901). Curran Associates. https://proceedings.neurips.cc/paper_files/paper/2021/hash/1fc5309ccc651bf6b5d22470f67561ea-Abstract.html

Liu, X., Kong, W., & Oh, S. (2022). Differential privacy and robust statistics in high dimensions. In P.-L. Loh, & M. Raginsky (Eds.,) *Proceedings of Thirty Fifth Conference on Learning Theory* (Vol. 178, pp. 1167–1246). Proceedings of Machine Learning Research. https://proceedings.mlr.press/v178/liu22b.html

Liu, X., & Oh, S. (2019). Minimax optimal estimation of approximate differential privacy on neighboring databases. In H. Wallach, H. Larochelle, A. Beygelzimer, F. d'Alché-Buc, E. Fox, & R. Garnett (Eds.), *Advances in neural information processing systems* (Vol. 32, pp. 2417–2428). Curran Associates. https://proceedings.neurips.cc/paper_files/paper/2019/hash/7a674153c63cff1ad7f0e261c369ab2c-Abstract.html

Liu, Y., Ott, M., Goyal, N., Du, J., Joshi, M., Chen, D., Levy, O., Lewis, M., Zettlemoyer, L., & Stoyanov, V. (2019). RoBERTa: A robustly optimized BERT pretraining approach. ArXiv. https://doi.org/10.48550/arXiv.1907.11692

Liu, Y., Zhao, S., Xiong, L., Liu, Y., & Chen, H. (2023). Echo of neighbors: Privacy amplification for personalized private federated learning with shuffle model. *Proceedings of the AAAI Conference on Artificial Intelligence, 37*(10), 11865–11872). https://doi.org/10.1609/aaai.v37i10.26400

Liu, Y., Suresh, A. T., Yu, F., Kumar, S., & Riley, M. (2020). Learning discrete distributions: User vs item-level privacy. In H. Larochelle, M. Ranzato, R. Hadsell, M.F. Balcan, & H. Lin (Eds.), *Advances in neural information processing systems* (Vol. 33, pp. 20965–20976). Curran Associates. https://proceedings.neurips.cc/paper_files/paper/2020/hash/f06edc8ab534b2c7ecbd4c2051d9cb1e-Abstract.html

Lovejoy, B. (2017, July 7). *As Apple starts analyzing web browsing and health data, how comfortable are you with differential privacy?* 9to5Mac. https://9to5mac.com/2017/07/07/what-is-differential-privacy/

Ma, Y., Zhu, X., & Hsu, J. (2019). Data poisoning against differentially-private learners: Attacks and defenses. In Sarit Kraus (Ed.), *Proceedings of the Twenty-Eighth International Joint Conference on Artificial Intelligence* (pp. 4732-4738). International Joint Conferences on Artificial Intelligence Organization. https://doi.org/10.24963/ijcai.2019%2F657

Machanavajjhala, A., Kifer, D., Gehrke, J., & Venkitasubramaniam, M. (2007). *L*-diversity: Privacy beyond *k*-anonymity. *ACM Transactions on Knowledge Discovery from Data*, *1*(1), 3–es. https://doi.org/10.1145/1217299.1217302

Mahloujifar, S., Sablayrolles, A., Cormode, G., & Jha, S. (2022). *Optimal membership inference bounds for adaptive composition of sampled Gaussian mechanisms*. arXiv. https://doi.org/10.48550/arXiv.2204.06106

Mangold, P., Bellet, A., Salmon, J., & Tommasi, M. (2022). Differentially private coordinate descent for composite empirical risk minimization. In K. Chaudhuri, S. Jegelka, L. Song, C. Szepesvari, G. Niu, & S. Sabato (Eds.), *Proceedings of the 39th International Conference on Machine Learning* (Vol. 162, pp.14948–14978). Proceedings of Machine Learning Research. https://proceedings.mlr.press/v162/mangold22a.html

Mangold, P., Perrot, M., Bellet, A., & Tommasi, M. (2022). Differential privacy has bounded impact on fairness in classification. In A. Krause, E. Brunskill, K. Cho, B. Engelhardt, S. Sabato, & J. Scarlett (Eds.), *Proceedings of the 40th International Conference on Machine Learning* (Vol. 202, pp. 23681–26705). Proceedings of Machine Learning Research. https://proceedings.mlr.press/v202/mangold23a.html

McMahan, B., Moore, E., Ramage, D., Hampson, S., & Agüera y Arcas, B. A. (2017). Communication-efficient learning of deep networks from decentralized data. In A. Singh & J. Zhu (Eds.), *Proceedings of the 20th International Conference on Artificial Intelligence and Statistics* (Vol. 54, pp. 1273–1282). Proceedings of Machine Learning Research. https://proceedings.mlr.press/v54/mcmahan17a.html

McMahan, H. B., Ramage, D., Talwar, K., & Zhang, L. (2018, April 30–May 3). *Learning differentially private recurrent language models* [Poster session]. ICLR 2018: 6th International Conference on Learning Representations, Vancouver, BC, Canada. https://openreview.net/forum?id=BJ0hF1Z0b

McSherry, F. (2016, June 14). Statistical inference considered harmful. *GitHub*. https://github.com/frankmcsherry/blog/blob/master/posts/2016-06-14.md

Mehta, H., Thakurta, A., Kurakin, A., & Cutkosky, A. (2022). *Large scale transfer learning for differentially private image classification*. ArXiv. https://doi.org/10.48550/arXiv.2205.02973

Milionis, J., Kalavasis, A., Fotakis, D., & Ioannidis, S. (2022). Differentially private regression with unbounded covariates. In G. Camps-Valls, F. J. R. Ruiz, & I. Valera (Eds.), *Proceedings of The 25th International Conference on Artificial Intelligence and Statistics* (Vol. 151, pp. 3242–3273). Proceedings of Machine Learning Research. https://proceedings.mlr.press/v151/milionis22a.html

Mironov, I. (2012). On significance of the least significant bits for differential privacy. In *CCS '12: Proceedings of the 2012 ACM Conference on Computer and Communications Security* (pp. 650–661). Association for Computing Machinery. https://doi.org/10.1145/2382196.2382264

Mironov, I. (2017). Rényi differential privacy. In *2017 IEEE 30th Computer Security Foundations Symposium* (pp. 263–275). IEEE. https://doi.org/10.1109/CSF.2017.11

Mohapatra, S., Sasy, S., He, X., Kamath, G., & Thakkar, O. (2022). The role of adaptive optimizers for honest private hyperparameter selection. *Proceedings of the AAAI Conference on Artificial Intelligence, 36*(7), 7806-7813. https://doi.org/10.1609/aaai.v36i7.20749

Nanayakkara, P., Smart, M. A., Cummings, R., Kaptchuk, G., & Redmiles, E. M. (2023). *What are the chances? Explaining the epsilon parameter in differential privacy*. ArXiv. https://doi.org/10.48550/arXiv.2303.00738

Nasirigerdeh, R., Torkzadehmahani, J., Rueckert, D., & Kaissis, G. (2023). Kernel normalized convolutional networks for privacy-preserving machine learning. In *2023 IEEE Conference on Secure and Trustworthy Machine Learning* (pp. 107–118). IEEE. https://doi.org/10.1109/SaTML54575.2023.00016

Nasr, M., Mahloujifar, S., Tang, X., Mittal, P., & Houmansadr, A. (2022). Effectively using public data in privacy preserving machine learning. In A. Krause, E. Brunskill, K. Cho, B. Engelhardt, S. Sabato, & J. Scarlett (Eds.), Proceedings of the 40th International Conference on Machine Learning (Vol. 202, pp. 25718–25732). Proceedings of Machine Learning Research. https://proceedings.mlr.press/v202/nasr23a.html

Nasr, M., Song, S., Thakurta, A., Papernot, N., & Carlini, N. (2021). *Adversary instantiation: Lower bounds for differentially private machine learning*. ArXiv. https://doi.org/10.48550/arXiv.2101.04535

Neel, S., Roth, A., & Sharifi-Malvajerdi, S. (2021). Descent-to-delete: Gradient-based methods for machine unlearning. In V. Feldman, K. Ligett, & S. Sabato (Eds.), *Proceedings of the 32nd International Conference on Algorithmic Learning* Theory (Vol. 132, pp. 931–962). Proceedings of Machine Learning Research. https://proceedings.mlr.press/v132/neel21a.html

Nikolov, A., & Tang, H. (2023). General Gaussian noise mechanisms and their optimality for unbiased mean estimation. ArXiv. https://doi.org/10.48550/arXiv.2301.13850

Nissenbaum, H. (2004). Privacy as contextual integrity. *Washington Law Review*, *79*(1), 119–158. https://digitalcommons.law.uw.edu/wlr/vol79/iss1/10/

Nissim, K., Bembenek, A., Wood, A., Bun, M., Gaboardi, M., Gasser, U., O'Brien, D., Steinke, T., & Vadhan, S. (2018). Bridging the gap between computer science and legal approaches to privacy. *Harvard Journal of Law & Technology*, *31*(2), 687–780.

Nissim, K., Raskhodnikova, S., & Smith, A. (2007). Smooth sensitivity and sampling in private data analysis. In *Proceedings of the 39th Annual ACM Symposium on the Theory of Computing* (pp. 75–84). Association for Computer Machinery. https://doi.org/10.1145/1250790.1250803

Noe, F., Herskind, R., & Søgaard, A. (2022). *Exploring the unfairness of DP-SGD across settings*. ArXiv. https://doi.org/10.48550/arXiv.2202.12058

Nuñez von Voigt, S., Pauli, M., Reichert, J., & Tschorsch, F. (2020). Every query counts: Analyzing the privacy loss of exploratory data analyses. In J. Garcia-Alfaro, G. Navarro-Arribas, & J. Herrera-Joancomarti (Eds), Lecture notes in computer science: Vol. 12484. *Data privacy management, cryptocurrencies and blockchain technology* (pp. 258–266). Springer. https://doi.org/10.1007/978-3-030-66172-4_17

OpenDP. (2022, November). Retrieved from https://opendp.org/

Oyallon, E., & Mallat, S. (2015). Deep roto-translation scattering for object classification. In *2015 IEEE Computer Society Conference on Computer Vision and Pattern Recognition* (pp. 2865–2873). IEEE. https://doi.org/10.1109/CVPR.2015.7298904

Papernot, N., Abadi, M., Erlingsson, Ú., Goodfellow, I., & Talwar, K. (2017, April 24–26). *Semi-supervised knowledge transfer for deep learning from private training data* [Paper presentation]. ICLR 2017: 5th International Conference on Learning Representations. https://openreview.net/forum?id=HkwoSDPgg

Papernot, N., Chien, S., Song, S., Thakurta, A., & Erlingsson, Ú. (2019, September 25). *Making the shoe fit: Architectures, initializations, and tuning for learning with privacy* [Conference blind submission]. ICLR 2020: The Eighth International Conference on Learning Representations, Virtual Event. https://openreview.net/forum?id=rJg851rYwH

Papernot, N., Song, S., Mironov, I., Raghunathan, A., Talwar, K., & Erlingsson, Ú. (2018, April 30–May 3). *Scalable private learning with PATE* [Poster session]. ICLR 2018: 6th International Conference on Learning Representations, Vancouver, BC, Canada. https://openreview.net/forum?id=rkZB1XbRZ

Papernot, N., & Steinke, T. (2022, April 25–29). *Hyperparameter tuning with Renyi differential privacy* [Conference presentation]. ICLR 2022: The Tenth International Conference on Learning Representations, Virtual Event. https://openreview.net/forum?id=-70L8lpp9DF

Papernot, N., Thakurta, A., Song, S., Chien, S., & Erlingsson, Ú. (2021). Tempered sigmoid activations for deep learning with differential privacy. *Proceedings of the AAAI Conference on Artificial Intelligence*, *35*(10), 9312–9321. https://doi.org/10.1609/aaai.v35i10.17123

Phan, H., Thai, M. T., Hu, H., Jin, R., Sun, T., & Dou, D. (2020). Scalable differential privacy with certified robustness in adversarial learning. In H. Daumé III, & A. Singh (Eds.), *Proceedings of the 37th International Conference on Machine Learning* (Vol. 119, pp. 7683–7694). Proceedings of Machine Learning Research. https://proceedings.mlr.press/v119/phan20a.html

Pujol, D., McKenna, R., Kuppam, S., Hay, M., Machanavajjhala, A., & Miklau, G. (2020). Fair decision making using privacy-protected data. In *Proceedings of the 2020 Conference on Fairness, Accountability, and Transparency* (pp. 189–199). Association for Computing Machinery. https://doi.org/10.1145/3351095.3372872

Pyrgelis, A., Troncoso, C., & De Cristofaro, E. (2017). Knock knock, who's there? Membership inference on aggregate location data. ArXiv. https://doi.org/10.48550/arXiv.1708.06145

Radford, A., Wu, J., Child, R., Luan, D., Amodei, D., & Sutskever, I. (2019). *Language models are unsupervised multitask learners*. OpenAI. https://cdn.openai.com/better-language-models/language_models_are_unsupervised_multitask_learners.pdf

Ramaswamy, S., Thakkar, O., Mathews, R., Andrew, G., McMahan, H. B., & Beaufays, F. (2020). *Training production language models without memorizing user data*. ArXiv. https://doi.org/10.48550/arXiv.2009.10031

Ramsay, K., & Chenouri, S. (2021). *Differentially private depth functions and their associated medians*. ArXiv. https://doi.org/10.48550/arXiv.2101.02800

Ramsay, K., Jagannath, A., & Chenouri, S. (2022). *Concentration of the exponential mechanism and differentially private multivariate medians*. ArXiv. https://doi.org/10.48550/arXiv.2210.06459

Redberg, R., Zhu, Y., & Wang, Y.-X. (2022). *Generalized PTR: User-friendly recipes for data-adaptive algorithms with differential privacy*. ArXiv. https://doi.org/10.48550/arXiv.2301.00301

Rogers, R., Subramaniam, S., Peng, S., Durfee, D., Lee, S., Kancha, S. K., Sahay, S., & Ahammad, P. (2021). LinkedIn's audience engagements API: A privacy preserving data analytics system at scale. *Journal of Privacy and Confidentiality*, *11*(3). https://doi.org/10.29012/jpc.782

Rogers, R. M., Roth, A., Ullman, J., & Vadhan, S. (2016). Privacy odometers and filters: Pay-as-you-go composition. In D. Lee, M. Sugiyama, U. Luxburg, I. Guyon, & R. Garnett (Eds.), *Advances in neural information processing systems* (Vol. 29, pp. 1921–1929). Curran Associates. https://papers.nips.cc/paper_files/paper/2016/hash/58c54802a9fb9526cd0923353a34a7ae-Abstract.html

Sablayrolles, A., Douze, M., Schmid, C., Ollivier, Y., & Jégou, H. (2019). White-box vs black-box: Bayes optimal strategies for membership inference. In K. Chaudhuri, & R. Salakhutdinov (Eds.), *Proceedings of the 36th International Conference on Machine Learning* (Vol. 97, pp. 5558–5567). Proceedings of Machine Learning Research. https://proceedings.mlr.press/v97/sablayrolles19a.html

Salem, A., Cherubin, G., Evans, D., Kopf, B., Paverd, A., Suri, A., Tople, S., & Zanella-Beguelin, S. (2022). *SoK: Let the privacy games begin! A unified treatment of data inference privacy in machine learning.* ArXiv. https://doi.org/10.48550/arXiv.2212.10986

Salem, A., Bhattacharyya, A., Backes, M., Fritz, M., & Zhang, Y. (2020). Updates-leak: Data set inference and reconstruction attacks in online learning. In *Proceedings of the 29th USENIX Conference on Security Symposium* (pp. 1291–1308). USENIX Association. https://www.usenix.org/conference/usenixsecurity20/presentation/salem

Samarati, P., & Sweeney, L. (1998). Generalizing data to provide anonymity when disclosing information. In *Proceedings of the 17th ACM SIGMOD-SIGACT-SIGART Symposium on Principles of Database Systems* (p. 188). Association for Computer Machinery. https://doi.org/10.1145/275487.275508

Sanyal, A., Hu, Y., & Yang, F. (2022). *How unfair is private learning?* ArXiv. https://doi.org/10.48550/arXiv.2206.03985

Sarathy, J., Song, S., Haque, A., Schlatter, T., & Vadhan, S. (2023). Don't look at the data! how differential privacy reconfigures the practices of data science. In A. Schmidt, K. Väänänen, T. Goyal, P. Ola Kristensson, A. Peters, S. Mueller, J. R. Williamson, & M. L. Wilson (Eds.), *Proceedings of the 2022 CHI Conference on Human Factors in Computing Systems* (pp. 1–19). Association for Computing Machinery. https://doi.org/10.1145/3544548.3580791

Shariff, R., & Sheffet, O. (2018). Differentially private contextual linear bandits. In S. Bengio, H. Wallach, H. Larochelle, K. Grauman, N. Cesa-Bianchi, & R. Garnett (Eds.), *Advances in neural information processing systems* (Vol. 31, (pp. 4296–4306). Curran Associates. https://papers.nips.cc/paper_files/paper/2018/hash/a1d7311f2a312426d710e1c617fcbc8c-Abstract.html

Sheffet, O. (2018). Locally private hypothesis testing. In J. Dy, & A. Krause (Eds.), *Proceedings of the 35th International Conference on Machine Learning* (Vol. 80, pp. 4605–4614). Proceedings of Machine Learning Research. https://proceedings.mlr.press/v80/sheffet18a.html

Shokri, R., Stronati, M., Song, C., & Shmatikov, V. (2017). Membership inference attacks against machine learning models. In 2017 *IEEE Symposium on Security and Privacy* (pp. 3–18). IEEE. https://doi.org/10.1109/SP.2017.41

Singhal, V., & Steinke, T. (2021). Privately learning subspaces. In M. Ranzato, A. Beygelzimer, Y. Dauphin, P.S. Liang, & J. Wortman Vaughan (Eds.), *Advances in neural information processing systems* (Vol. 34, pp. 1312–1324). Curran Associates. https://proceedings.neurips.cc/paper_files/paper/2021/hash/09b69adcd7cbae914c6204984097d2da-Abstract.html

Smart, M. A., Nanayakkara, P., Cummings, R., Kaptchuk, G., & Redmiles, E. M. (2020, October 15–16). *Improving explanations to end users about differential privacy* [Conference presentation]. 2020 USENIX Conference on Privacy Engineering Practice and Respect, Virtual Event. https://www.usenix.org/conference/pepr20/presentation/perera

Smith, A. (2011). Privacy-preserving statistical estimation with optimal convergence rates. In *Proceedings of the 43rd Annual ACM Symposium on the Theory of Computing* (pp. 813–822). Association for Computer Machinery. https://doi.org/10.1145/1993636.1993743

Song, S., Chaudhuri, K., & Sarwate, A. D. (2013). Stochastic gradient descent with differentially private updates. *In 2013 IEEE Global Conference on Signal and Information Processing* (pp. 245–248). IEEE. https://doi.org/10.1109/GlobalSIP.2013.6736861

Song, S., Steinke, T., Thakkar, O., & Thakurta, A. (2021). Evading the curse of dimensionality in unconstrained private GLMs. In A. Banerjee, & K. Fukumizu (Eds.), *Proceedings of the 24th International Conference on Artificial Intelligence and Statistics* (Vol. 130, pp. 2638–2646). Proceedings of Machine Learning Research. https://proceedings.mlr.press/v130/song21a.html

Song, S., Wang, Y., & Chaudhuri, K. (2017). Pufferfish privacy mechanisms for correlated data. In *Proceedings of the 2017 ACM International Conference on Management of Data* (pp. 1291–1306). Association for Computing Machinery. https://doi.org/10.1145/3035918.3064025

The State of Alabama et al v. United States Department of Commerce et al, Case No. 3:21-cv-211-RAH-ECM-KCN (WO) (M.D. Ala. Mar 10, 2021).

Steed, R., Liu, T., Wu, Z. S., & Acquisti, A. (2022). Policy impacts of statistical uncertainty and privacy. *Science*, *377*(6609), 928–931. https://doi.org/10.1126/science.abq4481

Steinke, T., & Ullman, J. (2015). Interactive fingerprinting codes and the hardness of preventing false discovery. In Peter Grünwald, Elad Hazan, Satyen Kale (Eds.), *Proceedings of the 28th Annual Conference on Learning Theory* (Vol. 40, pp. 1588–1628). Proceedings of Machine Learning Research. https://proceedings.mlr.press/v40/Steinke15.html

Steinke, T., & Ullman, J. (2017a). Between pure and approximate differential privacy. *The Journal of Privacy and Confidentiality*, *7*(2), 3–22. https://doi.org/10.29012/jpc.v7i2.648

Steinke, T., & Ullman, J. (2017b). Tight lower bounds for differentially private selection. In *2017 IEEE 58th Annual Symposium on Foundations of Computer Science* (pp. 552–563). IEEE. https://doi.org/10.1109/FOCS.2017.57

Suri, A., & Evans, D. (2022). Formalizing and estimating distribution inference risks. *Proceedings on Privacy Enhancing Technologies, 2022*(4), 528–551. https://doi.org/10.56553/popets-2022-0121

Suriyakumar, V. M., Papernot, N., Goldenberg, A., & Ghassemi, M. (2021). Chasing your long tails: Differentially private prediction in health care settings. In *Proceedings of the 2021 ACM Conference on Fairness, Accountability, and Transparency* (pp. 723–734). Association for Computer Machinery. https://doi.org/10.1145/3442188.3445934

Swanberg, M., Globus-Harris, I., Griffith, I., Ritz, A., Groce, A., & Bray, A. (2019). Improved differentially private analysis of variance. *Proceedings on Privacy Enhancing Technologies, 2019*(3), 310–330. https://doi.org/10.2478/POPETS-2019-0049

Talwar, K., Thakurta, A. G., & Zhang, L. (2015). Nearly optimal private LASSO. *Advances in neural information processing systems* (Vol. 28, pp. 3025–3033). Curran Associates. https://papers.nips.cc/paper_files/paper/2015/hash/52d080a3e172c33fd6886a37e7288491-Abstract.html

Thakkar, O., Andrew, G., McMahan, H. B., Ramaswamy, S. (2019). Differentially private learning with adaptive clipping. ArXiv. https://doi.org/10.48550/arXiv.1905.03871

Thakkar, O., Ramaswamy, S., Mathews, R., & Beaufays, F. (2020). *Understanding unintended memorization in federated learning*. ArXiv. https://doi.org/10.48550/arXiv.2006.07490

Thakurta, A. G., & Smith, A. (2013). Differentially private feature selection via stability arguments, and the robustness of the LASSO. In S. Shalev-Shwartz, & I. Steinwart (Eds.), *Proceedings of the 26th Annual Conference on Learning Theory* (Vol. 30, pp. 819–850. Proceedings of Machine Learning Research. https://proceedings.mlr.press/v30/Guha13.html

Thudi, A., Shumailov, I., Boenisch, F., & Papernot, N. (2022). *Bounding membership inference*. ArXiv. https://doi.org/10.48550/arXiv.2202.12232

Tramèr, F., Terzis, A., Steinke, T., Song, S., Jagielski, M., & Carlini, N. (2022). *Debugging differential privacy: A case study for privacy auditing*. ArXiv. https://doi.org/10.48550/arXiv.2202.12219

Tramèr, F., & Boneh, D. (2021, May 3–7). *Differentially private learning needs better features (or much more data)* [Spotlight presentation]. ICLR 2021: 9th International Conference on Learning Representations, Virtual Event, Austria. https://openreview.net/forum?id=YTWGvpFOQD-

Tramèr, F., Kamath, G., & Carlini, N. (2022). *Considerations for differentially private learning with large-scale public pretraining.* ArXiv. https://doi.org/10.48550/arXiv.2212.06470

Tran, C., Dinh, M., & Fioretto, F. (2021). Differentially private empirical risk minimization under the fairness lens. In M. Ranzato, A. Beygelzimer, Y. Dauphin, P.S. Liang, & J. Wortman Vaughan (Eds.), *Advances in neural information processing systems* (Vol. 34, pp. 27555–27565). Curran Associates. https://proceedings.neurips.cc/paper/2021/hash/e7e8f8e5982b3298c8addedf6811d500-Abstract.html

Tran, C., Fioretto, F., Van Hentenryck, P., & Yao, Z. (2021). Decision making with differential privacy under a fairness lens. In Z.-H. Zhou (Ed.), *Proceedings of the Thirtieth International Joint Conference on Artificial Intelligence* (pp. 560–566). International Joint Conferences on Artificial Intelligence Organization. https://doi.org/10.24963/ijcai.2021/78

Tsfadia, E., Cohen, E., Kaplan, H., Mansour, Y., & Stemmer, U. (2022). FriendlyCore: Practical differentially private aggregation. In K. Chaudhuri, S. Jegelka, L. Song, C. Szepesvari, G. Niu, & S. Sabato (Eds.), *Proceedings of the 39th International Conference on Machine Learning* (Vol. 162, pp. 21828–21863). Proceedings of Machine Learning Research. https://proceedings.mlr.press/v162/tsfadia22a.html

Tukey, J. W. (1960). A survey of sampling from contaminated distributions. In I. Oklin (Ed.), *Contributions to probability and statistics: Essays in honor of Harold Hotelling* (pp. 448–485). Stanford University Press.

Tversky, A., & Kahneman, D. (1974). Judgment under uncertainty: Heuristics and biases. *Science*, *185*(4157), 1124–1131. https://doi.org/10.1126/science.185.4157.1124

Uhlerop, C., Slavković, A., & Fienberg, S. E. (2013). Privacy-preserving data sharing for genome-wide association studies. *The Journal of Privacy and Confidentiality*, *5*(1), 137–166. https://doi.org/10.29012/jpc.v5i1.629

Ullman, J. (2016). Answering $n^2+o(1)$ counting queries with differential privacy is hard. *SIAM Journal on Computing*, *45*(2), 473–496. https://doi.org/10.1137/130928121

United States Census Bureau (n.d.). Federal statistical research data centers [Retrieved 3/17/23]. https://www.census.gov/about/adrm/fsrdc.html

Uniyal, A., Naidu, R., Kotti, S., Singh, S., Kenfack, P. J., Mireshghallah, F., & Trask, A. (2021). *DP-SGD vs PATE: Which has less disparate impact on model accuracy?* ArXiv. https://doi.org/10.48550/arXiv.2106.12576

Vadhan, S. (2017). The complexity of differential privacy. In Y. Lindell (Ed.), *Tutorials on the foundations of cryptography: Dedicated to Oded Goldreich* (pp. 347–450). Springer . https://doi.org/10.1007/978-3-319-57048-8_7

Vadhan, S., & Wang, T. (2021). Concurrent composition of differential privacy. In K. Nissim & B. Waters (Eds.), *Theory of cryptography* (pp. 582–604). Springer.

Vadhan, S. P., & Zhang, W. (2022). Concurrent composition theorems for all standard variants of differential privacy. In *Proceedings of the 55th Annual ACM Symposium on Theory of Computing* (pp. 507–519). Association for Computer Machinery. https://doi.org/10.1145/3564246.3585241

Varshney, P., Thakurta, A., & Jain, P. (2022). (Nearly) optimal private linear regression for sub-Gaussian data via adaptive clipping. In P.-L. Loh, & M. Raginsky (Eds.), *Proceedings of Thirty Fifth Conference on Learning Theory* (Vol. 178, pp. 1126–1166). Proceedings of Machine Learning Research. https://proceedings.mlr.press/v178/varshney22a.html

Vu, D., & Slavkovic, A. (2009). Differential privacy for clinical trial data: Preliminary evaluations. In *2009 IEEE International Conference on Data Mining Workshops* (pp. 138–143). IEEE. https://doi.org/10.1109/ICDMW.2009.52

Wang, D., Xiao, H., Devadas, S., & Xu, J. (2020). On differentially private stochastic convex optimization with heavy-tailed data. In H. Daumé III, & A. Singh (Eds.), *Proceedings of the 37th International Conference on Machine Learning* (Vol. 119, pp. 10081–10091). Proceedings of Machine Learning Research. . https://proceedings.mlr.press/v119/wang20y.html

Wang, D., Ye, M., & Xu, J. (2017). Differentially private empirical risk minimization revisited: Faster and more general. In I. Guyon, U. Von Luxburg, S. Bengio, H. Wallach, R. Fergus, S. Vishwanathan, & R. Garnett (Eds.,) *Advances in neural information processing systems* (Vol. 30, pp. 2722–2731). Curran Associates. https://papers.nips.cc/paper_files/paper/2017/hash/f337d999d9ad116a7b4f3d409fcc6480-Abstract.html

Wang, H., Zhang, Z., Wang, T., He, S., Backes, M., Chen, J., & Zhang, Y. (2023). PrivTrace: Differentially private trajectory synthesis by adaptive Markov model. In J. Calandrino, & C. Troncoso (Eds.), *Proceedings of the 32nd USENIX Conference on Security* Symposium (pp. 1649–1666). USENIX Association. https://www.usenix.org/conference/usenixsecurity23/presentation/wang-haiming

Wang, H., Hong, H., Xiong, L., Qin, Z., & Hong, Y. (2022). L-SRR: Local differential privacy for location-based services with staircase randomized response. In *CCS '22: Proceedings of the 2022 ACM SIGSAC Conference on Computer and Communications Security* (pp. 2809–2823). Association for Computing Machinery. https://doi.org/10.1145/3548606.3560636

Wang, H., Gao, S., Zhang, H., Su, W. J., & Shen, M. (2023, December 12). *DP-HyPO: An adaptive private framework for hyperparameter optimization* [Paper presentation]. Thirty-seventh Conference on Neural Information Processing Systems, New Orleans, LA, United States.

Wang, L., & Gu, Q. (2019). Differentially private iterative gradient hard thresholding for sparse learning. In S. Kraus (Ed.), *Proceedings of the Twenty-Eighth International Joint Conference on Artificial Intelligence* (pp. 3740–3747). International Joint Conferences on Artificial Intelligence Organization. https://doi.org/10.24963/ijcai.2019/519

Wang, M., Ji, Z., Kim, H.-E., Wang, S., Xiong, L., & Jiang, X. (2017). Selecting optimal subset to release under differentially private m-estimators from hybrid datasets. *IEEE Transactions on Knowledge and Data Engineering*, *30*(3), 573–584. https://doi.org/10.1109%2FTKDE.2017.2773545

Wang, Y., Lee, J., & Kifer, D. (2015). *Revisiting differentially private hypothesis tests for categorical data*. ArXiv. https://doi.org/10.48550/arXiv.1511.03376

Wang, Y.-X. (2018). *Revisiting differentially private linear regression: Optimal and adaptive prediction & estimation in unbounded domain*. ArXiv. https://doi.org/10.48550/arXiv.1803.02596

Wang, Y.-X. (2019). Per-instance differential privacy. *The Journal of Privacy and Confidentiality*, *9*(1). https://doi.org/10.29012/jpc.662

Wang, Y.-X., Fienberg, S., & Smola, A. (2015). Privacy for free: Posterior sampling and stochastic gradient Monte Carlo. In F. Bach, & D. Blei (Eds.), *Proceedings of the 32nd International Conference on Machine Learning* (Vol. 37, pp. 2493–2502). Proceedings of Machine Learning Research. https://proceedings.mlr.press/v37/wangg15.html

Watson, L., Guo, C., Cormode, G., & Sablayrolles, A. (2021). *On the importance of difficulty calibration in membership inference attacks*. ArXiv. https://doi.org/10.48550/arXiv.2111.08440

Whitehouse, J., Ramdas, A., Rogers, R., & Wu, Z. S. (2023). Fully-adaptive composition in differential privacy. In A. Krause, E. Brunskill, K. Cho, B. Engelhardt, S. Sabato, & J. Scarlett (Eds.), *Proceedings of the 40th International Conference on Machine Learning* (Vol. 202, pp. 36990–37007). Proceedings of Machine Learning Research. https://proceedings.mlr.press/v202/whitehouse23a.html

Whitehouse, J., Wu, Z. S., Ramdas, A., & Rogers, R. (2022). *Brownian noise reduction: Maximizing privacy subject to accuracy constraints*. ArXiv. https://doi.org/10.48550/arXiv.2206.07234

Wood, A., Altman, M., Bembenek, A., Bun, M., Gaboardi, M., Honaker, J., Nissim, K., O'Brien, D., Steinke, T., & Vadhan, S. (2018). Differential privacy: A primer for a non-technical audience. *Vanderbilt Journal of Entertainment and Technology Law*, *21*(1), 209–276. https://scholarship.law.vanderbilt.edu/jetlaw/vol21/iss1/4/

Wu, X., Li, F., Kumar, A., Chaudhuri, K., Jha, S., & Naughton, J. (2017). Bolt-on differential privacy for scalable stochastic gradient descent-based analytics. In *Proceedings of the 2017 ACM SIGMOD International*

*Conference on Management of Data* (pp. 1307–1322). Association for Computer Machinery. https://doi.org/10.1145/3035918.3064047

Xiao, Y., & Xiong, L. (2015). Protecting locations with differential privacy under temporal correlations. In *CCS '15: Proceedings of the 22nd ACM SIGSAC Conference on Computer and Communications Security* (pp. 1298–1309). Association for Computing Machinery. https://doi.org/10.1145/2810103.2813640

Xie, L., Lin, K., Wang, S., Wang, F., & Zhou, J. (2018). *Differentially private generative adversarial network*. ArXiv. https://doi.org/10.48550/arXiv.1802.06739

Xiong, A., Wu, C., Wang, T., Proctor, R., Blocki, J., Li, N., & Jha, S. (2022). *Using illustrations to communicate differential privacy trust models: An investigation of users' comprehension, perception, and data sharing decision*. ArXiv. https://doi.org/10.48550/arXiv.2202.10014

Xu, D., Du, W., & Wu, X. (2020). *Removing disparate impact of differentially private stochastic gradient descent on model accuracy*. ArXiv. https://doi.org/10.48550/arXiv.2003.03699

Ye, M., & Barg, A. (2018). Optimal schemes for discrete distribution estimation under locally differential privacy. *IEEE Transactions on Information Theory*, *64*(8), 5662–5676. https://doi.org/10.1109/ISIT.2017.8006630

Yeom, S., Giacomelli, I., Fredrikson, M., & Jha, S. (2018). Privacy risk in machine learning: Analyzing the connection to overfitting. In *2018 IEEE 31st Computer Security Foundations Symposium* (pp. 268–282). IEEE. https://doi.org/10.1109/CSF.2018.00027

Yin, H., Mallya, A., Vahdat, A., Alvarez, J. M., Kautz, J., & Molchanov, P. (2021). See through gradients: Image batch recovery via GradInversion. In *2021 IEEE/CVF Conference on Computer Vision and Pattern Recognition* (pp. 16337–16346). IEEE. https://doi.org/10.1109/CVPR46437.2021.01607

Yousefpour, A., Shilov, I., Sablayrolles, A., Testuggine, D., Prasad, K., Malek, M., Nguyen, J., Gosh, S., Bharadwaj, A., Zhao, J., Cormode, G., & Mironov, I. (2021). *Opacus: User-friendly differential privacy library in PyTorch*. ArXiv. https://doi.org/10.48550/arXiv.2109.12298

Yu, D., Naik, S., Backurs, A., Gopi, S., Inan, H. A., Kamath, G., Kulkarni, J., Lee, Y. T., Manoel, A., Wutschitz, L., Yekhanin, S., & Zhang, H. (2022, April 25–29). *Differentially private fine-tuning of language models* [Poster presentation]. ICLR 2022: The Tenth International Conference on Learning Representations, Virtual Event. https://openreview.net/forum?id=Q42f0dfjECO

Yu, D., Zhang, H., Chen, W., & Liu, T.-Y. (2021, May 3–7). *Do not let privacy overbill utility: Gradient embedding perturbation for private learning* [Poster presentation]. ICLR 2021: 9th International Conference on Learning Representations, Virtual Event, Austria. https://openreview.net/forum?id=7aogOj_VYO0

Yu, D., Zhang, H., Chen, W., Liu, T.-Y., & Yin, J. (2019). *Gradient perturbation is underrated for differentially private convex optimization*. ArXiv. https://doi.org/10.48550/arXiv.1911.11363

Yu, D., Zhang, H., Chen, W., Yin, J., & Liu, T.-Y. (2021). Large scale private learning via low-rank reparametrization. In M. Meila & T. Zhang (Eds.), *Proceedings of the 38th International Conference on Machine Learning* (Vol. 139, pp. 12208–12218). Proceedings of Machine Learning Research. https://proceedings.mlr.press/v139/yu21f.html

Yu, L., Liu, L., Pu, C., Gursoy, M. E., & Truex, S. (2019). Differentially private model publishing for deep learning. In *2019 IEEE Symposium on Security and Privacy* (pp. 332–349). IEEE. https://doi.org/10.1109/SP.2019.00019

Zanella-Béguelin, S., Wutschitz, L., Tople, S., Rühle, V., Paverd, A., Ohrimenko, O., Köpf, B., & Brockschmidt, M. (2020). Analyzing information leakage of updates to natural language models. In *CCS '20: Proceedings of the 2020 ACM SIGSAC conference on computer and communications security* (pp. 363–375). Association for Computing Machinery. https://doi.org/10.1145/3372297.3417880

Zanella-Béguelin, S., Wutschitz, L., Tople, S., Salem, A., Rühle, V., Paverd, A., Naseri, M., Köpf, B., & Jones, D. (2022). *Bayesian estimation of differential privacy*. ArXiv. https://doi.org/10.48550/ARXIV.2206.05199

Zhang, H., Kamath, G., Kulkarni, J., & Wu, Z. S. (2020). Privately learning Markov random fields. In H. Daumé III, & A. Singh (Eds.), *Proceedings of the 37th International Conference on Machine Learning* (Vol. 119, pp. 11129–11140). Proceedings of Machine Learning Research. https://proceedings.mlr.press/v119/zhang20l.html

Zhang, J., Cormode, G., Procopiuc, C. M., Srivastava, D., & Xiao, X. (2017). Privbayes: Private data release via Bayesian networks. *ACM Transactions on Database Systems (TODS), 42*(4), 1–41. https://doi.org/10.1145/3134428

Zhang, Q., Ma, J., Lou, J., & Xiong, L. (2021). Private stochastic non-convex optimization with improved utility rates. In *Proceedings of the Thirtieth International Joint Conference on Artificial Intelligence* (pp. 3370–3376). International Joint Conferences on Artificial Intelligence Organization. https://doi.org/10.24963/ijcai.2021/464

Zhang, W., Ohrimenko, O., & Cummings, R. (2022). Attribute privacy: Framework and mechanisms. In *Proceedings of the 2022 ACM Conference on Fairness, Accountability, and Transparency* (pp. 757–766). Association for Computer Machinery. https://doi.org/10.1145/3531146.3533139

Zhang, W., Tople, S., & Ohrimenko, O. (2021). Leakage of dataset properties in multi-party machine learning. In *Proceedings of the 30th USENIX Security Symposium* (pp. 2687–2704). USENIX Association. https://www.usenix.org/system/files/sec21-zhang-wanrong.pdf

Zhao, B., Mopuri, K. R., & Bilen, H. (2020). iDLG: Improved deep leakage from gradients. ArXiv. https://doi.org/10.48550/arXiv.2001.02610

Zhou, Y., Chen, X., Hong, M., Wu, Z. S., & Banerjee, A. (2020). *Private stochastic non-convex optimization: Adaptive algorithms and tighter generalization bounds*. ArXiv. https://doi.org/10.48550/arXiv.2006.13501

Zhou, Y., Wu, Z. S., & Banerjee, A. (2021, May 3–7). *Bypassing the ambient dimension: Private SGD with gradient subspace identification* [Poster presentation]. ICLR 2021: 9th International Conference on Learning Representations, Virtual Event, Austria. https://openreview.net/forum?id=7dpmlkBuJFC

Zhu, L., Liu, Z., & Han, S. (2019). Deep leakage from gradients. In H. Wallach, H. Larochelle, A. Beygelzimer, F. d'Alché-Buc, E. Fox, & R. Garnett (Eds.), *Advances in neural information processing systems* (Vol. 32, pp. 14774–14784). Curran Associates. https://papers.nips.cc/paper_files/paper/2019/hash/60a6c4002cc7b29142def8871531281a-Abstract.html

Zhu, Y., & Wang, Y.-X. (2022). Adaptive private-k-selection with adaptive k and application to multi-label PATE. In G. Camps-Valls, F. J. R. Ruiz, & I. Valera (Eds.), *Proceedings of The 25th International Conference on Artificial Intelligence and Statistics* (Vol. *151*, pp. 5622–5635). Proceedings of Machine Learning Research. https://proceedings.mlr.press/v151/zhu22e.html

Zhu, Y., Yu, X., Chandraker, M., & Wang, Y.-X. (2020). Private-kNN: Practical differential privacy for computer vision. In *2020 IEEE/CVF Conference on Computer Vision and Pattern Recognition* (pp. 11854–11862). IEEE. https://doi.org/10.1109/CVPR42600.2020.01187

## Footnotes

1. As an example of how adding DP noise is perceived, consider this text from Alabama's lawsuit against the Census Bureau (which was decided in favor of the Census Bureau): "Thus, while the Bureau touts its mission 'to count everyone once, only once, and in the right place,' it will force Alabama to redistrict using results that purposefully count people in the wrong place" (The State of Alabama et al v. United States Department of Commerce et al, 2005). ↩

2. GPU (Graphics Processing Unit) is a processor designed for fast math calculations, used to accelerate the training and execution of neural networks due to its parallel processing abilities. TPU (Tensor Processing Unit) is a specialized chip created by Google to accelerate TensorFlow-based deep learning tasks, offering high-speed computation tailored specifically for neural network operations. ↩

3. Table scrolls horizontally for additional columns. ↩

4. We note that the JFT data set is proprietary to Google and has not been made public. ↩

5. https://github.com/zykls/folktables ↩

6. https://github.com/jinyuan-jia/MemGuard/tree/master/data/location ↩

7. https://codalab.lisn.upsaclay.fr/competitions/8553 ↩

8. https://github.com/bargavj/Texas-100X ↩

9. https://physionet.org/content/mimiciv/2.2/ ↩

10. https://www.microsoft.com/en-us/research/publication/geolife-gps-trajectory-dataset-user-guide/ ↩

11. The term 'data anonymization' is used loosely here, in order to be consistent with colloquial usage. The term 'anonymization' does not have any specific technical meaning associated to it. ↩

# References

- Abadi, M., Chu, A., Goodfellow, I., McMahan, H. B., Mironov, I., Talwar, K., & Zhang, L. (2016). Deep learning with differential privacy. In *CCS '16: Proceedings of the 2016 ACM Conference on Computer and Communications Security* (pp. 308–318). Association for Computing Machinery. https://doi.org/10.1145/2976749.2978318

  ↩

- Abowd, J. M. (2018). The U.S. Census Bureau adopts differential privacy. In *Proceedings of the 24th ACM SIGKDD International Conference on Knowledge Discovery & Data Mining* (p. 2867). Association for Computing Machinery. https://doi.org/10.1145/3219819.3226070

  ↩

- Abowd, J., Ashmead, R., Cumings-Menon, R., Garfinkel, S., Heineck, M., Heiss, C., Johns, R., Kifer, D., Leclerc, P., Machanavajjhala, A., Moran, B., Sexton, W., Spence, M., & Zhuravlev, P. (2022). The 2020 Census disclosure avoidance system TopDown Algorithm. *Harvard Data Science Review*, (Special Issue 2). https://doi.org/10.1162/99608f92.529e3cb9

  ↩

- Acharya, J., & Sun, Z. (2019). Communication complexity in locally private distribution estimation and heavy hitters. In K. Chaudhuri, & R. Salakhutdinov (Eds.), *Proceedings of the 36th International Conference on Machine Learning* (Vol. 97, pp. 51–60). Proceedings of Machine Learning Research https://proceedings.mlr.press/v97/acharya19c.html

    ↩

- Acharya, J., Canonne, C. L., & Tyagi, H. (2019). Inference under information constraints: Lower bounds from chi-square contraction. In A. Beygelzimer, & D. Hsu (Eds.),*Proceedings of the 32nd Annual Conference on Learning Theory* (Vol. 99, pp. 3–17). Proceedings of Machine Learning Research. http://proceedings.mlr.press/v99/acharya19a.html

    ↩

- Acharya, J., Canonne, C. L., Freitag, C., & Tyagi, H. (2019). Test without trust: Optimal locally private distribution testing. In K. Chaudhuri, & M. Sugiyama (Eds.), *Proceedings of the 22nd International Conference on Artificial Intelligence and Statistics* (Vol. 89, pp. 2067–2076). Proceedings of Machine Learning Research. https://proceedings.mlr.press/v89/acharya19b.html

    ↩

- Acharya, J., Canonne, C. L., Liu, Y., Sun, Z., & Tyagi, H. (2021). Interactive inference under information constraints. *IEEE Transactions on Information Theory*, *68*(1), 502–516. https://doi.org/10.1109/TIT.2021.3123905

    ↩

- Acharya, J., Canonne, C. L., Sun, Z., & Tyagi, H. (2022). The role of interactivity in structured estimation. In P.-L. Loh, & M. Raginsky (Eds.), *Proceedings of Thirty Fifth Conference on Learning Theory* (Vol. 178, pp. 1328–1355). Proceedings of Machine Learning Research. https://proceedings.mlr.press/v178/acharya22b.html

    ↩

- Acharya, J., Kairouz, P., Liu, Y., & Sun, Z. (2021). Estimating sparse discrete distributions under privacy and communication constraints. In V. Feldman, K. Ligett, & S. Sabato (Eds.), *Proceedings of the 32nd International Conference on Algorithmic Learning Theory* (Vol. 132, pp. 79–98). Proceedings of Machine Learning Research. https://proceedings.mlr.press/v132/acharya21b.html

    ↩

- Acharya, J., Kamath, G., Sun, Z., & Zhang, H. (2018). INSPECTRE: Privately estimating the unseen. In J. Dy, & A. Krause (Eds.), *Proceedings of the 35th International Conference on Machine Learning* (Vol. 80, pp. 30–39). Proceedings of Machine Learning Research. https://proceedings.mlr.press/v80/acharya18a.html

    ↩

- Acharya, J., Sun, Z., & Zhang, H. (2018). Differentially private testing of identity and closeness of discrete distributions. In S. Bengio, H. Wallach, H. Larochelle, K. Grauman, N. Cesa-Bianchi, & R. Garnett (Eds.), *Advances in neural information processing systems* (Vol. 31, pp. 6878–6891). Curran Associates. https://proceedings.neurips.cc/paper/2018/hash/7de32147a4f1055bed9e4faf3485a84d-Abstract.html

    ↩

- Acharya, J., Sun, Z., & Zhang, H. (2021). Robust testing and estimation under manipulation attacks. In M. Meila & T. Zhang (Eds.), *Proceedings of the 38th International Conference on Machine Learning* (Vol. 139, pp. 43–53). Proceedings of Machine Learning Research. https://proceedings.mlr.press/v139/acharya21a.html

    ↩

- Agarwal, S. (2020). *Trade-offs between fairness, interpretability, and privacy in machine learning* (Master's thesis, University of Waterloo). http://hdl.handle.net/10012/15861

    ↩

- Alabi, D., Kothari, P. K., Tankala, P., Venkat, P., & Zhang, F. (2023). Privately estimating a Gaussian: Efficient, robust, and optimal. In *Proceedings of the 55th Annual ACM Symposium on the Theory of Computing* (pp. 483–496). Association for Computing Machinery. https://doi.org/10.1145/3564246.3585194

    ↩

- Aliakbarpour, M., Diakonikolas, I., & Rubinfeld, R. (2018). Differentially private identity and equivalence testing of discrete distributions. In J. Dy, & A. Krause (Eds.), *Proceedings of the 35th International Conference on Machine Learning* (Vol. 80, pp. 169–178). Proceedings of Machine Learning Research. https://proceedings.mlr.press/v80/aliakbarpour18a.html

    ↩

- Alon, N., Bassily, R., & Moran, S. (2019). Limits of private learning with access to public data. In H. Wallach, H. Larochelle, A. Beygelzimer, F. d'Alché-Buc, E. Fox, & R. Garnett (Eds.), *Advances in neural information processing systems* (Vol. 32, pp. 10342–10352). Curran Associates. https://papers.nips.cc/paper_files/paper/2019/hash/9a6a1aaafe73c572b7374828b03a1881-Abstract.html

    ↩

- Altschuler, J. M., & Talwar, K. (2022). *Privacy of noisy stochastic gradient descent: More iterations without more privacy loss*. ArXiv. https://doi.org/10.48550/arXiv.2205.13710

    ↩

- Amid, E., Ganesh, A., Mathews, R., Ramaswamy, S., Song, S., Steinke, T., Suriyakumar, V. M., Thakkar, O., & Thakurta, A. (2021). *Public data-assisted mirror descent for private model training*. ArXiv. https://doi.org/10.48550/arXiv.2112.00193

↩

- Amin, K., Dick, T., Khodak, M., & Vassilvitskii, S. (2022). *Private algorithms with private predictions.* ArXiv. https://doi.org/10.48550/arXiv.2210.11222

    ↩

- Amin, K., Dick, T., Kulesza, A., Munoz, A., & Vassilvitskii, S. (2019). Differentially private covariance estimation. In H. Wallach, H. Larochelle, A. Beygelzimer, F. d'Alché-Buc, E. Fox, & R. Garnett (Eds.), *Advances in neural information processing systems* (Vol. *32*, pp. 14190–14199). Curran Associates. https://papers.nips.cc/paper_files/paper/2019/hash/4158f6d19559955bae372bb00f6204e4-Abstract.html

    ↩

- Amin, K., Gillenwater, J., Joseph, M., Kulesza, A., & Vassilvitskii, S. (2022). *Plume: Differential privacy at scale.* ArXiv. https://doi.org/10.48550/arXiv.2201.11603

    ↩

- Amin, K., Joseph, M., Ribero, M., & Vassilvitskii, S. (2022). *Easy differentially private linear regression.* ArXiv. https://doi.org/10.48550/arXiv.2208.07353

    ↩

- Amin, K., Kulesza, A., Munoz, A., & Vassilvtiskii, S. (2019). Bounding user contributions: A bias-variance trade-off in differential privacy. In K. Chaudhuri, & R. Salakhutdinov (Eds.), *Proceedings of the 36th International Conference on Machine Learning* (Vol. 97, pp. 263–271). Proceedings of Machine Learning Research. https://proceedings.mlr.press/v97/amin19a.html

    ↩

- Andrés, M. E., Bordenabe, N. E., Chatzikokolakis, K., & Palamidessi, C. (2013). Geo-indistinguishability: Differential privacy for location-based systems. In *CCS'13: Proceedings of the 2013 ACM SIGSAC Conference on Computer & Communications Security*, (pp. 901–914). Association for Computer Machinery. https://doi.org/10.1145/2508859.2516735

    ↩

- Andrew, G., Thakkar, O., McMahan, B., & Ramaswamy, S. (2021). Differentially private learning with adaptive clipping. In M. Ranzato, A. Beygelzimer, Y. Dauphin, P.S. Liang, & J. Wortman Vaughan (Eds.), *Advances in neural information processing systems* (Vol. 34, pp. 17455–17466). Curran Associates. https://proceedings.neurips.cc/paper/2021/hash/91cff01af640a24e7f9f7a5ab407889f-Abstract.html

    ↩

- Andrysco, M., Kohlbrenner, D., Mowery, K., Jhala, R., Lerner, S., & Shacham, H. (2015). On subnormal floating point and abnormal timing. In *2015 IEEE Symposium on Security and Privacy* (pp. 623–639). IEEE. https://doi.org/10.1109/SP.2015.44

↩

- Apple Differential Privacy Team (2017, December 6). Learning with privacy at scale. *Apple Machine Learning Research*. https://machinelearning.apple.com/research/learning-with-privacy-at-scale

  ↩

- Arora, R., Bassily, R., González, T., Guzmán, C., Menart, M., & Ullah, E. (2022). *Faster rates of convergence to stationary points in differentially private optimization*. ArXiv. https://doi.org/10.48550/arXiv.2206.00846

  ↩

- Asi, H., & Duchi, J. C. (2020). *Near instance-optimality in differential privacy*. ArXiv. https://doi.org/10.48550/arXiv.2005.10630

  ↩

- Asi, H., Duchi, J., Fallah, A., Javidbakht, O., & Talwar, K. (2021). Private adaptive gradient methods for convex optimization. In M. Meila, & T. Zhang (Eds.), *Proceedings of the 38th International Conference on Machine Learning* (Vol. 139, pp. 383–392). Proceedings of Machine Learning Research. https://proceedings.mlr.press/v139/asi21a.html

  ↩

- Asi, H., Feldman, V., Koren, T., & Talwar, K. (2021). *Private stochastic convex optimization: Optimal rates in ℓ1 geometry*. ArXiv. https://doi.org/10.48550/arXiv.2103.01516

  ↩

- Asi, H., Ullman, J., & Zakynthinou, L. (2023). *From robustness to privacy and back*. ArXiv. https://doi.org/10.48550/arXiv.2302.01855

  ↩

- Ateniese, G., Mancini, L. V., Spognardi, A., Villani, A., Vitali, D., & Felici, G. (2015). Hacking smart machines with smarter ones: How to extract meaningful data from machine learning classifiers. *International Journal of Security and Networks*, *10*(3), 137–150. https://doi.org/10.1504/IJSN.2015.071829

  ↩

- Augenstein, S., McMahan, H. B., Ramage, D., Ramaswamy, S., Kairouz, P., Chen, M., Mathews, R., et al. (2019). *Generative models for effective ML on private, decentralized datasets*. ArXiv. https://doi.org/10.48550/arXiv.1911.06679

  ↩

- autodp contributors. (2019). *Autodp: Automating differential privacy computation*. GitHub. https://github.com/yuxiangw/autodp

↩

- Avella-Medina, M., & Brunel, V.-E. (2019). *Differentially private sub-Gaussian location estimators*. ArXiv. https://doi.org/10.48550/arXiv.1906.11923

↩

- Bagdasaryan, E., Poursaeed, O., & Shmatikov, V. (2019). Differential privacy has disparate impact on model accuracy. In H. Wallach, H. Larochelle, A. Beygelzimer, F. d'Alché-Buc, E. Fox, & R. Garnett (Eds.), *Advances in neural information processing systems* (Vol. 32, pp. 15479–15488). Curran Associates. https://proceedings.neurips.cc/paper/2019/hash/fc0de4e0396fff257ea362983c2dda5a-Abstract.html

↩

- Balle, B., Cherubin, G., & Hayes, J. (2022). Reconstructing training data with informed adversaries. *2022 IEEE Symposium on Security and Privacy* (pp. 1138–1156). IEEE. https://doi.org/10.1109/SP46214.2022.9833677

↩

- Barber, R. F., & Duchi, J. C. (2014). *Privacy and statistical risk: Formalisms and minimax bounds*. ArXiv. https://doi.org/10.48550/arXiv.1412.4451

↩

- Bassily, R., Cheu, A., Moran, S., Nikolov, A., Ullman, J., & Wu, S. (2020). Private query release assisted by public data. In H. Daumé III, & A. Singh (Eds.), *Proceedings of the 37th International Conference on Machine Learning* (Vol. 119, pp. 695–703). Proceedings of Machine Learning Research. https://proceedings.mlr.press/v119/bassily20a.html

↩

- Bassily, R., Feldman, V., Talwar, K., & Thakurta, A. G. (2019). Private stochastic convex optimization with optimal rates. In H. Wallach, H. Larochelle, A. Beygelzimer, F. d'Alché-Buc, E. Fox, & R. Garnett (Eds.), *Advances in neural information processing systems* (Vol. 32, pp. 11282–11291). Curran Associates. https://proceedings.neurips.cc/paper_files/paper/2019/hash/3bd8fdb090f1f5eb66a00c84dbc5ad51-Abstract.html

↩

- Bassily, R., Guzmán, C., & Menart, M. (2021). Differentially private stochastic optimization: New results in convex and non-convex settings. In M. Ranzato, A. Beygelzimer, Y. Dauphin, P.S. Liang, & J. Wortman Vaughan (Eds.), *Advances in neural information processing systems* (Vol. 34, pp. 9317–9329). Curran Associates. https://proceedings.neurips.cc/paper/2021/hash/4ddb5b8d603f88e9de689f3230234b47-Abstract.html

↩

- Bassily, R., Moran, S., & Nandi, A. (2020). Learning from mixtures of private and public populations. In H. Larochelle, M. Ranzato, R. Hadsell, M.F. Balcan, & H. Lin (Eds.), *Advances in neural information processing systems* (Vol. 33, pp. 2947–2957). Curran Associates. https://proceedings.neurips.cc/paper/2020/hash/1ee942c6b182d0f041a2312947385b23-Abstract.html

    ↩

- Bassily, R., Smith, A., & Thakurta, A. (2014). Private empirical risk minimization: Efficient algorithms and tight error bounds. In *Proceedings of the 55th Annual IEEE Symposium on Foundations of Computer Science* (pp. 464–473). IEEE. https://doi.org/10.1109/FOCS.2014.56

    ↩

- Bassily, R., Thakkar, O., & Guha Thakurta, A. (2018). Model-agnostic private learning. In S. Bengio, H. Wallach, H. Larochelle, K. Grauman, N. Cesa-Bianchi, & R. Garnett (Eds.), *Advances in neural information processing systems* (Vol. 31, pp. 7102–7112). Curran Associates. https://papers.nips.cc/paper_files/paper/2018/hash/aa97d584861474f4097cf13ccb5325da-Abstract.html

    ↩

- Beimel, A., Brenner, H., Kasiviswanathan, S. P., & Nissim, K. (2014). Bounds on the sample complexity for private learning and private data release. *Machine Learning*, *94*(3), 401–437. https://doi.org/10.1007/s10994-013-5404-1

    ↩

- Beimel, A., Nissim, K., & Stemmer, U. (2013). Private learning and sanitization: Pure vs. approximate differential privacy. In *Lecture notes in computer science: Vol. 8096 Approximation, randomization, and combinatorial optimization. Algorithms and techniques* (pp. 363–378). Springer. https://doi.org/10.1007/978-3-642-40328-6_26

    ↩

- Beimel, A., Nissim, K., & Stemmer, U. (2015). Learning privately with labeled and unlabeled examples. In *Proceedings of the 26th Annual ACM-SIAM Symposium on Discrete Algorithms* (pp. 461–477). Society for Industrial and Applied Mathematics. https://dl.acm.org/doi/abs/10.5555/2722129.2722161

    ↩

- Bie, A., Kamath, G., & Singhal, V. (2022). *Private estimation with public data*. ArXiv. https://doi.org/10.48550/arXiv.2208.07984

    ↩

- Biswas, S., Dong, Y., Kamath, G., & Ullman, J. (2020). Coinpress: Practical private mean and covariance estimation. In H. Larochelle, M. Ranzato, R. Hadsell, M.F. Balcan, & H. Lin (Eds.), *Advances in neural information processing systems* (Vol. 33, pp. 14475–14485). Curran Associates.

↵

- Błasiok, J., Bun, M., Nikolov, A., & Steinke, T. (2019). Towards instance-optimal private query release. In *Proceedings of the Thirtieth Annual ACM-SIAM Symposium on Discrete Algorithms* (pp. 2480–2497). Society for Industrial and Applied Mathematics. https://doi.org/10.1137/1.9781611975482.152

↵

- Boenisch, F., Dziedzic, A., Schuster, R., Shamsabadi, A. S., Shumailov, I., & Papernot, N. (2021). *When the curious abandon honesty: Federated learning is not private*. ArXiv. https://doi.org/10.48550/arXiv.2112.02918

↵

- Bolukbasi, T., Chang, K.-W., Zou, J. Y., Saligrama, V., & Kalai, A. T. (2016). Man is to computer programmer as woman is to homemaker? In D. Lee, M. Sugiyama, U. Luxburg, I. Guyon, & R. Garnett (Eds.), Debiasing word embeddings. *Advances in Neural Information Processing Systems 29* (pp. 4349–4357). https://papers.nips.cc/paper_files/paper/2016/hash/a486cd07e4ac3d270571622f4f316ec5-Abstract.html

↵

- Bommasani, R., Hudson, D. A., Adeli, E., Altman, R., Arora, S., von Arx, S., Bernstein, M. S., Bohg, J., Bosselut, A., Brunskill, E., Brynjolfsson, E., Buch, S., Card, D., Castellon, R., Chatterji, N., Chen, A., Creel, K., Davis, J. Q., Demszky, D., . . . Liang, P. (2021). *On the opportunities and risks of foundation models*. ArXiv. https://doi.org/10.48550/arXiv.2108.07258

↵

- Bonawitz, K., Kairouz, P., Mcmahan, B., & Ramage, D. (2022). Federated learning and privacy. *Communications of the ACM*, *65*(4), 90–97. https://doi.org/10.1145/3500240

↵

- Bourtoule, L., Chandrasekaran, V., Choquette-Choo, C. A., Jia, H., Travers, A., Zhang, B., Lie, D., & Papernot, N. (2021). Machine unlearning. In *2021 IEEE Symposium on Security and Privacy* (pp. 141–159). IEEE. https://doi.org/10.1109/SP40001.2021.00019

↵

- Brown, G., Gaboardi, M., Smith, A., Ullman, J., & Zakynthinou, L. (2021). Covariance-aware private mean estimation without private covariance estimation. In M. Ranzato, A. Beygelzimer, Y. Dauphin, P.S. Liang, & J. Wortman Vaughan (Eds.), *Advances in neural information processing systems* (Vol. 43, pp. 7950–7964). Curran Associates. https://proceedings.neurips.cc/paper/2021/hash/42778ef0b5805a96f9511e20b5611fce-Abstract.html

↵

- Brown, G., Hopkins, S. B., & Smith, A. (2023). *Fast, sample-efficient, affine-invariant private mean and covariance estimation for subgaussian distributions*. ArXiv. https://doi.org/10.48550/arXiv.2301.12250

    ↩

- Brown, H., Lee, K., Mireshghallah, F., Shokri, R., & Tramèr, F. (2022). What does it mean for a language model to preserve privacy? In *Proceedings of the 2022 ACM Conference on Fairness, Accountability, and Transparency* (pp. 2280–2292). Association for Computer Machinery. https://doi.org/10.1145/3531146.3534642

    ↩

- Brunel, V.-E., & Avella-Medina, M. (2020). *Propose, test, release: Differentially private estimation with high probability*. ArXiv. https://doi.org/10.48550/arXiv.2002.08774

    ↩

- Bu, Z., Dong, J., Long, Q., & Su, W. J. (2020). Deep learning with Gaussian differential privacy. *Harvard Data Science Review*, *2*(3). https://doi.org/10.1162/99608f92.cfc5dd25

    ↩

- Bun, M., & Steinke, T. (2016). Concentrated differential privacy: Simplifications, extensions, and lower bounds. *Proceedings of the 14th Conference on Theory of Cryptography*, 635–658. ↩
- Bun, M., & Steinke, T. (2019). Average-case averages: Private algorithms for smooth sensitivity and mean estimation. *Advances in Neural Information Processing Systems 32*, 181–191. ↩
- Bun, M., Desfontaines, D., Dwork, C., Naor, M., Nissim, K., Roth, A., Smith, A., Steinke, T., Ullman, J., & Vadhan, S. (2021, June 3). *Statistical inference is not a privacy violation*. DifferentialPrivacy.org. https://differentialprivacy.org/inference-is-not-a-privacy-violation/.

    ↩

- Bun, M., Dwork, C., Rothblum, G. N., & Steinke, T. (2018). Composable and versatile privacy via truncated CDP. *Proceedings of the 50th Annual ACM SIGACT Symposium on Theory of Computing* (pp. 74–86). Association for Computer Machinery. https://doi.org/10.1145/3188745.3188946

    ↩

- Bun, M., Kamath, G., Steinke, T., & Wu, Z. S. (2019). Private hypothesis selection. In H. Wallach, H. Larochelle, A. Beygelzimer, F. d'Alché-Buc, E. Fox, & R. Garnett (Eds.), *Advances in neural information processing systems* (Vol. 32, pp. 156–167). Curran Associates. https://proceedings.neurips.cc/paper_files/paper/2019/hash/9778d5d219c5080b9a6a17bef029331c-Abstract.html

    ↩

- 

↩

- Bun, M., Ullman, J., & Vadhan, S. (2014). Fingerprinting codes and the price of approximate differential privacy. In *Proceedings of the 46th Annual ACM Symposium on the Theory of Computing* (pp. 1–10). Association for Computer Machinery. https://doi.org/10.1145/2591796.2591877

↩

- Buolamwini, J., & Gebru, T. (2018). Gender shades: Intersectional accuracy disparities in commercial gender classification. In S. A. Friedler, & C. Wilson (Eds.), *Proceedings of the 1st Conference on Fairness, Accountability, and Transparency* (Vol. 81, pp. 77–91). Proceedings of Machine Learning Research. https://proceedings.mlr.press/v81/buolamwini18a.html

↩

- Cai, B., Daskalakis, C., & Kamath, G. (2017). Priv'IT: Private and sample efficient identity testing. In D. Precup, & Y. W. Teh (Eds.), *Proceedings of the 34th International Conference on Machine Learning* (Vol. 70, pp. 635–644). Proceedings of Machine Learning Research. https://proceedings.mlr.press/v70/cai17a.html

↩

- Cai, T. T., Wang, Y., & Zhang, L. (2019). *The cost of privacy: Optimal rates of convergence for parameter estimation with differential privacy*. ArXiv. https://doi.org/10.48550/arXiv.1902.04495

↩

- Cai, T. T., Wang, Y., & Zhang, L. (2020). *The cost of privacy in generalized linear models: Algorithms and minimax lower bounds*. ArXiv. https://doi.org/10.48550/arXiv.2011.03900

↩

- Caliskan, A., Bryson, J. J., & Narayanan, A. (2017). Semantics derived automatically from language corpora contain human-like biases. *Science*, *356*(6334), 183–186. https://doi.org/10.1126/science.aal4230

↩

- Campbell, Z., Bray, A., Ritz, A., & Groce, A. (2018). Differentially private ANOVA testing. In *2018 International Conference on Data Intelligence and Security* (pp. 281–285). IEEE. https://doi.org/10.1109/ICDIS.2018.00052

↩

- Canonne, C. L., Kamath, G., McMillan, A., Smith, A., & Ullman, J. (2019). The structure of optimal private tests for simple hypotheses. In *Proceedings of the 51st Annual ACM Symposium on the Theory of Computing* (pp. 310–321). Association for Computer Machinery. https://doi.org/10.1145/3313276.3316336

↩

-

↵

- Cao, Y., & Yang, J. (2015). Towards making systems forget with machine unlearning. In *2015 IEEE Symposium on Security and Privacy* (pp. 463–480). IEEE. https://doi.org/10.1109/SP.2015.35

  ↵

- Carlini, N., Chien, S., Nasr, M., Song, S., Terzis, A., & Tramer, F. (2021). *Membership inference attacks from first principles*. ArXiv. https://doi.org/10.48550/arXiv.2112.03570

  ↵

- Carlini, N., Hayes, J., Nasr, M., Jagielski, M., Sehwag, V., Tramèr, F., Balle, B., Ippolito, D., & Wallace, E. (2023). *Extracting training data from diffusion models*. ArXiv. https://doi.org/10.48550/arXiv.2301.13188

  ↵

- Carlini, N., Jagielski, M., & Mironov, I. (2020). Cryptanalytic extraction of neural network models. In D. Micciancio, & T. Ristenpart (Eds.), *Lecture notes in computer science: Vol. 12172. Advances in Cryptology – CRYPTO 2020* (pp. 189–218). Springer. https://doi.org/10.1007/978-3-030-56877-1_7

  ↵

- Carlini, N., Jagielski, M., Papernot, N., Terzis, A., Tramer, F., & Zhang, C. (2022). *The privacy onion effect: Memorization is relative*. ArXiv. https://doi.org/10.48550/arXiv.2206.10469

  ↵

- Carlini, N., Liu, C., Erlingsson, Ú., Kos, J., & Song, D. (2019). The secret sharer: Evaluating and testing unintended memorization in neural networks. In *Proceedings of the 28th USENIX Conference on Security Symposium* (pp. 267–284). USENIX Association. https://www.usenix.org/system/files/sec19-carlini.pdf

  ↵

- Carlini, N., Tramèr, F., Wallace, E., Jagielski, M., Herbert-Voss, A., Lee, K., Roberts, A., Brown, T. B., Song, D., Erlingsson, Ú., Oprea, A., Raffel, C. (2021). Extracting training data from large language models. In *Proceedings of the 30th USENIX Security Symposium* (pp. 2633–2650). USENIX Association. https://www.usenix.org/system/files/sec21-carlini-extracting.pdf

  ↵

- Cattan, Y., Choquette-Choo, C. A., Papernot, N., & Thakurta, A. (2022). *Fine-tuning with differential privacy necessitates an additional hyperparameter search*. ArXiv. https://doi.org/10.48550/arXiv.2210.02156

  ↵

- Chaudhuri, K., & Hsu, D. (2011). Sample complexity bounds for differentially private learning. In S. M. Kakade, & U. von Luxburg (Eds.), *Proceedings of the 24th Annual Conference on Learning Theory* (Vol. 19,

↩

- Chaudhuri, K., & Hsu, D. (2012). Convergence rates for differentially private statistical estimation. In J. Langford, & J. Pineau (Eds.), *Proceedings of the 29th International Conference on International Conference on Machine Learning* (pp. 1715–1722). Omnipress.

↩

- Chaudhuri, K., Monteleoni, C., & Sarwate, A. D. (2011). Differentially private empirical risk minimization. *Journal of Machine Learning Research*, *12*(29), 1069–1109. https://www.jmlr.org/papers/v12/chaudhuri11a.html

↩

- Chen, M., Zhang, Z., Wang, T., Backes, M., Humbert, M., & Zhang, Y. (2021). When machine unlearning jeopardizes privacy. In *CCS '21: Proceedings of the 2021 ACM SIGSAC Conference on Computer and Communications Security* (pp. 896–911). Association for Computing Machinery. https://doi.org/10.1145/3460120.3484756

↩

- Cheng, A., Wang, J., Zhang, X. S., Chen, Q., Wang, P., & Cheng, J. (2022). DPNAS: Neural architecture search for deep learning with differential privacy. *Proceedings of the AAAI Conference on Artificial Intelligence*, *36*(6), 6358–6366. hhttps://doi.org/10.1609/aaai.v36i6.20586

↩

- Cheu, A., Smith, A., & Ullman, J. (2021). Manipulation attacks in local differential privacy. In *2021 IEEE Symposium on Security and Privacy* (pp. 883–900). IEEE. https://doi.org/10.1109/SP40001.2021.00001

↩

- Choquette-Choo, C. A., Tramer, F., Carlini, N., & Papernot, N. (2021). Label-only membership inference attacks. In M Meila, & T Zhang (Eds.), *Proceedings of the 38th International Conference on Machine Learning* (Vol. 139, pp.1964–1974). Proceedings of Machine Learning Research. https://proceedings.mlr.press/v139/choquette-choo21a.html

↩

- Cohen, A., & Nissim, K. (2020). Towards formalizing the GDPR's notion of singling out. *Proceedings of the National Academy of Sciences*, *117*(15), 8344–8352. https://doi.org/10.1073/pnas.1914598117

↩

- Cormode, G. (2010). *Individual privacy vs population privacy: Learning to attack anonymization*. ArXiv. https://doi.org/10.48550/ARXIV.1011.2511

↩

- Couch, S., Kazan, Z., Shi, K., Bray, A., & Groce, A. (2019). Differentially private nonparametric hypothesis testing. In *CCS '19: Proceedings of the 2019 ACM Conference on Computer and Communications Security* (pp. 737–751). Association for Computing Machinery. https://doi.org/10.1145/3319535.3339821

    ↩

- Cummings, R., Gupta, V., Kimpara, D., & Morgenstern, J. (2019). On the compatibility of privacy and fairness. In *Adjunct Publication of the 27th Conference on User Modeling, Adaptation and Personalization* (pp. 309–315). Association for Computing Machinery. https://doi.org/10.1145/3314183.3323847

    ↩

- Cummings, R., Kaptchuk, G., & Redmiles, E. M. (2021). "I need a better description": An investigation into user expectations for differential privacy. In *CCS '21: Proceedings of the 2021 ACM SIGSAC Conference on Computer and Communications Security* (pp. 3037–3052). Association for Computing Machinery. https://doi.org/10.1145/3460120.3485252

    ↩

- Cummings, R., Krehbiel, S., Mei, Y., Tuo, R., & Zhang, W. (2018). Differentially private change-point detection. In S. Bengio, H. Wallach, H. Larochelle, K. Grauman, N. Cesa-Bianchi, & R. Garnett (Eds.), *Advances in neural information processing systems* (Vol. 31, pp. 10825–10834). Curran Associates. https://papers.nips.cc/paper_files/paper/2018/hash/f19ec2b84181033bf4753a5a51d5d608-Abstract.html

    ↩

- Cummings, R., Ligett, K., Pai, M. M., & Roth, A. (2016). The strange case of privacy in equilibrium models. In *Proceedings of the 2016 ACM Conference on Economics and Computation* (pp. 659–659). Association for Computer Machinery. https://doi.org/10.1145/2940716.2940740

    ↩

- Damaskinos, G., Mendler-Dünner, C., Guerraoui, R., Papandreou, N., & Parnell, T. (2021). Differentially private stochastic coordinate descent. *Proceedings of the AAAI Conference on Artificial Intelligence*, *35*(8), 7176–7184. https://doi.org/10.1609/aaai.v35i8.16882

    ↩

- De, S., Berrada, L., Hayes, J., Smith, S. L., & Balle, B. (2022). *Unlocking high-accuracy differentially private image classification through scale*. ArXiv. https://doi.org/10.48550/arXiv.2204.13650

    ↩

- Decarolis, C., Ram, M., Esmaeili, S., Wang, Y.-X., & Huang, F. (2020). An end-to-end differentially private latent Dirichlet allocation using a spectral algorithm. In H. Daumé III, & A. Singh (Eds.), *Proceedings of the 37th International Conference on Machine Learning* (Vol. 119, pp. 2421–2431). Proceedings of Machine Learning Research. https://proceedings.mlr.press/v119/decarolis20a.html

↩

- Desfontaines, D. (2021, October, 1). A list of real-world uses of differential privacy. *Ted is writing things*. https://desfontain.es/privacy/real-world-differential-privacy.html

↩

- Devlin, J., Chang, M.-W., Lee, K., & Toutanova, K. (2019). BERT: Pre-training of deep bidirectional transformers for language understanding. In J. Burstein, C. Doran, & T. Solorio (Eds.), *Proceedings of the 2019 Conference of the North American Chapter of the Association for Computational Linguistics: Human Language Technologies, Volume 1 (Long and Short Papers)* (pp. 4171–4186). Association for Computational Linguistics. https://doi.org/10.18653/v1%2FN19-1423

↩

- Diakonikolas, I., Kamath, G., Kane, D. M., Li, J., Moitra, A., & Stewart, A. (2016). Robust estimators in high dimensions without the computational intractability. In *Proceedings of the 57th Annual IEEE Symposium on Foundations of Computer Science* (pp. 655–664). IEEE. https://doi.org/10.1109/FOCS.2016.85

↩

- Dick, T., Kulesza, A., Sun, Z., & Suresh, A. T. (2023). *Subset-based instance optimality in private estimation*. ArXiv. https://doi.org/10.48550/arXiv.2303.01262

↩

- Dimitrov, D. I., Balunović, M., Jovanović, N., & Vechev, M. (2022*). LAMP: Extracting text from gradients with language model priors*. ArXiv. https://doi.org/10.48550/arXiv.2202.08827

↩

- Ding, B., Kulkarni, J., & Yekhanin, S. (2017). Collecting telemetry data privately. In I. Guyon, U. Von Luxburg, S. Bengio, H. Wallach, R. Fergus, S. Vishwanathan, & R. Garnett (Eds.), *Advances in neural information processing systems* (Vol. 30, pp. 3571–3580). Curran Associates. https://papers.nips.cc/paper_files/paper/2017/hash/253614bbac999b38b5b60cae531c4969-Abstract.html

↩

- Ding, J., Zhang, X., Li, X., Wang, J., Yu, R., & Pan, M. (2020). Differentially private and fair classification via calibrated functional mechanism. *Proceedings of the AAAI Conference on Artificial Intelligence*, *34*(1), 622–629. https://doi.org/10.1609/aaai.v34i01.5402

↩

- Ding, Z., Wang, Y., Wang, G., Zhang, D., & Kifer, D. (2018). Detecting violations of differential privacy. In *CCS '18: Proceedings of the 2018 ACM SIGSAC Conference on Computer and Communications Security* (pp. 475–489). Association for Computing Machinery. https://doi.org/10.1145/3243734.3243818

↩

- Dinur, I., & Nissim, K. (2003). Revealing information while preserving privacy. In *Proceedings of the 22nd ACM SIGMOD-SIGACT-SIGART Symposium on Principles of Database Systems* (pp. 202–210). Association for Computer Machinery. https://doi.org/10.1145/773153.773173

  ↩

- Dong, J., Roth, A., & Su, W. J. (2022). Gaussian differential privacy. *Journal of the Royal Statistical Society: Series B (Statistical Methodology)*, *84*(1), 3–37. https://doi.org/10.1111/rssb.12454

  ↩

- Dong, J., Su, W., & Zhang, L. (2021). A central limit theorem for differentially private query answering. In M. Ranzato, A. Beygelzimer, Y. Dauphin, P.S. Liang, & J. Wortman Vaughan (Eds.), *Advances in neural information processing systems* (Vol. 34, pp. 14759–14770). Curran Associates. https://proceedings.neurips.cc/paper/2021/hash/7c2c48a32443ad8f805e48520f3b26a4-Abstract.html

  ↩

- Dong, W., & Yi, K. (2021). *Universal private estimators*. ArXiv. https://doi.org/10.48550/arXiv.2111.02598

  ↩

- Du, M., Jia, R., & Song, D. (2020, April 26–May 1). *Robust anomaly detection and backdoor attack detection via differential privacy* [Poster session]. ICLR 2020: The Eighth International Conference on Learning Representations, Virtual Event. https://openreview.net/forum?id=SJx0q1rtvS

  ↩

- Du, W., Foot, C., Moniot, M., Bray, A., & Groce, A. (2020). *Differentially private confidence intervals*. ArXiv. https://doi.org/10.48550/arXiv.2001.02285

  ↩

- Duchi, J. C., & Ruan, F. (2018). *The right complexity measure in locally private estimation: It is not the Fisher information*. ArXiv. https://doi.org/10.48550/arXiv.1806.05756

  ↩

- Duchi, J. C., Jordan, M. I., & Wainwright, M. J. (2017). Minimax optimal procedures for locally private estimation. *Journal of the American Statistical Association*.
  Duchi, J. C., Jordan, M. I., & Wainwright, M. J. (2018). Minimax optimal procedures for locally private estimation. *Journal of the American Statistical Association, 113*(521), 182–201. https://doi.org/10.1080/01621459.2017.1389735

  ↩

-

↩

- Duchi, J., & Rogers, R. (2019). Lower bounds for locally private estimation via communication complexity. In A. Beygelzimer, & D. Hsu (Eds.), *Proceedings of the 32nd Annual Conference on Learning Theory* (Vol. 99, pp. 1161–1191). Proceedings of Machine Learning Research. https://proceedings.mlr.press/v99/duchi19a.html

↩

- Duchi, J., Haque, S., & Kuditipudi, R. (2023). *A fast algorithm for adaptive private mean estimation*. ArXiv. https://doi.org/10.48550/arXiv.2301.07078

↩

- Durfee, D., & Rogers, R. M. (2019). Practical differentially private top-k selection with pay-what-you-get composition. In H. Wallach, H. Larochelle, A. Beygelzimer, F. d'Alché-Buc, E. Fox, & R. Garnett (Eds.), *Advances in neural information processing systems* (Vol. 32, pp. 3532–3542). Curran Associates. https://papers.nips.cc/paper_files/paper/2019/hash/b139e104214a08ae3f2ebcce149cdf6e-Abstract.html

↩

- Dwork, C. (2008). Differential privacy: A survey of results. In M. Agrawal, D. Du, Z. Duan, A. Li, (Eds), *Lecture notes in computer science: Vol. 4978. Theory and Applications of Models of Computation* (pp. 1–19). Springer. https://doi.org/10.1007/978-3-540-79228-4_1

↩

- Dwork, C., & Lei, J. (2009). Differential privacy and robust statistics. In *Proceedings of the 41st Annual ACM Symposium on the Theory of Computing* (pp. 371–380). Association for Computer Machinery. https://doi.org/10.1145/1536414.1536466

↩

- Dwork, C., & Ullman, J. (2018). The Fienberg Problem: How to allow human interactive data analysis in the age of differential privacy. *Journal of Privacy and Confidentiality*, *8*(1). https://doi.org/10.29012/jpc.687

↩

- Dwork, C., Kenthapadi, K., McSherry, F., Mironov, I., & Naor, M. (2006). Our data, ourselves: Privacy via distributed noise generation. In S. Vaudenay (Ed.), *Proceedings of the 24th Annual International Conference on the Theory and Applications of Cryptographic Techniques* (pp. 486–503). Springer. https://doi.org/10.1007/11761679_29

↩

- Dwork, C., Kohli, N., & Mulligan, D. (2019). Differential privacy in practice: Expose your epsilons! *Journal of Privacy and Confidentiality*, *9*(2). https://doi.org/10.29012/jpc.689

↩

- Dwork, C., McSherry, F., Nissim, K., & Smith, A. (2006). Calibrating noise to sensitivity in private data analysis. In S. Halevi, & T. Rabin (Eds.), *Proceedings of the 3rd Conference on Theory of Cryptography* (pp. 265–284). Springer. https://doi.org/10.1007/11681878_14

↩

- Dwork, C., Naor, M., Reingold, O., Rothblum, G. N., & Vadhan, S. (2009). On the complexity of differentially private data release: Efficient algorithms and hardness results. In *Proceedings of the 41st Annual ACM Symposium on the Theory of Computing* (pp. 381–390). Association for Computer Machinery. https://doi.org/10.1145/1536414.1536467

↩

- Dwork, C., Smith, A., Steinke, T., & Ullman, J. (2017). Exposed! A survey of attacks on private data. *Annual Review of Statistics and Its Application*, *4*(1), 61–84. https://doi.org/10.1146/annurev-statistics-060116-054123

↩

- Dwork, C., Smith, A., Steinke, T., Ullman, J., & Vadhan, S. (2015). Robust traceability from trace amounts. In *Proceedings of the 56th Annual IEEE Symposium on Foundations of Computer Science* (pp. 650–669). IEEE. https://doi.org/10.1109/FOCS.2015.46

↩

- Epasto, A., Mahdian, M., Mao, J., Mirrokni, V., & Ren, L. (2020). Smoothly bounding user contributions in differential privacy. In H. Larochelle, M. Ranzato, R. Hadsell, M.F. Balcan, & H. Lin (Eds.), *Advances in neural information processing systems* (Vol. *33*, pp. 13999–14010). Curran Associates. https://proceedings.neurips.cc/paper_files/paper/2020/hash/a0dc078ca0d99b5ebb465a9f1cad54ba-Abstract.html

↩

- Erlingsson, Ú., Mironov, I., Raghunathan, A., & Song, S. (2019). *That which we call private*. ArXiv. https://doi.org/10.48550/arXiv.1908.03566

↩

- Erlingsson, Ú., Pihur, V., & Korolova, A. (2014). RAPPOR: Randomized aggregatable privacy-preserving ordinal response. In *CCS '14: Proceedings of the 2014 ACM Conference on Computer and Communications Security* (pp. 1054–1067). Association for Computing Machinery. https://doi.org/10.1145/2660267.2660348

↩

- Esipova, M. S., Ghomi, A. A., Luo, Y., & Cresswell, J. C. (2022). *Disparate impact in differential privacy from gradient misalignment*. ArXiv. https://doi.org/10.48550/arXiv.2206.07737

- ↩

- Evfimievski, A., Gehrke, J., & Srikant, R. (2003). Limiting privacy breaches in privacy preserving data mining. In *Proceedings of the 22nd ACM SIGMOD-SIGACT-SIGART Symposium on Principles of Database Systems* (pp. 211–222). Association for Computer Machinery. https://doi.org/10.1145/773153.773174

- ↩

- Farrand, T., Mireshghallah, F., Singh, S., & Trask, A. (2020). Neither private nor fair: Impact of data imbalance on utility and fairness in differential privacy. In *Proceedings of the 2020 Workshop on Privacy-Preserving Machine Learning in Practice* (pp. 15–19). Association for Computing Machinery. https://doi.org/10.1145/3411501.3419419

- ↩

- Feldman, V. (2020). Does learning require memorization? A short tale about a long tail. In *Proceedings of the 52nd Annual ACM Symposium on the Theory of Computing* (pp. 954–959). Association for Computer Machinery. https://doi.org/10.1145/3357713.3384290

- ↩

- Feldman, V., & Steinke, T. (2018). Calibrating noise to variance in adaptive data analysis. In S. Bubeck, V. Perchet, & P. Rigollet (Eds.), *Proceedings of the 31st Conference On Learning Theory* (Vol. 75, pp. 535–544). Proceedings of Machine Learning Research. https://proceedings.mlr.press/v75/feldman18a.html

- ↩

- Feldman, V., & Zrnic, T. (2021). Individual privacy accounting via a Rényi filter. In M. Ranzato, A. Beygelzimer, Y. Dauphin, P.S. Liang, & J. Wortman Vaughan (Eds.), *Advances in neural information processing systems* (Vol. 34, pp. 28080–28091). Curran Associates. https://proceedings.neurips.cc/paper/2021/hash/ec7f346604f518906d35ef0492709f78-Abstract.html

- ↩

- Feldman, V., Mironov, I., Talwar, K., & Thakurta, A. (2018). Privacy amplification by iteration. In *2018 IEEE 59th Annual Symposium on Foundations of Computer Science* (pp. 521–532). IEEE. https://doi.org/10.1109/FOCS.2018.00056

- ↩

- Fichtenberger, H., Henzinger, M., & Ost, W. (2021). *Differentially private algorithms for graphs under continual observation*. ArXiv. https://doi.org/10.48550/arXiv.2106.14756

- ↩

- Fioretto, F., Tran, C., Van Hentenryck, P., & Zhu, K. (2022). Differential privacy and fairness in decisions and learning tasks: A survey. In L. De Raedt (Ed.), *Proceedings of the Thirty-First International Joint*

↵

- Fowl, L., Geiping, J., Czaja, W., Goldblum, M., & Goldstein, T. (2021). *Robbing the fed: Directly obtaining private data in federated learning with modified models*. ArXiv. https://doi.org/10.48550/arXiv.2110.13057

↵

- Fowl, L., Geiping, J., Reich, S., Wen, Y., Czaja, W., Goldblum, M., & Goldstein, T. (2022). *Decepticons: Corrupted transformers breach privacy in federated learning for language models*. ArXiv. https://doi.org/10.48550/arXiv.2201.12675

↵

- Fredrikson, M., Jha, S., & Ristenpart, T. (2015). Model inversion attacks that exploit confidence information and basic countermeasures. In *CCS '15: Proceedings of the 22nd ACM SIGSAC conference on computer and communications security* (pp. 1322–1333). Association for Computing Machinery. https://doi.org/10.1145/2810103.2813677

↵

- Gaboardi, M., Honaker, J., King, G., Murtagh, J., Nissim, K., Ullman, J., & Vadhan, S. (2016). *PSI (Ψ): A private data sharing interface*. ArXiv. https://doi.org/10.48550/arXiv.1609.04340

↵

- Gaboardi, M., Lim, H., Rogers, R. M., & Vadhan, S. P. (2016). Differentially private chi-squared hypothesis testing: Goodness of fit and independence testing. In M. F. Balcan, & K. Q. Weinberger (Eds.), *Proceedings of the 33rd International Conference on Machine Learning* (Vol. 48, pp. 1395–1403). Proceedings of Machine Learning Research. https://proceedings.mlr.press/v48/rogers16.html

↵

- Ganesh, A., Haghifam, M., Nasr, M., Oh, S., Steinke, T., Thakkar, O., Thakurta, A., & Wang, L. (2023). *Why is public pretraining necessary for private model training?* ArXiv. https://doi.org/10.48550/arXiv.2302.09483

↵

- Ganesh, A., Liu, D., Oh, S., & Thakurta, A. (2023). *Private (stochastic) non-convex optimization revisited: Second-order stationary points and excess risks*. ArXiv. https://doi.org/10.48550/arXiv.2302.09699

↵

- Ganev, G., Oprisanu, B., & De Cristofaro, E. (2022). Robin Hood and Matthew effects: Differential privacy has disparate impact on synthetic data. In K. Chaudhuri, S. Jegelka, L. Song, C. Szepesvari, G. Niu, & S. Sabato (Eds.), *Proceedings of the 39th International Conference on Machine Learning* (Vol. 162, pp. 6944–6959). Proceedings of Machine Learning Research. https://proceedings.mlr.press/v162/ganev22a.html

Harvard Data Science Review • Issue 6.1, Winter 2024    Advancing Differential Privacy: Where We Are Now and Future Directions for Real-World Deployment

x
[↩](#)

- Ganju, K., Wang, Q., Yang, W., Gunter, C. A., & Borisov, N. (2018). Property inference attacks on fully connected neural networks using permutation invariant representations. In *CCS '18: Proceedings of the 2018 ACM SIGSAC Conference on Computer and Communications Security* (pp. 619–633). Association for Computing Machinery. https://doi.org/10.1145/3243734.3243834

[↩](#)

- Gao, C., & Wright, S. J. (2023). *Differentially private optimization for smooth nonconvex ERM*. ArXiv. https://doi.org/10.48550/arXiv.2302.04972

[↩](#)

- Garfinkel, S., Abowd, J. M., & Martindale, C. (2019). Understanding database reconstruction attacks on public data. *Communications of the ACM*, *62*(3), 46–53. https://doi.org/10.1145/3287287

[↩](#)

- Garg, S., Raz, R., & Tal, A. (2018). Extractor-based time-space lower bounds for learning. In *Proceedings of the 50th Annual ACM Symposium on the Theory of Computing* (pp. 990–1002). Association for Computer Machinery. https://doi.org/10.1145/3188745.3188962

[↩](#)

- Ge, J., Wang, Z., Wang, M., & Liu, H. (2018). Minimax-optimal privacy-preserving sparse PCA in distributed systems. In A. Storkey, & F. Perez-Cruz *Proceedings of the Twenty-First International Conference on Artificial Intelligence and Statistics* (Vol. 84, pp. 1589–1598). Proceedings of Machine Learning Research. https://proceedings.mlr.press/v84/ge18a.html

[↩](#)

- Geiping, J., Bauermeister, H., Dröge, H., & Moeller, M. (2020). Inverting gradients - How easy is it to break privacy in federated learning? In H. Larochelle, M. Ranzato, R. Hadsell, M.F. Balcan, & H. Lin (Eds.), *Advances in neural information processing systems* (Vol. *33*, pp. 16937–16947). Curran Associates. https://proceedings.neurips.cc/paper_files/paper/2020/hash/c4ede56bbd98819ae6112b20ac6bf145-Abstract.html

[↩](#)

- Georgiev, K., & Hopkins, S. B. (2022). Privacy induces robustness: Information-computation gaps and sparse mean estimation. In S. Koyejo, S. Mohamed, A. Agarwal, D. Belgrave, K. Cho, & A. Oh (Eds.), *Advances in neural information processing systems* (Vol. *35*, pp. 6829–6842). Curran Associates. https://proceedings.neurips.cc/paper_files/paper/2022/hash/2d76b6a9f96181ab717c1a15ab9302e1-Abstract-Conference.html

[↩](#)

- Ghazi, B., Golowich, N., Kumar, R., Manurangsi, P., & Zhang, C. (2021). Deep learning with label differential privacy. In M. Ranzato, A. Beygelzimer, Y. Dauphin, P. S. Liang, & J. Wortman Vaughan (Eds.), *Advances in neural information processing systems* (Vol. 34, pp. 27131–27145). Curran Associates. https://proceedings.neurips.cc/paper/2021/hash/e3a54649aeec04cf1c13907bc6c5c8aa-Abstract.html

    ↩

- Gilbert, A., & McMillan, A. (2018). Property testing for differential privacy. In 2018 *56th Annual Allerton Conference on Communication, Control, and Computing* (pp. 249–258). IEEE. https://doi.org/10.1109/ALLERTON.2018.8636068

    ↩

- Golatkar, A., Achille, A., Wang, Y.-X., Roth, A., Kearns, M., & Soatto, S. (2022). Mixed differential privacy in computer vision. In *2022 IEEE/CVF Conference on Computer Vision and Pattern Recognition* (pp. 8366-8376). IEEE. https://doi.org/10.1109/CVPR52688.2022.00819

    ↩

- Goodfellow, I. (2015). *Efficient per-example gradient computations*. ArXiv. https://doi.org/10.48550/arXiv.1510.01799

    ↩

- Google. (2023). *Google's differential privacy libraries*. GitHub. Retreived March 17, 2023, from https://github.com/google/differential-privacy

    ↩

- Gopi, S. G., Gulhane, P., Kulkarni, J., Shen, J., Shokouhi, M., & Yekhanin, S. (2020). Differentially private set union. In H. Daumé III, & A. Singh (Eds.), *Proceedings of the 37th International Conference on Machine Learning* (Vol. 119, pp. 3627–3636). Proceedings of Machine Learning Research. https://proceedings.mlr.press/v119/gopi20a.html

    ↩

- Gu, X., Kamath, G., & Wu, Z. S. (2023). *Choosing public datasets for private machine learning via gradient subspace distance*. ArXiv. https://doi.org/10.48550/arXiv.2303.01256

    ↩

- Gupta, S., Huang, Y., Zhong, Z., Gao, T., Li, K., & Chen, D. (2022). Recovering private text in federated learning of language models. In S. Koyejo, S. Mohamed, A. Agarwal, D. Belgrave, K. Cho, & A. Oh (Eds.), *Advances in neural information processing systems* (Vol. 35, pp. 8130–8143). Curran Associates. https://proceedings.neurips.cc/paper_files/paper/2022/hash/35b5c175e139bff5f22a5361270fce87-Abstract-Conference.html

↩

- Haeberlen, A., Pierce, B. C., & Narayan, A. (2011). Differential privacy under fire. In *Proceedings of the 20th USENIX Conference on Security* (pp. 33). USENIX Association. https://www.usenix.org/conference/usenix-security-11/differential-privacy-under-fire

    ↩

- Haim, N., Vardi, G., Yehudai, G., Shamir, O., & Irani, M. (2022). *Reconstructing training data from trained neural networks*. ArXiv. https://doi.org/10.48550/arXiv.2206.07758

    ↩

- Haney, S., Desfontaines, D., Hartman, L., Shrestha, R., & Hay, M. (2022). *Precision-based attacks and interval refining: How to break, then fix, differential privacy on finite computers*. ArXiv. https://doi.org/10.48550/arXiv.2207.13793

    ↩

- Harder, F., Bauer, M., & Park, M. (2020). Interpretable and differentially private predictions. *Proceedings of the AAAI Conference on Artificial Intelligence*, *34*(04), 4083–4090. https://doi.org/10.1609/aaai.v34i04.5827

    ↩

- Hardt, M., & Talwar, K. (2010). On the geometry of differential privacy. In *Proceedings of the 42nd Annual ACM Symposium on the Theory of Computing* (pp. 705–714). Association for Computer Machinery. https://doi.org/10.1145/1806689.1806786

    ↩

- Hartmann, F., & Kairouz, P. (2023, March 2). Distributed differential privacy for federated learning. *Google Research*. https://blog.research.google/2023/03/distributed-differential-privacy-for.html

    ↩

- Hay, M., Machanavajjhala, A., Miklau, G., Chen, Y., Zhang, D., & Bissias, G. (n.d.). Dpcomp.org. Retrieved March 17, 2023, from https://www.dpcomp.org/

    ↩

- He, J., Li, X., Yu, D., Zhang, H., Kulkarni, J., Lee, Y. T., Backurs, A., Yu, N., & Bian, J. (2023, May 1–5). *Exploring the limits of differentially private deep learning with group-wise clipping* [Poster session]. ICLR 2023: The Eleventh International Conference on Learning Representations, Kigali, Rwanda. https://openreview.net/pdf?id=oze0clVGPeX

    ↩

- Holohan, N., Braghin, S., Mac Aonghusa, P., & Levacher, K. (2019). Diffprivlib: The IBM differential privacy library. *ArXiv.* https://doi.org/10.48550/arXiv.1907.02444

↩

- Hoory, S., Feder, A., Tendler, A., Cohen, A. Erell, S., Laish, I., Nakhost, H., Stemmer, U., Benjamini, A., Hassidim, A., & Matias, Y. (2021). Learning and evaluating a differentially private pre-trained language model. In M.-F. Moens, X. Huang, L. Specia, & S. W.-t. Yih (Eds.), *Findings of the Association for Computational Linguistics: EMNLP 2021* (pp. 1178–1189). Association for Computational Linguistics. https://doi.org/10.18653/v1/2021.findings-emnlp.102

↩

- Hopkins, S. B., Kamath, G., & Majid, M. (2022). Efficient mean estimation with pure differential privacy via a sum-of-squares exponential mechanism. In *Proceedings of the 54th Annual ACM Symposium on the Theory of Computing* (pp. 1406–1417). Association for Computer Machinery. https://doi.org/10.1145/3519935.3519947

↩

- Hopkins, S. B., Kamath, G., Majid, M., & Narayanan, S. (2023). Robustness implies privacy in statistical estimation. In *Proceedings of the 55th Annual ACM Symposium on the Theory of Computing* (pp. 497–506). Association for Computer Machinery. https://doi.org/10.1145/3564246.3585115

↩

- Hu, L., Ni, S., Xiao, H., & Wang, D. (2022). High dimensional differentially private stochastic optimization with heavy-tailed data. In *Proceedings of the 41st ACM SIGMOD-SIGACT-SIGAI Symposium on Principles of Database Systems* (pp. 227–236). Association for Computer Machinery. https://doi.org/10.1145/3517804.3524144

↩

- Hu, L., Xiang, Z., Liu, J., & Wang, D. (2023). Privacy-preserving sparse Generalized Eigenvalue Problem. In F. Ruiz, J. Dy, & J.–W. van de Meent (Eds.), *Proceedings of The 26th International Conference on Artificial Intelligence and Statistics* (Vol. 206, pp. 5052–5062). Proceedings of Machine Learning Research. https://proceedings.mlr.press/v206/hu23a.html

↩

- Huang, W. R., Chien, S., Thakkar, O. D., & Mathews, R. (2022). Detecting unintended memorization in language-model-fused ASR. In H. Ko & J. H. L. Hansen (Eds.), *Interspeech 2022* (pp. 2808–2812). International Speech Communication Association. https://doi.org/10.21437/Interspeech.2022-10909

↩

- Huang, Y., Gupta, S., Song, Z., Li, K., & Arora, S. (2021). Evaluating gradient inversion attacks and defenses in federated learning. In M. Ranzato, A. Beygelzimer, Y. Dauphin, P.S. Liang, & J. Wortman Vaughan (Eds.), *Advances in neural information processing systems* (Vol. 34, pp. 7232–7241). Curran

↩

- Huang, Z., Liang, Y., & Yi, K. (2021). Instance-optimal mean estimation under differential privacy. In M. Ranzato, A. Beygelzimer, Y. Dauphin, P.S. Liang, & J. Wortman Vaughan (Eds.), *Advances in neural information processing systems* (Vol. 34, (pp. 25993–26004). Curran Associates. https://proceedings.neurips.cc/paper/2021/hash/da54dd5a0398011cdfa50d559c2c0ef8-Abstract.html

↩

- Ilvento, C. (2020). Implementing the exponential mechanism with base-2 differential privacy. In *CCS '20: Proceedings of the 2020 ACM SIGSAC Conference on Computer and Communications Security* (pp. 717–742). Association for Computing Machinery. https://doi.org/10.1145/3372297.3417269

↩

- Imola, J., Murakami, T., & Chaudhuri, K. (2021). Locally differentially private analysis of graph statistics. In *Proceedings of the 30th USENIX Security Symposium* (pp. 983–1000). USENIX Association. https://www.usenix.org/conference/usenixsecurity21/presentation/imola

↩

- Iyengar, R., Near, J. P., Song, D., Thakkar, O., Thakurta, A., & Wang, L. (2019). Towards practical differentially private convex optimization. In *2019 IEEE Symposium on Security and Privacy* (pp. 299–316). IEEE. https://doi.org/10.1109/SP.2019.00001

↩

- Jagielski, M., Thakkar, O., Tramèr, F., Ippolito, D., Lee, K., Carlini, N., Wallace, E., Song, S., Thakurta, A., Papernot, N., & Zhang, C. (2022). Measuring forgetting of memorized training examples.AarXiv. https://doi.org/10.48550/arXiv.2207.00099

↩

- Jagielski, M., Wu, S., Oprea, A., Ullman, J., & Geambasu, R. (2022). How to combine membership-inference attacks on multiple updated models. ArXiv. https://doi.org/10.48550/arXiv.2205.06369

↩

- Jain, P., Rush, J., Smith, A., Song, S., & Guha Thakurta, A. (2021). Differentially private model personalization. In M. Ranzato, A. Beygelzimer, Y. Dauphin, P.S. Liang, & J. Wortman Vaughan (Eds.), *Advances in neural information processing systems* (Vol. 34, pp. 29723–29735). Curran Associates. https://proceedings.neurips.cc/paper/2021/hash/f8580959e35cb0934479bb007fb241c2-Abstract.html

↩

- Jayaraman, B., & Evans, D. (2022). Are attribute inference attacks just imputation? In *Proceedings of the 2022 ACM SIGSAC Conference on Computer and Communications Security* (pp. 1569–1582). Association for Computing Machinery. https://doi.org/10.1145/3548606.3560663

↩

- Jayaraman, B., Wang, L., Evans, D., & Gu, Q. (2018). Distributed learning without distress: Privacy-preserving empirical risk minimization. In S. Bengio, H. Wallach, H. Larochelle, K. Grauman, N. Cesa-Bianchi, & R. Garnett (Eds.), *Advances in neural information processing systems* (Vol. 31, pp. 6343–6354). Curran Associates. https://papers.nips.cc/paper_files/paper/2018/hash/7221e5c8ec6b08ef6d3f9ff3ce6eb1d1-Abstract.html

↩

- Jayaraman, B., Wang, L., Knipmeyer, K., Gu, Q., & Evans, D. (2020). *Revisiting membership inference under realistic assumptions*. ArXiv. https://doi.org/10.48550/arXiv.2005.10881

↩

- Ji, Z., & Elkan, C. (2013). Differential privacy based on importance weighting. *Machine Learning*, *93*(1), 163–183. https://doi.org/10.1007/s10994-013-5396-x

↩

- Jin, J., McMurtry, E., Rubinstein, B. I., & Ohrimenko, O. (2022). Are we there yet? Timing and floating-point attacks on differential privacy systems. In *2022 IEEE Symposium on Security and Privacy* (pp. 473–488). IEEE. https://doi.org/10.1109/SP46214.2022.9833672

↩

- Jordon, J., Yoon, J., & Van Der Schaar, M. (2019, May 6–9). *PATE-GAN: Generating synthetic data with differential privacy guarantees* [Poster presentation]. ICLR 2019: 7th International Conference on Learning Representations, New Orleans, LA, United States. https://openreview.net/forum?id=S1zk9iRqF7

↩

- Jorgensen, Z., Yu, T., & Cormode, G. (2015). Conservative or liberal? personalized differential privacy. In *2015 IEEE 31St International Conference on Data Engineering* (pp. 1023–1034). IEEE. https://doi.org/10.1109/ICDE.2015.7113353

↩

- Joseph, M., Mao, J., Neel, S., & Roth, A. (2019). The role of interactivity in local differential privacy. In *Proceedings of the 60th Annual IEEE Symposium on Foundations of Computer Science* (pp. 94–105). IEEE. https://doi.org/10.1109/FOCS.2019.00015

↩

- Kairouz, P., McMahan, B., Song, S., Thakkar, O., Thakurta, A., & Xu, Z. (2021). Practical and private (deep) learning without sampling or shuffling. In M. Meila, & T. Zhang (Eds.), *Proceedings of the 38th International Conference on Machine Learning* (Vol. 139, pp. 5213–5225). https://proceedings.mlr.press/v139/kairouz21b.html

↵

- Kairouz, P., McMahan, H. B., Avent, B., Bellet, A., Bennis, M., Bhagoji, A. N., Bonawitz, K., Charles, Z., Cormode, G., Cummings, R., D'Oliveira, R. G. L., Eichner, H., El Rouayheb, S., Evans, D., Gardner, J., Garrett, Z., Gascón, A., Ghazi, B., Gibbons, P. B., … Zhao, S. (2021). Advances and open problems in federated learning. *Foundations and Trends® in Machine Learning*, *14*(1–2), 1–210. http://doi.org/10.1561/2200000083

↵

- Kairouz, P., Ribero, M., Rush, K., & Thakurta, A. (2020). *Fast dimension independent private AdaGrad on publicly estimated subspaces*. ArXiv. https://doi.org/10.48550/arXiv.2008.06570

↵

- Kairouz, P., Ribero, M., Rush, K., & Thakurta, A. (2021). (Nearly) dimension independent private ERM with AdaGrad rates via publicly estimated subspaces. In M. Belkin, & S. Kpotufe (Eds.), *Proceedings of the 34th Annual Conference on Learning Theory* (Vol. 134, pp. 2717–2746). Proceedings of Machine Learning Research. https://proceedings.mlr.press/v134/kairouz21a.html

↵

- Kakizaki, K., Sakuma, J., & Fukuchi, K. (2017). Differentially private chi-squared test by unit circle mechanism. In D. Precup, & Y. W. Teh (Eds.), *Proceedings of the 34th International Conference on Machine Learning* (Vol. 40, pp. 1761–1770). Proceedings of Machine Learning Research. https://proceedings.mlr.press/v70/kakizaki17a.html

↵

- Kamath, G., & Ullman, J. (2020). *A primer on private statistics*. ArXiv. https://doi.org/10.48550/arXiv.2005.00010

↵

- Kamath, G., Li, J., Singhal, V., & Ullman, J. (2019). Privately learning high-dimensional distributions. In A. Beygelzimer, & D. Hsu (Eds.), *Proceedings of the 32nd Annual Conference on Learning Theory* (Vol. 99, pp. 1853–1902). Proceedings of Machine Learning Research. https://proceedings.mlr.press/v99/kamath19a.html

↵

- Kamath, G., Liu, X., & Zhang, H. (2022). Improved rates for differentially private stochastic convex optimization with heavy-tailed data. In K. Chaudhuri, S. Jegelka, L. Song, C. Szepesvari, G. Niu, & S. Sabato (Eds.), *Proceedings of the 39th International Conference on Machine Learning* (Vol. 162, pp. 10633–10660). Proceedings of Machine Learning Research. https://proceedings.mlr.press/v162/kamath22a.html

↵

- Kamath, G., Mouzakis, A., & Singhal, V. (2022). New lower bounds for private estimation and a generalized fingerprinting lemma. In S. Koyejo, S. Mohamed, A. Agarwal, D. Belgrave, K. Cho, & A. Oh (Eds.), *Advances in neural information processing systems,* (Vol. 35, pp. 24405–24418). Curran Associates. https://proceedings.neurips.cc/paper_files/paper/2022/hash/9a6b278218966499194491f55ccf8b75-Abstract-Conference.html

    ↩

- Kamath, G., Mouzakis, A., Singhal, V., Steinke, T., & Ullman, J. (2021). *A private and computationally-efficient estimator for unbounded Gaussians*. ArXiv. https://doi.org/10.48550/arXiv.2111.04609

    ↩

- Kamath, G., Singhal, V., & Ullman, J. (2020). Private mean estimation of heavy-tailed distributions. In J. Abernethy, & S. Agarwal (Eds.), *Proceedings of the 33rd Annual Conference on Learning Theory* (Vol. 125, 2204–2235). Proceedings of Machine Learning Research. https://proceedings.mlr.press/v125/kamath20a.html

    ↩

- Karwa, V., & Vadhan, S. (2018). Finite sample differentially private confidence intervals. In A. R. Karlin (Ed.), *Leibniz International Proceedings in Informatics* (Vol. 49). *9th Conference on Innovations in Theoretical Computer Science*, (pp. 44:1–44:9). Dagstuhl Publishing. https://drops.dagstuhl.de/opus/volltexte/2018/8344/pdf/LIPIcs-ITCS-2018-44.pdf

    ↩

- Kearns, M., Pai, M., Roth, A., & Ullman, J. (2014). Mechanism design in large games: Incentives and privacy. *American Economic Review, 104*(5), 431–435.

    ↩

- Kelley, P. G., Bresee, J., Cranor, L. F., & Reeder, R. W. (2009). A "nutrition label" for privacy. In L. F. Cranor (Ed.), *Proceedings of the 5th Symposium on Usable Privacy and Security*. Association for Computing Machinery. https://doi.org/10.1145/1572532.1572538

    ↩

- Kenny, C. T., Kuriwaki, S., McCartan, C., Rosenman, E. T., Simko, T., & Imai, K. (2021). The use of differential privacy for census data and its impact on redistricting: The case of the 2020 US Census. *Science Advances, 7*(41), Article eabk3283. https://doi.org/10.1126/sciadv.abk3283

    ↩

- Kifer, D., & Machanavajjhala, A. (2014). Pufferfish: A framework for mathematical privacy definitions. *ACM Transactions Database Systems, 39*(1), Article 3. https://doi.org/10.1145/2514689

↩

- Kifer, D., & Rogers, R. M. (2017). A new class of private chi-square tests. In A. Singh, & J. Zhu (Eds.), *Proceedings of the 20th International Conference on Artificial Intelligence and Statistics* (Vol. 54, pp. 991–1000). Proceedings of Machine Learning Research. https://proceedings.mlr.press/v54/rogers17a.html

↩

- Kifer, D., Smith, A., & Thakurta, A. (2012). Private convex empirical risk minimization and high-dimensional regression. In S. Mannor, N. Srebro, & R. C. Williamson (Eds.), *Proceedings of the 25th Annual Conference on Learning Theory* (Vol. 23, pp. 25.1–25.40). Proceedings of Machine Learning Research. https://proceedings.mlr.press/v23/kifer12.html

↩

- Kim, K., Gopi, S., Kulkarni, J., & Yekhanin, S. (2021). Differentially private n-gram extraction. In M. Ranzato, A. Beygelzimer, Y. Dauphin, P.S. Liang, & J. Wortman Vaughan (Eds.), *Advances in neural information processing systems* (Vol. 34, pp. 5102–5111). Curran Associates. https://proceedings.neurips.cc/paper/2021/hash/28ce9bc954876829eeb56ff46da8e1ab-Abstract.html

↩

- Kothari, P. K., Manurangsi, P., & Velingker, A. (2021). *Private robust estimation by stabilizing convex relaxations*. ArXiv. https://doi.org/10.48550/arXiv.2112.03548

↩

- Kurakin, A., Chien, S., Song, S., Geambasu, R., Terzis, A., & Thakurta, A. (2022). Toward training at ImageNet scale with differential privacy. ArXiv. https://doi.org/10.48550/arXiv.2201.12328

↩

- Lai, K. A., Rao, A. B., & Vempala, S. (2016). Agnostic estimation of mean and covariance. In *Proceedings of the 57th Annual IEEE Symposium on Foundations of Computer Science* (pp. 665–674). IEEE. https://doi.org/10.1109/FOCS.2016.76

↩

- Lee, D., Yu, H., Jiang, X., Rogith, D., Gudala, M., Tejani, M., Zhang, Q., & Xiong, L. (2020). Generating sequential electronic health records using dual adversarial autoencoder. *Journal of the American Medical Informatics Association*, *27*(9), 1411–1419. https://doi.org/10.1093/jamia/ocaa119

↩

- Lee, J., & Kifer, D. (2018). Concentrated differentially private gradient descent with adaptive per-iteration privacy budget. *Proceedings of the 24th ACM SIGKDD International Conference on Knowledge Discovery & Data Mining*, 1656–1665. ↩
-

↩

- Lee, J., & Kifer, D. (2021). Scaling up differentially private deep learning with fast per-example gradient clipping. *Proceedings on Privacy Enhancing Technologies*, *2021*(1),128–144 . https://doi.org/10.2478/popets-2021-0008

  ↩

- Levy, D., Sun, Z., Amin, K., Kale, S., Kulesza, A., Mohri, M., & Suresh, A. T. (2021). Learning with user-level privacy. In M. Ranzato, A. Beygelzimer, Y. Dauphin, P.S. Liang, & J. Wortman Vaughan (Eds.), *Advances in neural information processing systems* (Vol. 34, pp. 12466–12479). Curran Associates. https://proceedings.neurips.cc/paper_files/paper/2021/hash/67e235e7f2fa8800d8375409b566e6b6-Abstract.html

  ↩

- Li, T., Zaheer, M., Reddi, S. J., & Smith, V. (2022). Private adaptive optimization with side information. *Proceedings of the 39th International Conference on Machine Learning*. ↩

- Li, T., Zaheer, M., Reddi, S. J., & Smith, V. (2022). Private adaptive optimization with side information. In K. Chaudhuri, S. Jegelka, L. Song, C. Szepesvari, G. Niu, & S. Sabato (Eds.), *Proceedings of the 39th International Conference on Machine Learning* (Vol. 162, pp. 13086–13105). Proceedings of Machine Learning Research. https://proceedings.mlr.press/v162/li22x.html

  ↩

- Li, X., Liu, D., Hashimoto, T., Inan, H. A., Kulkarni, J., Lee, Y. T., & Thakurta, A. G. (2022). *When does differentially private learning not suffer in high dimensions?* ArXiv. https://doi.org/10.48550/arXiv.2207.00160

  ↩

- Li, X., Tramèr, F., Kulkarni, J., & Hashimoto, T. (2022, March 15). *Differentially private deep learning can be effective with self-supervised models*. DifferentialPrivacy.org. https://differentialprivacy.org/dp-fine-tuning/

  ↩

- Li, X., Tramer, F., Liang, P., & Hashimoto, T. (2021). *Large language models can be strong differentially private learners*. ArXiv. https://doi.org/10.48550/arXiv.2110.05679

  ↩

- Li, X., Tramèr, F., Liang, P., & Hashimoto, T. (2022, April 25–29). *Large language models can be strong differentially private learners* [Conference presentation]. ICLR 2022: The Tenth International Conference on Learning Representations, online. https://openreview.net/forum?id=bVuP3ltATMz

  ↩

- Li, Z., & Zhang, Y. (2021). Membership leakage in label-only exposures. *Proceedings of the 2021 ACM SIGSAC Conference on Computer and Communications Security*, 880–895. ↩
- Li, Z., & Zhang, Y. (2021). Membership leakage in label-only exposures. In *CCS '21: Proceedings of the 2021 ACM SIGSAC Conference on Computer and Communications Security* (pp. 880–895). Association for Computing Machinery. https://doi.org/10.1145/3460120.3484575

    ↩
- Ligett, K., Neel, S., Roth, A., Waggoner, B., & Wu, S. Z. (2017). Accuracy first: Selecting a differential privacy level for accuracy constrained ERM. In I. Guyon, U. Von Luxburg, S. Bengio, H. Wallach, R. Fergus, S. Vishwanathan, & R. Garnett (Eds.), *Advances in neural information processing systems* (Vol. 30, pp. 1566–2575). Curran Associates. https://papers.nips.cc/paper_files/paper/2017/hash/86df7dcfd896fcaf2674f757a2463eba-Abstract.html

    ↩
- Liu, C., Zhu, Y., Chaudhuri, K., & Wang, Y.-X. (2021). Revisiting model-agnostic private learning: Faster rates and active learning. *The Journal of Machine Learning Research*, *22*(1), 11936–11979. https://jmlr.org/papers/volume22/20-1251/20-1251.pdf

    ↩
- Liu, D., & Lu, Z. (2021). Lower bounds for differentially private ERM: Unconstrained and non-Euclidean. ArXiv. https://doi.org/10.48550/arXiv.2105.13637

    ↩
- Liu, J., & Talwar, K. (2019). Private selection from private candidates. In *Proceedings of the 51st Annual ACM Symposium on the Theory of Computing* (pp. 298–309). Association for Computer Machinery. https://doi.org/10.1145/3313276.3316377

    ↩
- Liu, J., Lou, J., Xiong, L., Liu, J., & Meng, X. (2021). Projected federated averaging with heterogeneous differential privacy. *Proceedings of the VLDB Endowment*, *15*(4), 828–840. https://doi.org/10.14778/3503585.3503592

    ↩
- Liu, T., Vietri, G., Steinke, T., Ullman, J., & Wu, S. (2021). Leveraging public data for practical private query release. In M. Meila & T. Zhang (Eds.), *Proceedings of the 38th International Conference on Machine Learning* (Vol. 139, pp. 6968–6977). Proceedings of Machine Learning Research. https://proceedings.mlr.press/v139/liu21v.html

    ↩
- 

↵

- Liu, X., Jain, P., Kong, W., Oh, S., & Suggala, A. S. (2023). *Near optimal private and robust linear regression*. ArXiv. https://doi.org/10.48550/arXiv.2301.13273

  ↵

- Liu, X., Kong, W., & Oh, S. (2022). Differential privacy and robust statistics in high dimensions. *Conference on Learning Theory*, 1167–1246. ↵
- Liu, X., Kong, W., & Oh, S. (2022). Differential privacy and robust statistics in high dimensions. In P.-L. Loh, & M. Raginsky (Eds.,) *Proceedings of Thirty Fifth Conference on Learning Theory* (Vol. 178, pp. 1167–1246). Proceedings of Machine Learning Research. https://proceedings.mlr.press/v178/liu22b.html

  ↵

- Liu, X., Kong, W., Jain, P., & Oh, S. (2022). DP-PCA: Statistically optimal and differentially private PCA. In S. Koyejo, S. Mohamed, A. Agarwal, D. Belgrave, K. Cho, & A. Oh (Eds.), *Advances in neural information processing systems* (Vol. 35, pp. 29929–29943). Curran Associates https://proceedings.neurips.cc/paper_files/paper/2022/hash/c150ebe1b9d1ca0eb61502bf979fa87d-Abstract-Conference.html

  ↵

- Liu, X., Kong, W., Kakade, S., & Oh, S. (2021). Robust and differentially private mean estimation. In M. Ranzato, A. Beygelzimer, Y. Dauphin, P.S. Liang, & J. Wortman Vaughan (Eds.), *Advances in neural information processing systems* (Vol. 34, pp. 3887–3901). Curran Associates. https://proceedings.neurips.cc/paper_files/paper/2021/hash/1fc5309ccc651bf6b5d22470f67561ea-Abstract.html

  ↵

- Liu, Y., Ott, M., Goyal, N., Du, J., Joshi, M., Chen, D., Levy, O., Lewis, M., Zettlemoyer, L., & Stoyanov, V. (2019). RoBERTa: A robustly optimized BERT pretraining approach. ArXiv. https://doi.org/10.48550/arXiv.1907.11692

  ↵

- Liu, Y., Ott, M., Goyal, N., Du, J., Joshi, M., Chen, D., Levy, O., Lewis, M., Zettlemoyer, L., & Stoyanov, V. (2019). RoBERTa: A robustly optimized BERT pretraining approach. ArXiv. https://doi.org/10.48550/arXiv.1907.11692

  ↵

- Liu, Y., Suresh, A. T., Yu, F., Kumar, S., & Riley, M. (2020). Learning discrete distributions: User vs item-level privacy. In H. Larochelle, M. Ranzato, R. Hadsell, M.F. Balcan, & H. Lin (Eds.), *Advances in neural information processing systems* (Vol. 33, pp. 20965–20976). Curran Associates.

↩

- Liu, Y., Zhao, S., Xiong, L., Liu, Y., & Chen, H. (2023). Echo of neighbors: Privacy amplification for personalized private federated learning with shuffle model. *Proceedings of the AAAI Conference on Artificial Intelligence, 37*(10), 11865–11872). https://doi.org/10.1609/aaai.v37i10.26400

↩

- Lovejoy, B. (2017, July 7). *As apple starts analyzing web browsing and health data, how comfortable are you with differential privacy?* 9to5Mac. https://9to5mac.com/2017/07/07/what-is-differential-privacy/.

↩

- Ma, Y., Zhu, X., & Hsu, J. (2019). Data poisoning against differentially-private learners: Attacks and defenses. In Sarit Kraus (Ed.), *Proceedings of the Twenty-Eighth International Joint Conference on Artificial Intelligence* (pp. 4732-4738). International Joint Conferences on Artificial Intelligence Organization. https://doi.org/10.24963/ijcai.2019%2F657

↩

- Machanavajjhala, A., Kifer, D., Gehrke, J., & Venkitasubramaniam, M. (2007). *L*-diversity: Privacy beyond *k*-anonymity. *ACM Transactions on Knowledge Discovery from Data, 1*(1), 3–es. https://doi.org/10.1145/1217299.1217302

↩

- Mahloujifar, S., Sablayrolles, A., Cormode, G., & Jha, S. (2022). *Optimal membership inference bounds for adaptive composition of sampled Gaussian mechanisms*. arXiv. https://doi.org/10.48550/arXiv.2204.06106

↩

- Mangold, P., Bellet, A., Salmon, J., & Tommasi, M. (2022). Differentially private coordinate descent for composite empirical risk minimization. In K. Chaudhuri, S. Jegelka, L. Song, C. Szepesvari, G. Niu, & S. Sabato (Eds.), *Proceedings of the 39th International Conference on Machine Learning* (Vol. 162, pp.14948–14978). Proceedings of Machine Learning Research. https://proceedings.mlr.press/v162/mangold22a.html

↩

- Mangold, P., Perrot, M., Bellet, A., & Tommasi, M. (2022). Differential privacy has bounded impact on fairness in classification. In A. Krause, E. Brunskill, K. Cho, B. Engelhardt, S. Sabato, & J. Scarlett (Eds.), *Proceedings of the 40th International Conference on Machine Learning* (Vol. 202, pp. 23681–26705). Proceedings of Machine Learning Research. https://proceedings.mlr.press/v202/mangold23a.html

↩

- McMahan, B., Moore, E., Ramage, D., Hampson, S., & Agüera y Arcas, B. A. (2017). Communication-efficient learning of deep networks from decentralized data. In A. Singh & J. Zhu (Eds.), *Proceedings of the*

↩

- McMahan, H. B., Ramage, D., Talwar, K., & Zhang, L. (2018, April 30–May 3). *Learning differentially private recurrent language models* [Poster session]. ICLR 2018: 6th International Conference on Learning Representations, Vancouver, BC, Canada. https://openreview.net/forum?id=BJ0hF1Z0b

  ↩

- McSherry, F. (2016, June 14). Statistical inference considered harmful. *GitHub*. https://github.com/frankmcsherry/blog/blob/master/posts/2016-06-14.md

  ↩

- Mehta, H., Thakurta, A., Kurakin, A., & Cutkosky, A. (2022). *Large scale transfer learning for differentially private image classification*. ArXiv. https://doi.org/10.48550/arXiv.2205.02973

  ↩

- Milionis, J., Kalavasis, A., Fotakis, D., & Ioannidis, S. (2022). Differentially private regression with unbounded covariates. In G. Camps-Valls, F. J. R. Ruiz, & I. Valera (Eds.), *Proceedings of The 25th International Conference on Artificial Intelligence and Statistics* (Vol. 151, pp. 3242–3273). Proceedings of Machine Learning Research. https://proceedings.mlr.press/v151/milionis22a.html

  ↩

- Mironov, I. (2012). On significance of the least significant bits for differential privacy. In *CCS '12: Proceedings of the 2012 ACM Conference on Computer and Communications Security* (pp. 650–661). Association for Computing Machinery. https://doi.org/10.1145/2382196.2382264

  ↩

- Mironov, I. (2017). Rényi differential privacy. In *2017 IEEE 30th Computer Security Foundations Symposium* (pp. 263–275). IEEE. https://doi.org/10.1109/CSF.2017.11

  ↩

- Mohapatra, S., Sasy, S., He, X., Kamath, G., & Thakkar, O. (2022). The role of adaptive optimizers for honest private hyperparameter selection. *Proceedings of the AAAI Conference on Artificial Intelligence, 36*(7), 7806-7813. https://doi.org/10.1609/aaai.v36i7.20749

  ↩

- Nanayakkara, P., Smart, M. A., Cummings, R., Kaptchuk, G., & Redmiles, E. M. (2023). *What are the chances? Explaining the epsilon parameter in differential privacy*. ArXiv. https://doi.org/10.48550/arXiv.2303.00738

  ↩

- 

↩

- Nasr, M., Mahloujifar, S., Tang, X., Mittal, P., & Houmansadr, A. (2022). Effectively using public data in privacy preserving machine learning. In A. Krause, E. Brunskill, K. Cho, B. Engelhardt, S. Sabato, & J. Scarlett (Eds.), Proceedings of the 40th International Conference on Machine Learning (Vol. 202, pp. 25718–25732). Proceedings of Machine Learning Research. https://proceedings.mlr.press/v202/nasr23a.html

↩

- Nasr, M., Song, S., Thakurta, A., Papernot, N., & Carlini, N. (2021). *Adversary instantiation: Lower bounds for differentially private machine learning*. ArXiv. https://doi.org/10.48550/arXiv.2101.04535

↩

- Neel, S., Roth, A., & Sharifi-Malvajerdi, S. (2021). Descent-to-delete: Gradient-based methods for machine unlearning. In V. Feldman, K. Ligett, & S. Sabato (Eds.), *Proceedings of the 32nd International Conference on Algorithmic Learning* Theory (Vol. 132, pp. 931–962). Proceedings of Machine Learning Research. https://proceedings.mlr.press/v132/neel21a.html

↩

- Nikolov, A., & Tang, H. (2023). General Gaussian noise mechanisms and their optimality for unbiased mean estimation. ArXiv. https://doi.org/10.48550/arXiv.2301.13850

↩

- Nissenbaum, H. (2004). Privacy as contextual integrity. *Washington Law Review*, *79*(1), 119–158. https://digitalcommons.law.uw.edu/wlr/vol79/iss1/10/

↩

- Nissim, K., Bembenek, A., Wood, A., Bun, M., Gaboardi, M., Gasser, U., O'Brien, D., Steinke, T., & Vadhan, S. (2018). Bridging the gap between computer science and legal approaches to privacy. *Harvard Journal of Law & Technology*, *31*(2), 687–780.

↩

- Nissim, K., Raskhodnikova, S., & Smith, A. (2007). Smooth sensitivity and sampling in private data analysis. In *Proceedings of the 39th Annual ACM Symposium on the Theory of Computing* (pp. 75–84). Association for Computer Machinery. https://doi.org/10.1145/1250790.1250803

↩

- NIST. (2022, December 30) 2020 differential privacy temporal map challenge. https://www.nist.gov/ctl/pscr/open-innovation-prize-challenges/past-prize-challenges/2020-differential-privacy-temporal

↩

- Noe, F., Herskind, R., & Søgaard, A. (2022). *Exploring the unfairness of DP-SGD across settings*. ArXiv. https://doi.org/10.48550/arXiv.2202.12058

    ↩

- Nuñez von Voigt, S., Pauli, M., Reichert, J., & Tschorsch, F. (2020). Every query counts: Analyzing the privacy loss of exploratory data analyses. In J. Garcia-Alfaro, G. Navarro-Arribas, & J. Herrera-Joancomarti (Eds), Lecture notes in computer science: Vol. 12484. *Data privacy management, cryptocurrencies and blockchain technology* (pp. 258–266). Springer. https://doi.org/10.1007/978-3-030-66172-4_17

    ↩

- OpenDP. (2022, November). Retrieved from https://opendp.org/

    ↩

- Oyallon, E., & Mallat, S. (2015). Deep roto-translation scattering for object classification. In *2015 IEEE Computer Society Conference on Computer Vision and Pattern Recognition* (pp. 2865–2873). IEEE. https://doi.org/10.1109/CVPR.2015.7298904

    ↩

- Papernot, N., & Steinke, T. (2022, April 25–29). *Hyperparameter tuning with Renyi differential privacy* [Conference presentation]. ICLR 2022: The Tenth International Conference on Learning Representations, Virtual Event. https://openreview.net/forum?id=-70L8lpp9DF

    ↩

- Papernot, N., Abadi, M., Erlingsson, Ú., Goodfellow, I., & Talwar, K. (2017, April 24–26). *Semi-supervised knowledge transfer for deep learning from private training data* [Paper presentation]. ICLR 2017: 5th International Conference on Learning Representations. https://openreview.net/forum?id=HkwoSDPgg

    ↩

- Papernot, N., Chien, S., Song, S., Thakurta, A., & Erlingsson, Ú. (2019, September 25). *Making the shoe fit: Architectures, initializations, and tuning for learning with privacy* [Conference blind submission]. ICLR 2020: The Eighth International Conference on Learning Representations, Virtual Event. https://openreview.net/forum?id=rJg851rYwH

    ↩

- Papernot, N., Song, S., Mironov, I., Raghunathan, A., Talwar, K., & Erlingsson, Ú. (2018, April 30–May 3). *Scalable private learning with PATE* [Poster session]. ICLR 2018: 6th International Conference on Learning Representations, Vancouver, BC, Canada. https://openreview.net/forum?id=rkZB1XbRZ

    ↩

-

↩

- Phan, H., Thai, M. T., Hu, H., Jin, R., Sun, T., & Dou, D. (2020). Scalable differential privacy with certified robustness in adversarial learning. In H. Daumé III, & A. Singh (Eds.), *Proceedings of the 37th International Conference on Machine Learning* (Vol. 119, pp. 7683–7694). Proceedings of Machine Learning Research. https://proceedings.mlr.press/v119/phan20a.html

↩

- Pujol, D., McKenna, R., Kuppam, S., Hay, M., Machanavajjhala, A., & Miklau, G. (2020). Fair decision making using privacy-protected data. In *Proceedings of the 2020 Conference on Fairness, Accountability, and Transparency* (pp. 189–199). Association for Computing Machinery. https://doi.org/10.1145/3351095.3372872

↩

- Pyrgelis, A., Troncoso, C., & De Cristofaro, E. (2017). Knock knock, who's there? Membership inference on aggregate location data. ArXiv. https://doi.org/10.48550/arXiv.1708.06145

↩

- Radford, A., Wu, J., Child, R., Luan, D., Amodei, D., & Sutskever, I. (2019). *Language models are unsupervised multitask learners*. OpenAI. https://cdn.openai.com/better-language-models/language_models_are_unsupervised_multitask_learners.pdf

↩

- Ramaswamy, S., Thakkar, O., Mathews, R., Andrew, G., McMahan, H. B., & Beaufays, F. (2020). *Training production language models without memorizing user data*. ArXiv. https://doi.org/10.48550/arXiv.2009.10031

↩

- Ramsay, K., & Chenouri, S. (2021). *Differentially private depth functions and their associated medians*. ArXiv. https://doi.org/10.48550/arXiv.2101.02800

↩

- Ramsay, K., Jagannath, A., & Chenouri, S. (2022). *Concentration of the exponential mechanism and differentially private multivariate medians*. ArXiv. https://doi.org/10.48550/arXiv.2210.06459

↩

- Redberg, R., Zhu, Y., & Wang, Y.-X. (2022). *Generalized PTR: User-friendly recipes for data-adaptive algorithms with differential privacy*. ArXiv. https://doi.org/10.48550/arXiv.2301.00301

↩

- 

↩

- Rogers, R., Subramaniam, S., Peng, S., Durfee, D., Lee, S., Kancha, S. K., Sahay, S., & Ahammad, P. (2021). LinkedIn's audience engagements API: A privacy preserving data analytics system at scale. *Journal of Privacy and Confidentiality*, *11*(3). https://doi.org/10.29012/jpc.782

↩

- Sablayrolles, A., Douze, M., Schmid, C., Ollivier, Y., & Jégou, H. (2019). White-box vs black-box: Bayes optimal strategies for membership inference. In K. Chaudhuri, & R. Salakhutdinov (Eds.), *Proceedings of the 36th International Conference on Machine Learning* (Vol. 97, pp. 5558–5567). Proceedings of Machine Learning Research. https://proceedings.mlr.press/v97/sablayrolles19a.html

↩

- Salem, A., Bhattacharyya, A., Backes, M., Fritz, M., & Zhang, Y. (2020). Updates-leak: Data set inference and reconstruction attacks in online learning. In *Proceedings of the 29th USENIX Conference on Security Symposium* (pp. 1291–1308). USENIX Association. https://www.usenix.org/conference/usenixsecurity20/presentation/salem

↩

- Salem, A., Cherubin, G., Evans, D., Kopf, B., Paverd, A., Suri, A., Tople, S., & Zanella-Beguelin, S. (2022). *SoK: Let the privacy games begin! A unified treatment of data inference privacy in machine learning.* ArXiv. https://doi.org/10.48550/arXiv.2212.10986

↩

- Samarati, P., & Sweeney, L. (1998). Generalizing data to provide anonymity when disclosing information. In *Proceedings of the 17th ACM SIGMOD-SIGACT-SIGART Symposium on Principles of Database Systems* (p. 188). Association for Computer Machinery. https://doi.org/10.1145/275487.275508

↩

- Sanyal, A., Hu, Y., & Yang, F. (2022). *How unfair is private learning?* ArXiv. https://doi.org/10.48550/arXiv.2206.03985

↩

- Sarathy, J., Song, S., Haque, A., Schlatter, T., & Vadhan, S. (2023). Don't look at the data! how differential privacy reconfigures the practices of data science. In A. Schmidt, K. Väänänen, T. Goyal, P. Ola Kristensson, A. Peters, S. Mueller, J. R. Williamson, & M. L. Wilson (Eds.), *Proceedings of the 2022 CHI Conference on Human Factors in Computing Systems* (pp. 1–19). Association for Computing Machinery. https://doi.org/10.1145/3544548.3580791

↩

- 

↵

- Sheffet, O. (2018). Locally private hypothesis testing. In J. Dy, & A. Krause (Eds.), *Proceedings of the 35th International Conference on Machine Learning* (Vol. 80, pp. 4605–4614). Proceedings of Machine Learning Research. https://proceedings.mlr.press/v80/sheffet18a.html

↵

- Shokri, R., Stronati, M., Song, C., & Shmatikov, V. (2017). Membership inference attacks against machine learning models. In 2017 *IEEE Symposium on Security and Privacy* (pp. 3–18). IEEE. https://doi.org/10.1109/SP.2017.41

↵

- Singhal, V., & Steinke, T. (2021). Privately learning subspaces. In M. Ranzato, A. Beygelzimer, Y. Dauphin, P.S. Liang, & J. Wortman Vaughan (Eds.), *Advances in neural information processing systems (Vol.* 34, pp. 1312–1324). Curran Associates. https://proceedings.neurips.cc/paper_files/paper/2021/hash/09b69adcd7cbae914c6204984097d2da-Abstract.html

↵

- Smart, M. A., Nanayakkara, P., Cummings, R., Kaptchuk, G., & Redmiles, E. M. (2020, October 15–16). *Improving explanations to end users about differential privacy* [Conference presentation]. 2020 USENIX Conference on Privacy Engineering Practice and Respect, Virtual Event. https://www.usenix.org/conference/pepr20/presentation/perera

↵

- Smith, A. (2011). Privacy-preserving statistical estimation with optimal convergence rates. In *Proceedings of the 43rd Annual ACM Symposium on the Theory of Computing* (pp. 813–822). Association for Computer Machinery. https://doi.org/10.1145/1993636.1993743

↵

- Song, S., Chaudhuri, K., & Sarwate, A. D. (2013). Stochastic gradient descent with differentially private updates. *In 2013 IEEE Global Conference on Signal and Information Processing* (pp. 245–248). IEEE. https://doi.org/10.1109/GlobalSIP.2013.6736861

↵

- Song, S., Steinke, T., Thakkar, O., & Thakurta, A. (2021). Evading the curse of dimensionality in unconstrained private GLMs. In A. Banerjee, & K. Fukumizu (Eds.), *Proceedings of the 24th International Conference on Artificial Intelligence and Statistics* (Vol. 130, pp. 2638–2646). Proceedings of Machine Learning Research. https://proceedings.mlr.press/v130/song21a.html

↵

- Song, S., Wang, Y., & Chaudhuri, K. (2017). Pufferfish privacy mechanisms for correlated data. In *Proceedings of the 2017 ACM International Conference on Management of Data* (pp. 1291–1306). Association for Computing Machinery. https://doi.org/10.1145/3035918.3064025

    ↩

- Steed, R., Liu, T., Wu, Z. S., & Acquisti, A. (2022). Policy impacts of statistical uncertainty and privacy. *Science*, *377*(6609), 928–931. https://doi.org/10.1126/science.abq4481

    ↩

- Steinke, T., & Ullman, J. (2015). Interactive fingerprinting codes and the hardness of preventing false discovery. In Peter Grünwald, Elad Hazan, Satyen Kale (Eds.), *Proceedings of the 28th Annual Conference on Learning Theory* (Vol. 40, pp. 1588–1628). Proceedings of Machine Learning Research. https://proceedings.mlr.press/v40/Steinke15.html

    ↩

- Steinke, T., & Ullman, J. (2017a). Between pure and approximate differential privacy. *The Journal of Privacy and Confidentiality*, *7*(2), 3–22. https://doi.org/10.29012/jpc.v7i2.648

    ↩

- Steinke, T., & Ullman, J. (2017b). Tight lower bounds for differentially private selection. In *2017 IEEE 58th Annual Symposium on Foundations of Computer Science* (pp. 552–563). IEEE. https://doi.org/10.1109/FOCS.2017.57

    ↩

- Suri, A., & Evans, D. (2022). Formalizing and estimating distribution inference risks. *Proceedings on Privacy Enhancing Technologies*, *2022*(4), 528–551. https://doi.org/10.56553/popets-2022-0121

    ↩

- Suriyakumar, V. M., Papernot, N., Goldenberg, A., & Ghassemi, M. (2021). Chasing your long tails: Differentially private prediction in health care settings. In *Proceedings of the 2021 ACM Conference on Fairness, Accountability, and Transparency* (pp. 723–734). Association for Computer Machinery. https://doi.org/10.1145/3442188.3445934

    ↩

- Swanberg, M., Globus-Harris, I., Griffith, I., Ritz, A., Groce, A., & Bray, A. (2019). Improved differentially private analysis of variance. *Proceedings on Privacy Enhancing Technologies*, *2019*(3), 310–330. https://doi.org/10.2478/POPETS-2019-0049

    ↩

-

↩

- Thakkar, O., Andrew, G., McMahan, H. B., Ramaswamy, S. (2019). Differentially private learning with adaptive clipping. ArXiv. https://doi.org/10.48550/arXiv.1905.03871

↩

- Thakkar, O., Ramaswamy, S., Mathews, R., & Beaufays, F. (2020). *Understanding unintended memorization in federated learning.* ArXiv. https://doi.org/10.48550/arXiv.2006.07490

↩

- Thakurta, A. G., & Smith, A. (2013). Differentially private feature selection via stability arguments, and the robustness of the LASSO. In S. Shalev-Shwartz, & I. Steinwart (Eds.), *Proceedings of the 26th Annual Conference on Learning Theory* (Vol. 30, pp. 819–850). Proceedings of Machine Learning Research. https://proceedings.mlr.press/v30/Guha13.html

↩

- Thudi, A., Shumailov, I., Boenisch, F., & Papernot, N. (2022). *Bounding membership inference.* ArXiv. https://doi.org/10.48550/arXiv.2202.12232

↩

- Tramèr, F., & Boneh, D. (2021, May 3–7). *Differentially private learning needs better features (or much more data)* [Spotlight presentation]. ICLR 2021: 9th International Conference on Learning Representations, Virtual Event, Austria. https://openreview.net/forum?id=YTWGvpFOQD-

↩

- Tramèr, F., Kamath, G., & Carlini, N. (2022). *Considerations for differentially private learning with large-scale public pretraining.* ArXiv. https://doi.org/10.48550/arXiv.2212.06470

↩

- Tramèr, F., Terzis, A., Steinke, T., Song, S., Jagielski, M., & Carlini, N. (2022). *Debugging differential privacy: A case study for privacy auditing.* ArXiv. https://doi.org/10.48550/arXiv.2202.12219

↩

- Tramèr, F., Terzis, A., Steinke, T., Song, S., Jagielski, M., & Carlini, N. (2022). *Debugging differential privacy: A case study for privacy auditing.* ArXiv. https://doi.org/10.48550/arXiv.2202.12219

↩

- Tran, C., Dinh, M., & Fioretto, F. (2021). Differentially private empirical risk minimization under the fairness lens. In M. Ranzato, A. Beygelzimer, Y. Dauphin, P.S. Liang, & J. Wortman Vaughan (Eds.), *Advances in neural information processing systems* (Vol. 34, pp. 27555–27565). Curran Associates. https://proceedings.neurips.cc/paper/2021/hash/e7e8f8e5982b3298c8addedf6811d500-Abstract.html

↩

- Tran, C., Fioretto, F., Van Hentenryck, P., & Yao, Z. (2021). Decision making with differential privacy under a fairness lens. In Z.-H. Zhou (Ed.), *Proceedings of the Thirtieth International Joint Conference on Artificial Intelligence* (pp. 560–566). International Joint Conferences on Artificial Intelligence Organization. https://doi.org/10.24963/ijcai.2021/78

  ↩

- Tsfadia, E., Cohen, E., Kaplan, H., Mansour, Y., & Stemmer, U. (2022). FriendlyCore: Practical differentially private aggregation. In K. Chaudhuri, S. Jegelka, L. Song, C. Szepesvari, G. Niu, & S. Sabato (Eds.), *Proceedings of the 39th International Conference on Machine Learning* (Vol. 162, pp. 21828–21863). Proceedings of Machine Learning Research. https://proceedings.mlr.press/v162/tsfadia22a.html

  ↩

- Tukey, J. W. (1960). A survey of sampling from contaminated distributions. In I. Oklin (Ed.), *Contributions to probability and statistics: Essays in honor of Harold Hotelling* (pp. 448–485). Stanford University Press.

  ↩

- Tversky, A., & Kahneman, D. (1974). Judgment under uncertainty: Heuristics and biases. *Science, 185*(4157), 1124–1131. https://doi.org/10.1126/science.185.4157.1124

  ↩

- Uhlerop, C., Slavković, A., & Fienberg, S. E. (2013). Privacy-preserving data sharing for genome-wide association studies. *The Journal of Privacy and Confidentiality, 5*(1), 137–166. https://doi.org/10.29012/jpc.v5i1.629

  ↩

- Ullman, J. (2016). Answering $n^2+o(1)$ counting queries with differential privacy is hard. *SIAM Journal on Computing, 45*(2), 473–496. https://doi.org/10.1137/130928121

  ↩

- United States Census Bureau (n.d.). Federal statistical research data centers [Retrieved 3/17/23]. https://www.census.gov/about/adrm/fsrdc.html

  ↩

- Uniyal, A., Naidu, R., Kotti, S., Singh, S., Kenfack, P. J., Mireshghallah, F., & Trask, A. (2021). *DP-SGD vs PATE: Which has less disparate impact on model accuracy?* ArXiv. https://doi.org/10.48550/arXiv.2106.12576

  ↩

-

↵

- Vadhan, S. P., & Zhang, W. (2022). Concurrent composition theorems for all standard variants of differential privacy. In *Proceedings of the 55th Annual ACM Symposium on Theory of Computing* (pp. 507–519). Association for Computer Machinery. https://doi.org/10.1145/3564246.3585241

    ↵

- Vadhan, S., & Wang, T. (2021). Concurrent composition of differential privacy. In K. Nissim & B. Waters (Eds.), *Theory of cryptography* (pp. 582–604). Springer.

    ↵

- Varshney, P., Thakurta, A., & Jain, P. (2022). (Nearly) optimal private linear regression for sub-Gaussian data via adaptive clipping. In P.-L. Loh, & M. Raginsky (Eds.), *Proceedings of Thirty Fifth Conference on Learning Theory* (Vol. 178, pp. 1126–1166). Proceedings of Machine Learning Research. https://proceedings.mlr.press/v178/varshney22a.html

    ↵

- Vu, D., & Slavkovic, A. (2009). Differential privacy for clinical trial data: Preliminary evaluations. In *2009 IEEE International Conference on Data Mining Workshops* (pp. 138–143). IEEE. https://doi.org/10.1109/ICDMW.2009.52

    ↵

- Wang, D., Xiao, H., Devadas, S., & Xu, J. (2020). On differentially private stochastic convex optimization with heavy-tailed data. In H. Daumé III, & A. Singh (Eds.), *Proceedings of the 37th International Conference on Machine Learning* (Vol. 119, pp. 10081–10091). Proceedings of Machine Learning Research. . https://proceedings.mlr.press/v119/wang20y.html

    ↵

- Wang, D., Ye, M., & Xu, J. (2017). Differentially private empirical risk minimization revisited: Faster and more general. In I. Guyon, U. Von Luxburg, S. Bengio, H. Wallach, R. Fergus, S. Vishwanathan, & R. Garnett (Eds.,) *Advances in neural information processing systems* (Vol. 30, pp. 2722–2731). Curran Associates. https://papers.nips.cc/paper_files/paper/2017/hash/f337d999d9ad116a7b4f3d409fcc6480-Abstract.html

    ↵

- Wang, H., Gao, S., Zhang, H., Su, W. J., & Shen, M. (2023, December 12). *DP-HyPO: An adaptive private framework for hyperparameter optimization* [Paper presentation]. Thirty-seventh Conference on Neural Information Processing Systems, New Orleans, LA, United States.

    ↵

-

↩

- Wang, H., Zhang, Z., Wang, T., He, S., Backes, M., Chen, J., & Zhang, Y. (2023). PrivTrace: Differentially private trajectory synthesis by adaptive Markov model. In J. Calandrino, & C. Troncoso (Eds.), *Proceedings of the 32nd USENIX Conference on Security* Symposium (pp. 1649–1666). USENIX Association. https://www.usenix.org/conference/usenixsecurity23/presentation/wang-haiming

  ↩

- Wang, L., & Gu, Q. (2019). Differentially private iterative gradient hard thresholding for sparse learning. In S. Kraus (Ed.), *Proceedings of the Twenty-Eighth International Joint Conference on Artificial Intelligence* (pp. 3740–3747). International Joint Conferences on Artificial Intelligence Organization. https://doi.org/10.24963/ijcai.2019/519

  ↩

- Wang, M., Ji, Z., Kim, H.-E., Wang, S., Xiong, L., & Jiang, X. (2017). Selecting optimal subset to release under differentially private m-estimators from hybrid datasets. *IEEE Transactions on Knowledge and Data Engineering, 30*(3), 573–584. https://doi.org/10.1109%2FTKDE.2017.2773545

  ↩

- Wang, Y.-X. (2018). *Revisiting differentially private linear regression: Optimal and adaptive prediction & estimation in unbounded domain*. ArXiv. https://doi.org/10.48550/arXiv.1803.02596

  ↩

- Wang, Y.-X. (2019). Per-instance differential privacy. *The Journal of Privacy and Confidentiality, 9*(1). https://doi.org/10.29012/jpc.662

  ↩

- Wang, Y.-X., Fienberg, S., & Smola, A. (2015). Privacy for free: Posterior sampling and stochastic gradient Monte Carlo. In F. Bach, & D. Blei (Eds.), *Proceedings of the 32nd International Conference on Machine Learning* (Vol. 37, pp. 2493–2502). Proceedings of Machine Learning Research. https://proceedings.mlr.press/v37/wangg15.html

  ↩

- Wang, Y., Lee, J., & Kifer, D. (2015). *Revisiting differentially private hypothesis tests for categorical data*. ArXiv. https://doi.org/10.48550/arXiv.1511.03376

  ↩

- Watson, L., Guo, C., Cormode, G., & Sablayrolles, A. (2021). *On the importance of difficulty calibration in membership inference attacks*. ArXiv. https://doi.org/10.48550/arXiv.2111.08440

  ↩

- Whitehouse, J., Ramdas, A., Rogers, R., & Wu, Z. S. (2023). Fully-adaptive composition in differential privacy. In A. Krause, E. Brunskill, K. Cho, B. Engelhardt, S. Sabato, & J. Scarlett (Eds.), *Proceedings of the 40th International Conference on Machine Learning* (Vol. 202, pp. 36990–37007). Proceedings of Machine Learning Research. https://proceedings.mlr.press/v202/whitehouse23a.html

    ↩

- Whitehouse, J., Wu, Z. S., Ramdas, A., & Rogers, R. (2022). *Brownian noise reduction: Maximizing privacy subject to accuracy constraints*. ArXiv. https://doi.org/10.48550/arXiv.2206.07234

    ↩

- Wood, A., Altman, M., Bembenek, A., Bun, M., Gaboardi, M., Honaker, J., Nissim, K., O'Brien, D., Steinke, T., & Vadhan, S. (2018). Differential privacy: A primer for a non-technical audience. *Vanderbilt Journal of Entertainment and Technology Law*, *21*(1), 209–276. https://scholarship.law.vanderbilt.edu/jetlaw/vol21/iss1/4/

    ↩

- Wu, X., Li, F., Kumar, A., Chaudhuri, K., Jha, S., & Naughton, J. (2017). Bolt-on differential privacy for scalable stochastic gradient descent-based analytics. In *Proceedings of the 2017 ACM SIGMOD International Conference on Management of Data* (pp. 1307–1322). Association for Computer Machinery. https://doi.org/10.1145/3035918.3064047

    ↩

- Xiao, Y., & Xiong, L. (2015). Protecting locations with differential privacy under temporal correlations. In *CCS '15: Proceedings of the 22nd ACM SIGSAC Conference on Computer and Communications Security* (pp. 1298–1309). Association for Computing Machinery. https://doi.org/10.1145/2810103.2813640

    ↩

- Xie, L., Lin, K., Wang, S., Wang, F., & Zhou, J. (2018). *Differentially private generative adversarial network*. ArXiv. https://doi.org/10.48550/arXiv.1802.06739

    ↩

- Xiong, A., Wu, C., Wang, T., Proctor, R., Blocki, J., Li, N., & Jha, S. (2022). *Using illustrations to communicate differential privacy trust models: An investigation of users' comprehension, perception, and data sharing decision*. ArXiv. https://doi.org/10.48550/arXiv.2202.10014

    ↩

- Xu, D., Du, W., & Wu, X. (2020). *Removing disparate impact of differentially private stochastic gradient descent on model accuracy*. ArXiv. https://doi.org/10.48550/arXiv.2003.03699

    ↩

- Ye, M., & Barg, A. (2018). Optimal schemes for discrete distribution estimation under locally differential privacy. *IEEE Transactions on Information Theory*, *64*(8), 5662–5676. https://doi.org/10.1109/ISIT.2017.8006630

    ↩

- Yeom, S., Giacomelli, I., Fredrikson, M., & Jha, S. (2018). Privacy risk in machine learning: Analyzing the connection to overfitting. In *2018 IEEE 31st Computer Security Foundations Symposium* (pp. 268–282). IEEE. https://doi.org/10.1109/CSF.2018.00027

    ↩

- Yin, H., Mallya, A., Vahdat, A., Alvarez, J. M., Kautz, J., & Molchanov, P. (2021). See through gradients: Image batch recovery via GradInversion. In *2021 IEEE/CVF Conference on Computer Vision and Pattern Recognition* (pp. 16337–16346). IEEE. https://doi.org/10.1109/CVPR46437.2021.01607

    ↩

- Yousefpour, A., Shilov, I., Sablayrolles, A., Testuggine, D., Prasad, K., Malek, M., Nguyen, J., Gosh, S., Bharadwaj, A., Zhao, J., Cormode, G., & Mironov, I. (2021). *Opacus: User-friendly differential privacy library in PyTorch*. ArXiv. https://doi.org/10.48550/arXiv.2109.12298

    ↩

- Yu, D., Naik, S., Backurs, A., Gopi, S., Inan, H. A., Kamath, G., Kulkarni, J., Lee, Y. T., Manoel, A., Wutschitz, L., Yekhanin, S., & Zhang, H. (2022, April 25–29). *Differentially private fine-tuning of language models* [Poster presentation]. ICLR 2022: The Tenth International Conference on Learning Representations, Virtual Event. https://openreview.net/forum?id=Q42f0dfjECO

    ↩

- Yu, D., Zhang, H., Chen, W., & Liu, T.-Y. (2021, May 3–7). *Do not let privacy overbill utility: Gradient embedding perturbation for private learning* [Poster presentation]. ICLR 2021: 9th International Conference on Learning Representations, Virtual Event, Austria. https://openreview.net/forum?id=7aogOj_VYO0

    ↩

- Yu, D., Zhang, H., Chen, W., Liu, T.-Y., & Yin, J. (2019). *Gradient perturbation is underrated for differentially private convex optimization*. ArXiv. https://doi.org/10.48550/arXiv.1911.11363

    ↩

- Yu, D., Zhang, H., Chen, W., Yin, J., & Liu, T.-Y. (2021). Large scale private learning via low-rank reparametrization. In M. Meila & T. Zhang (Eds.), *Proceedings of the 38th International Conference on Machine Learning* (Vol. 139, pp. 12208–12218). Proceedings of Machine Learning Research. https://proceedings.mlr.press/v139/yu21f.html

↩

- Yu, L., Liu, L., Pu, C., Gursoy, M. E., & Truex, S. (2019). Differentially private model publishing for deep learning. In *2019 IEEE Symposium on Security and Privacy* (pp. 332–349). IEEE. https://doi.org/10.1109/SP.2019.00019

  ↩

- Zanella-Béguelin, S., Wutschitz, L., Tople, S., Rühle, V., Paverd, A., Ohrimenko, O., Köpf, B., & Brockschmidt, M. (2020). Analyzing information leakage of updates to natural language models. In *CCS '20: Proceedings of the 2020 ACM SIGSAC conference on computer and communications security* (pp. 363–375). Association for Computing Machinery. https://doi.org/10.1145/3372297.3417880

  ↩

- Zanella-Béguelin, S., Wutschitz, L., Tople, S., Salem, A., Rühle, V., Paverd, A., Naseri, M., Köpf, B., & Jones, D. (2022). *Bayesian estimation of differential privacy*. ArXiv. https://doi.org/10.48550/ARXIV.2206.05199

  ↩

- Zhang, H., Kamath, G., Kulkarni, J., & Wu, Z. S. (2020). Privately learning Markov random fields. In H. Daumé III, & A. Singh (Eds.), *Proceedings of the 37th International Conference on Machine Learning* (Vol. 119, pp. 11129–11140). Proceedings of Machine Learning Research. https://proceedings.mlr.press/v119/zhang20l.html

  ↩

- Zhang, J., Cormode, G., Procopiuc, C. M., Srivastava, D., & Xiao, X. (2017). Privbayes: Private data release via Bayesian networks. *ACM Transactions on Database Systems (TODS), 42*(4), 1–41. https://doi.org/10.1145/3134428

  ↩

- Zhang, Q., Ma, J., Lou, J., & Xiong, L. (2021). Private stochastic non-convex optimization with improved utility rates. In *Proceedings of the Thirtieth International Joint Conference on Artificial Intelligence* (pp. 3370–3376). International Joint Conferences on Artificial Intelligence Organization. https://doi.org/10.24963/ijcai.2021/464

  ↩

- Zhang, W., Ohrimenko, O., & Cummings, R. (2022). Attribute privacy: Framework and mechanisms. In *Proceedings of the 2022 ACM Conference on Fairness, Accountability, and Transparency* (pp. 757–766). Association for Computer Machinery. https://doi.org/10.1145/3531146.3533139

  ↩

-

↩

- Zhao, B., Mopuri, K. R., & Bilen, H. (2020). iDLG: Improved deep leakage from gradients. ArXiv. https://doi.org/10.48550/arXiv.2001.02610

↩

- Zhou, Y., Chen, X., Hong, M., Wu, Z. S., & Banerjee, A. (2020). *Private stochastic non-convex optimization: Adaptive algorithms and tighter generalization bounds*. ArXiv. https://doi.org/10.48550/arXiv.2006.13501

↩

- Zhou, Y., Wu, Z. S., & Banerjee, A. (2021, May 3–7). *Bypassing the ambient dimension: Private SGD with gradient subspace identification* [Poster presentation]. ICLR 2021: 9th International Conference on Learning Representations, Virtual Event, Austria. https://openreview.net/forum?id=7dpmlkBuJFC

↩

- Zhu, L., Liu, Z., & Han, S. (2019). Deep leakage from gradients. In H. Wallach, H. Larochelle, A. Beygelzimer, F. d'Alché-Buc, E. Fox, & R. Garnett (Eds.), *Advances in neural information processing systems* (Vol. 32, pp. 14774–14784). Curran Associates. https://papers.nips.cc/paper_files/paper/2019/hash/60a6c4002cc7b29142def8871531281a-Abstract.html

↩

- Zhu, Y., & Wang, Y.-X. (2022). Adaptive private-k-selection with adaptive k and application to multi-label PATE. In G. Camps-Valls, F. J. R. Ruiz, & I. Valera (Eds.), *Proceedings of The 25th International Conference on Artificial Intelligence and Statistics* (Vol. *151*, pp. 5622–5635). Proceedings of Machine Learning Research. https://proceedings.mlr.press/v151/zhu22e.html

↩

- Zhu, Y., Yu, X., Chandraker, M., & Wang, Y.-X. (2020). Private-kNN: Practical differential privacy for computer vision. In *2020 IEEE/CVF Conference on Computer Vision and Pattern Recognition* (pp. 11854–11862). IEEE. https://doi.org/10.1109/CVPR42600.2020.01187

↩